\documentclass[aps,10pt,showpacs,amsmath,amssymb,prd,onecolumn,floatfix,a4,nofootinbib]{revtex4}
\preprint
\tighten
\usepackage{graphicx}
\usepackage{subfigure}     \usepackage{comment}    \usepackage{feynmf}
\usepackage{dcolumn}
\usepackage{bm}
\usepackage{epsfig} \usepackage{natbib}
 
\def\be{\begin{eqnarray}}  
\def\ee{\end{eqnarray}}
\def\ba{\begin{eqnarray}}
\def\ea{\end{eqnarray}}

\begin{document}

\author{Jamie  Portsmouth}  \affiliation{Department  of  Astrophysics,
  Oxford   University}   \email{jamiep@astro.ox.ac.uk}  \author{Edmund
  Bertschinger}     \affiliation{Department    of     Physics,    MIT}
\email{edbert@mit.edu}


\title{A tensor  formalism  for  transfer and Compton
 scattering of  polarized light \label{ch1}}

\begin{abstract}

A  novel covariant  formalism for  the treatment  of the  transfer and
Compton scattering of partially polarized light is presented.  
This was initially developed to aid in the computation of relativistic
corrections to the polarization generated by the Sunyaev-Zeldovich
effect (demonstrated in a companion paper), but it is of more general
utility. In this approach, the  polarization state  of a light  beam is
described  by a tensor constructed from the time  average of quadratic
products of the electric  field  components in  a  local  observer
frame. This  leads naturally to  a covariant description which is
ideal for calculations involving  the  boosting  of  polarized light
beams  between  Lorentz frames,  and is more  flexible than  the
traditional  Stokes parameter approach  in which  a separate  set of
polarization basis  vectors is required for each photon.
The  covariant kinetic  equation for  Compton scattering  of partially
polarized light  by relativistic electrons  is obtained in  the tensor
formalism by a heuristic semi-classical line of reasoning. The kinetic
equation is  derived first in the  electron rest frame  in the Thomson
limit,  and then  is generalized  to account  for electron  recoil and
allow for scattering from an arbitrary distribution of electrons.


\end{abstract}

\date{\today}

                              \maketitle

                       \setcounter{section}{0}

The theory of transfer and scattering of polarized light is 
fundamental to astrophysics and cosmology, and there is
an extensive literature dealing with the subject
\citep{Wolf1959, Chandrasekhar1960,
  Dialetis1969, 1974ApJ...191..567A, 1994ApJ...427..288H,
  1994MNRAS.267..303L, 1995ApJ...441..400C, 1999MNRAS.306..153H,
  2000PhRvD..62d3004C, 2000PhRvE..61.2024C}.  
Most treatments write the transfer equation using the four Stokes
parameters $I,Q,U,V$, which provide a complete description of the
radiation field (there also exist several other approaches to the
description of the polarization properties of radiation fields, for
example the Jones calculus, Mueller matrices, and coherency matrices; see \cite{Swindell}). These parameters have dimensions of specific
intensity, and are functions of time, photon propagation direction,
and frequency.  In the case of unpolarized photons, a complete
description of the radiation field is given by the total specific
intensity Stokes parameter $I_{\nu}$, or equivalently the phase space
density of photons. The Stokes parameters are defined with respect to
a set of two orthonormal polarization basis vectors, normal to the
photon direction, which must be specified for each possible photon
direction. 

However, in Comptonization calculations with polarized photons, 
this Stokes parameter formalism
becomes very cumbersome --- 
the elegance of reducing the description of the radiation field
to four functions is achieved at the expense of a very complicated
transfer equation.
Compton scattering involves a relativistic scattering
electron in general, and the complete transfer equation involves
Lorentz transformation of the Stokes parameters, which is somewhat
complicated.
To get around this difficulty, 
in the context of computing relativistic corrections to the
polarization generated by the Sunyaev-Zeldovich effect, 
we found it convenient and illuminating
to introduce a novel formalism for doing radiative transfer
calculations with polarized photons, the \emph{polarization tensor
  formalism}.  
In a companion paper (\citep{Portsmouth2004c}, hereafter referred to
as Paper II) we apply this formalism to a calculation of the
polarization generated in the Sunyaev-Zeldovich effect.
While the new description is superficially a little
more complicated than the Stokes parameter approach, it leads to
a form of transfer equation which is easier to manipulate and ideal
for calculations involving  the  boosting  of  polarized light  beams
between  Lorentz frames.  
In this paper we will restrict discussion to Compton scattering,
but our methods could be extended to other scattering processes
without great difficulty.

Our approach is closest in spirit to the ``coherency matrix''
formalism introduced by \cite{Wolf1959}
(a review of this, as a lead in to our formalism, is given
in \S\ref{ch1:sec1}).
The basic idea is that since the Stokes parameters are essentially 
time averages of certain quadratic products of the electric field
components of the electromagnetic field (as measured in a local
observer frame), by expressing the electric fields in terms of the
Maxwell field strength tensor components this leads naturally to
a classical formalism in which the four Stokes parameters are replaced
by a two index complex Hermitian tensor $I^{\mu\nu}$.
The trace of $I^{\mu\nu}$ is the usual total intensity Stokes parameter.
A tensor analogue of the phase space distribution function,
$f^{\mu\nu}$, is also easily defined.  These objects are collectively
termed the \emph{polarization tensor} (or matrix).
A similar formalism has been developed by Challinor
\cite{2000PhRvD..62d3004C}, but he did not provide a clear
physically motivated derivation of the form of the Boltzmann equation
for Compton scattering, which we provide here. 

In this  matrix formalism (in flat space) we associate a $3\times 3$
matrix with each photon, rather than a set of polarization basis
vectors and the associated Stokes parameters. For example, a beam of
partially-polarized light travelling in the $z$-direction is described
by the Hermitian matrix (termed the \emph{polarization matrix}):
\begin{equation}
\bm{I} = [I_{ij}] = \frac{1}{2} \left(
\begin{array}{ccc}
I+Q & U+iV & 0 \\ U-iV & I-Q & 0 \\ 0 & 0 & 0 \end{array} \right) \ ,
\end{equation}
where $I, Q, U, V$ are the Stokes parameters (with units of specific
intensity).  The trace of the matrix is the total beam intensity. For
a general photon direction $\bm{n}$, the beam is described by a matrix
$\bm{I}(\bm{n})$, and transversality of the polarization implies $n^i
I_{ij}=0$. In general, the polarization matrix is a function of photon
frequency and direction as well as spatial position and time,
$\bm{I}=\bm{I}(\nu,\bm{n},\bm{x},t)$.  

The real advantage of this
description is that there is no need to perform a complicated rotation
of axes when examining photons with different direction vectors (in
the angular integrations needed in the radiative transfer equation for
example).  In addition, it is simple to extend the $3\times 3$ matrix
description to a $4\times 4$ manifestly covariant tensor description
in which Lorentz transformation of polarized beams between frames is
easy. In the matrix approach, the radiative transfer equation for
scattering of polarized radiation is much more straightforward than in
the Stokes approach.  There is no need for rotation of axes to define
separate Stokes parameters for the incoming and outgoing beams.  Both
are described by a single polarization matrix.  The
transfer equation for Thomson scattering is elegantly expressed in
terms of a set of projection matrices $\bm{P}(\bm{n}_s)$ which project
out of the matrix $\bm{I}(\bm{n})$ the component of polarization
orthogonal to $\bm{n}_s$.
By contrast, when using
Stokes parameters one has a complicated angular integral involving
rotation matrices with Euler angles
\citep{1999MNRAS.306..153H,Chandrasekhar1960}.

We now outline the structure of the paper.
In \S\ref{ch1:sec1}, the Stokes and coherency matrix formalisms
are reviewed.
In \S\ref{ch1:sec2}, this notion is generalized and our tensor
description of polarized light described, first in a non-covariant
manner.  The covariant formalism is introduced in \S\ref{ch1:sec3},
and we discuss
the properties of the polarization tensors, their evolution in the
absence of scattering and in the geometrical optics limit, and their
relation to the Stokes description. In \S\ref{ch1:sec4},
the behaviour of the polarization tensor under Lorentz transformation
is discussed, and an explicit example of the computation of the
polarization of a boosted beam presented.
In \S\ref{ch3:sec1}, the
classical non-relativistic physics of the generation of polarization
by Thomson scattering in the electron rest frame is discussed using
the polarization tensor approach. An equation for the time evolution
of the distribution function polarization tensor in the electron rest
frame due to Thomson scattering is derived.  Then in \S\ref{ch3:sec2}
we derive the Boltzmann collision integral using a phenomenological
approach based on the master equation of kinetic theory, still in the
Thomson limit.  As a check, we construct the matrix analogue of the
radiative transfer equation in the case of a scattering medium
composed of stationary electrons, which agrees with the results of
\cite{Chandrasekhar1960}.  In \S\ref{ch3:sec3} the full relativistic
kinetic equation is obtained, working in the rest frame of the initial
electron -- following the procedure used in the Thomson limit, but
using the Klein-Nishina cross section and taking into account recoil.
The transformation to a common lab frame is then taken, to obtain the
kinetic equation for scattering from electrons with a general
distribution of velocities.  We check that this can be expressed in a
manifestly covariant form.

Note that throughout the paper, boldface quantities, e.g. $\bm{p}$,
denote 3-vectors, and quantities with vector arrows, e.g. $\vec{p}$,
denote 4-vectors. The indices of 3-vectors and tensors are denoted
with Roman indices, and those of 4-vectors and tensors with Greek
indices. Both $3\times 3$ and $4\times 4$ matrices are denoted with
boldface quantities.  Note also that $\Re$ denotes the set of real
numbers, $\mathcal{C}$ denotes the set of complex numbers, and
$\Re\mbox{e}$ denotes the operation of taking the real part.

\section{The coherency matrix \label{ch1:sec1}}

The classical description of partially polarized light uses the well
known Stokes parameters, which are defined operationally in terms of
experiments with polarizing plates (the most complete treatment of the
Stokes parameter formalism for polarized radiative transfer is
contained in the monograph \citep{Chandrasekhar1960}).
Physically the Stokes parameters
can be thought of as time averages of instantaneous products of
electric field components.  
There is a close relationship between the
Stokes parameters and the notion of the coherence of the two photon
polarization states, which is described mathematically by the
\emph{coherency matrix} introduced by \cite{Wolf1959}, based on the
work of \cite{Wiener1930}.  Additional work was done by
\cite{Barakat1963} to extend the concept to a spectral coherency
matrix.  It is worthwhile reviewing the notion of the coherency
matrix, since this leads naturally to the polarization tensor
description.

We will only consider electromagnetic fields which are superpositions
of plane electromagnetic waves. An idealized superposition of such
waves whose wave-vectors are all perfectly aligned will be termed a
\emph{beam}.  Consider first a beam propagating along the $z$-axis.
The transverse electric field components at a specified fixed spatial
point $(x,y,z)$ are real functions of time,
$E_x^{(r)}(t),\;E_y^{(r)}(t)$.  These functions can be expressed as a
superposition of an infinite number of monochromatic waves with
arbitrary phases, i.e. as Fourier transforms
\begin{equation} \label{fourier1}
E_j^{(r)}(t) = \frac{1}{\sqrt{2\pi}} \int_{-\infty}^{\infty}
\tilde{E}_j(\omega) e^{-i\omega t} d\omega \ .  \quad\quad\quad
E_j^{(r)}\in\Re, \;\;j \in \{x,y\}
\end{equation}
We have assumed here of course that the Fourier transform exists --
which is not true for all functions $E_x^{(r)}(t),\;E_y^{(r)}(t)$, but
we will gloss over this point (the existence of the Fourier transform
can be assured without difficulty by working with functions which are
truncated as $t\rightarrow \pm\infty$.  See for example
\cite{BornWolf}).
In order to ensure reality of $E_j^{(r)}(t)$, the Fourier transforms
must satisfy $\tilde{E}_j(-\omega)=\tilde{E}_j^*(\omega)$.  Now we
split the integral above into two parts:
\begin{eqnarray} \label{ch1:analyticsignal}
E_j^{(r)}(t) &=& \frac{1}{\sqrt{2\pi}} \int_0^{\infty}
\tilde{E}_j(\omega) e^{-i\omega t} d\omega +
\frac{1}{\sqrt{2\pi}}\int_{-\infty}^0 \tilde{E}_j(\omega) e^{-i\omega
  t} d\omega \nonumber \\ &=& \frac{1}{2} \left(E_j(t)+E_j^*(t)\right)
= \Re\mbox{e} \;E_j(t) \ ,
\end{eqnarray}
where we have defined the complex functions $E_j(t)$, conventionally
called the \emph{analytic signal} \citep{BornWolf} associated with
$E_j^{(r)}(t)$:
\begin{equation}\label{ch1:anlsig}
E_j(t) = \frac{2}{\sqrt{2\pi}} \int_0^{\infty} \tilde{E}_j(\omega)
e^{-i\omega t} d\omega \ . \quad\quad\quad E_j(t) \in \mathcal{C}
\end{equation}
We may decompose $\tilde{E}_j(\omega)$ uniquely into a real amplitude
and complex phase factor:
\begin{equation}
\tilde{E}_j(\omega) = a_j(\omega) e^{i\phi_j(\omega)} \ .
\quad\quad\quad a_j, \phi_j \in \Re
\end{equation}
The analytic signal is thus
\begin{equation} \label{ch1:alsft}
E_j(t) = \frac{2}{\sqrt{2\pi}} \int_0^{\infty} a_j(\omega)
e^{i\phi_j(\omega)-i\omega t} d\omega \ .
\end{equation}
How are all of these quantities related to what is measured by a real
polarimeter? Generally speaking, polarimeters measure the time average
of the intensity of the light beam at a fixed spatial point after it
has traveled through a combination of filters (See e.g.
\cite{2000ApJ...532.1240B} for a good general discussion of
astronomical polarimetry).  The two basic filter elements required to
measure the polarization state are a \emph{polarizing plate}, and a
\emph{compensator} \citep{Stone1963}.  We shall describe how the time
average is constructed from the quantities we have defined, and then
consider the effect of the two types of filter on the beam.

We first make the simplifying assumption that the beam is
\emph{quasi-monochromatic}, which means that the functions
$\tilde{E}_j(\omega)$ are assumed to be non-vanishing only in a narrow
frequency band
$\omega\in[\omega_0-\Delta\omega/2,\omega_0+\Delta\omega/2]$, with
$\Delta\omega \ll \omega_0$. Physically this means that the beam is a
wave-packet of spectral width $\Delta\omega$, centered roughly on
frequency $\omega_0$.  This implies that the functions $a_j(t),
\phi_j(t)$ vary slowly in comparison to $\cos(\omega_0 t)$.  To see
this, first note that we can always choose to write the analytic
signals in the form
\begin{equation}
E_j(t) = a_j(t) e^{i[\phi_j(t)-\omega_0 t]} \ .
\end{equation}
Then it follows from Eqn.~(\ref{ch1:anlsig}) that
\begin{eqnarray}
a_j(t) e^{i\phi_j(t)} &=& \frac{2}{\sqrt{2\pi}} \int_0^{\infty}
\tilde{E}_i(\omega) e^{-i(\omega-\omega_0)t} d\omega \nonumber \\ &=&
\frac{2}{\sqrt{2\pi}} \int_{-\omega_0}^{\infty}
\tilde{E}_i(\omega'+\omega_0) e^{-i\omega' t} d\omega' \ .
\end{eqnarray}
Then since $\tilde{E}_i(\omega'+\omega_0)$ vanishes by assumption for
$\vert\omega'\vert>\Delta\omega/2$, the left hand side is a
superposition of Fourier modes of low frequency $\vert\omega'\vert <
\Delta\omega/2 \ll \omega_0$.

Then the time average is defined by
\begin{equation} \label{taverage}
\left< E_j^{(r)}(t) \right> \equiv \frac{1}{2T} \int_{t-T}^{t+T}
E_j^{(r)}(t) dt \ ,
\end{equation}
where $T$ is chosen such that $\Delta \omega \ll \frac{2\pi}{T} \ll
\omega_0$.
The quantities measured by the detector will be some combination of
the following time averaged real quantities (expanding using
Eqn.~(\ref{ch1:analyticsignal})):
\begin{equation}
\left< E_i^{(r)}(t)E_j^{(r)}(t) \right> = \frac{1}{4}\left[\langle
  E_i(t) E_j(t) \rangle + \langle E_i(t) E^*_j(t) \rangle + \langle
  E^*_i(t) E_j(t) \rangle + \langle E^*_i(t) E^*_j(t) \rangle \right]
\ .
\end{equation}
With the assumption of quasi-monochromaticity we may now ignore time
averages which contain the rapidly varying phase factor $e^{i\omega_0
  t}$ and retain only those over the slowly varying functions
$a_j(t)$, $e^{i\phi_j(t)}$. Thus, for example
\begin{eqnarray}
\left<E_x(t) E_x(t)\right> &=& \left< a_x^2(t)
e^{2i\phi_x(t)}e^{-2i\omega_0 t}\right> \nonumber \\ &=& 0 \nonumber
\\ \left<E_x(t) E^*_x(t)\right> &=& \left< a_x^2(t) \right> \nonumber
\\ \left<E_x(t) E_y(t)\right> &=& 0 \nonumber \\ \left<E_x(t)
E^*_y(t)\right> &=& \left< a_x(t) a_y(t) e^{i(\phi_x(t)-\phi_y(t))}
\right> \ .
\end{eqnarray}
The non-vanishing elements are all of the form $J_{ij}=\langle E_i(t)
E^*_j(t)\rangle$.  We denote the Hermitian matrix of quantities
$J_{ij}$ the \emph{coherency matrix}:
\begin{equation} \label{ch1:def22coh}
\bm{J} = \left[\begin{array}{cc} \langle a_x^2(t) \rangle & \langle
    a_x(t) a_y(t) e^{i(\phi_x(t)-\phi_y(t))} \rangle \\ \langle a_x(t)
    a_y(t) e^{i(\phi_y(t)-\phi_x(t))} \rangle & \langle a_y^2(t)\rangle
\end{array} \right] \ .
\end{equation}
This matrix was introduced by \cite{Wolf1959}.  Now we relate the
elements of the coherency matrix to measurements with a polarimeter.
With an optical element known as a compensator, a coherent phase delay
between the $x$ and $y$ components of the beam can be
introduced. After passing through this device, the resulting analytic
signal has the form
\begin{equation}
\bar{E}_j(t) = a_j(t) e^{i[\phi_j(t)-\epsilon_j-\omega_0 t]} \ ,
\end{equation}
where the phase difference $\delta\equiv \epsilon_x-\epsilon_y$ is a
known constant.  Taking time averages of products of these quantities
yields
\begin{eqnarray}
\langle \bar{E}_i(t) \bar{E}_j^*(t)\rangle &=&
e^{-i(\epsilon_i-\epsilon_j)} \langle a_i(t) a_j(t)
e^{i(\phi_i(t)-\phi_j(t))} \rangle \nonumber \\ &=&
e^{-i(\epsilon_i-\epsilon_j)} J_{ij} \ .
\end{eqnarray}
The polarization is measured by passing the beam through a further
optical element, a polarizing plate oriented at angle $\theta$ to the
$x$--direction, and measuring the total intensity of the transmitted
light, $I(\theta)$. The transmitted electric field is
\begin{equation}
\bar{E}^{(r)}(\theta,t) = \bar{E}_x^{(r)}(t) \cos\theta +
\bar{E}_y^{(r)}(t) \sin\theta \ .
\end{equation}
The measured intensity is thus
\begin{eqnarray}
I(\theta) &=& 2\left<\bar{E}^{(r)}(\theta,t)^2\right> \nonumber \\ &=&
J_{xx} \cos^2\theta + J_{yy}\sin^2\theta + \sin(2\theta) \left[J_{xy}
  e^{-i\delta}+J_{yx}e^{i\delta}\right] \ .
\end{eqnarray}
The Stokes parameters are then identified as
\begin{eqnarray}\label{ch1:defstokes}
I &=& J_{xx}+J_{yy} = \langle a_x^2(t)\rangle + \langle
a_y^2(t)\rangle \ , \nonumber \\ Q &=& J_{xx}-J_{yy} = \langle
a_x^2(t)\rangle - \langle a_y^2(t)\rangle \ , \nonumber \\ U &=&
J_{xy}+J^*_{xy} = 2\langle a_x(t) a_y(t)
\cos(\phi_x(t)-\phi_y(t))\rangle \ , \nonumber \\ V &=&
-i\left[J_{xy}-J^*_{xy}\right] = 2\langle a_x(t) a_y(t)
\sin(\phi_x(t)-\phi_y(t))\rangle \ .
\end{eqnarray}
The measured intensity in terms of the Stokes parameters is:
\begin{equation}
I(\theta) = \cos^2\theta(I+Q) + \sin^2\theta (I-Q) + \frac{1}{2}
\sin(2\theta) \left[(U+iV)e^{-i\delta}+(U-iV)e^{+i\delta}\right] \ .
\end{equation}
%
The Stokes parameters can thus be determined by choosing various
combinations of $\delta$ and $\theta$ and measuring $I(\theta)$.
%
(Note that the assumption of quasi-monochromaticity is actually not
necessary to define the Stokes parameters, e.g. see \cite{Wolf1959}).

A few properties of the Stokes parameters and the associated coherency
matrix $\bm{J}$ are worth noting.  The Stokes $Q$ parameter measures
the amount of linear polarization in the beam in the $x$ or
$y$--directions.  $U$ measures the linear polarization in the
directions at an angle $\pi/4$ to the $x$ axis in the $x$--$y$
plane. $V$ measures the amount of circular polarization.  If the wave
is perfectly monochromatic, the amplitudes and phases of the electric
field components do not vary in time. Then we may remove the time
average brackets in Eqn.(\ref{ch1:defstokes}) and there is the
following relation between the Stokes parameters:
\begin{equation} \label{ch1:stokes_con_mono}
I^2 = Q^2 + U^2 + V^2 \ .
\end{equation}
For the general case, this constraint becomes an inequality instead:
\begin{equation} \label{ch1:stokes_con_part}
I^2 \ge Q^2 + U^2 + V^2 \ .
\end{equation}
The matrix $\bm{J}$ is obviously Hermitian, $J^*_{ij}=J_{ji}$.  The
determinant of $\bm{J}$ is:
\begin{equation} \label{ch1:detJ}
\mbox{det}[\bm{J}] = \frac{1}{4}\left[I^2-(Q^2+U^2+V^2)\right] \ge 0 \
.
\end{equation}
The \emph{polarization magnitude} $\Pi$ (or degree of polarization) is
a dimensionless quantity defined by
\begin{equation}
\Pi^2 \equiv \frac{Q^2+U^2+V^2}{I^2} = 1 - 4\;\mbox{det}[\bm{J}]/I^2 \
.
\end{equation}
Note that most authors use the dimensionless polarization magnitude
$\Pi$ as defined here, but some prefer to use dimensions of specific
intensity (by multiplication by the total intensity) or brightness
temperature.  A beam with $\Pi=0$ is said to be \emph{unpolarized}.  A
beam with $\Pi=1$ is said to be a \emph{pure state} (this terminology
stems from the analogy
\footnote{See for example \cite{Simmons1970} and
  \cite{1994PhDT........24K}.}  between coherency matrices and density
matrices in quantum mechanics). By Eqn.~(\ref{ch1:stokes_con_mono}), a
perfectly monochromatic beam (as opposed to a quasi-monochromatic
beam) is a pure state.


If several quasi-monochromatic beams all with the same mean frequency
are superimposed, and the electric fields of each beam have phases
which are varying completely independently of the phases of the other
beams, then the coherency matrix of the total beam is simply the sum
of the coherency matrices of the separate beams.  An elementary proof
may be found in \cite{Rybicki1979} --- the gist is that in the forming
the time average of the quadratic products of the sum of the electric
fields, the cross terms between separate beams vanish (by the
assumption of the independence of the phases). Beams with electric
fields with no permanent phase relations are said to be
\emph{incoherent}.  We will always assume, in summing two beams with
the same direction and frequency, that the beams are incoherent and
thus that the coherency matrices may be summed.

A general polychromatic beam can be constructed by superimposing an
arbitrary number of quasi-monochromatic beams. 
The coherency matrix elements and Stokes parameters may then 
be considered to be functions of frequency (spectral Stokes
parameters). It is worth mentioning here that there is an alternative
approach which yields spectral Stokes parameters without going via the
route of the assumption of quasi-monochromaticity,
developed by \citep{Barakat1963}.
This uses some of the machinery of the theory of stochastic
properties (see for example \citep{Ochi}).
We will give a brief description.
Consider again the example of a beam propagating
in the $z$-direction with (real) electric field components 
$E^{(r)}_i(t), i,j \in \{x,y\}$.
First the \emph{auto-correlation}
($R_{ii}(\tau)$) and 
\emph{cross-correlation} ($R_{ij}(\tau)), \;i\ne j$) functions are defined:
\be
R_{ij}(\tau) \equiv \langle E^{(r)}_i(t) E^{(r)}_j(t+\tau) \rangle \quad
\quad\quad i,j \in \{x,y\} \ .
\ee
Here the electric field components  are assumed to
be \emph{stationary stochastic processes}, and the correlation
functions are accordingly functions only of the time difference $\tau$.
The angle brackets denote the expectation value. With the assumption
of ergodicity (essentially that the expectation value is equivalent to
a time average over a sufficiently long time interval), the expectation value is given by Eqn.~(\ref{taverage}). 
We now define the \emph{auto-spectral density functions}
$S_{ii}(\omega)$ by a Fourier transform:
\be \label{WKT}
S_{ii}(\omega) \equiv \frac{1}{\pi} \int_{\infty}^{\infty} R_{ii}(\tau)
e^{-i\omega \tau} d\tau \ .
\ee
It follows from the assumption of ergodicity that
these give the power in a given Fourier mode of the corresponding
electric field component:
\be
S_{ii}(\omega) = \lim_{T\rightarrow \infty} \frac{1}{2\pi T} \vert \tilde{E}_i(\omega)\vert^2 \ ,
\ee
where $\tilde{E}_i(\omega)$ are the Fourier transforms defined in
Eqn.~(\ref{fourier1}). Thus Eqn.~(\ref{WKT}) is just the
Wiener-Khintchine theorem. The $S_{ii}$ are real.
Similarly the \emph{cross-spectral density functions} $S_{ij}(\omega),
\;i\ne j$ are defined by
\be
S_{ij}(\omega) \equiv \frac{1}{\pi} \int_{-\infty}^{\infty} R_{ij}(\tau)
e^{-i\omega \tau} d\tau \quad (i\ne j) \ .
\ee
The $S_{ij} (i\ne j)$ are complex in general, and satisfy
$S_{xy}(\omega) = S_{yx}^*(\omega)$.
In Barakat's scheme, the functions $S_{ij}(\omega)$ are related to 
the spectral Stokes parameters in a completely analagous fashion to
the coherency matrix 
elements, except there is no need to invoke quasi-monochromaticity.
This is perhaps a more satisfactory approach than the original
treatment of Wolf.

The polarization state and intensity of the beam associated with each
frequency may also be considered to be a function of time.  One can
imagine decomposing the beam into a time series and Fourier analyzing
successive segments of the time series to obtain the time dependence
of each Fourier mode (this is what is actually done in polarimetric
measurements of the time dependence of spectral Stokes parameters, see
e.g. \cite{2001Natur.411..662C}.).
    
In the next section, we describe a generalization of these
coherency matrix methods to photon beams propagating in arbitrary
spatial directions, which is the basis of our radiative transfer
formalism.

\section{A tensor generalization of the coherency matrix \label{ch1:sec2}}
The polarization state and intensity of a beam of light propagating in
the $z$--direction is characterized completely by the $2\times 2$
Hermitian coherency matrix $J_{ij}$, with $(i,j)\in\{x,y\}$.  There are several
papers which study a description of polarized radiation transfer using
the $2\times 2$ coherency matrix
\citep{1974ApJ...191..567A,1978AN....299...13D, 1989A&A...219...25B,
  1989CQGra...6.1171B, 1990CQGra...7.2367B, 1994PhDT........24K}.  An
obvious generalization is to allow $(i,j)$ to become Cartesian tensor
indices and to run over all of $\{x,y,z\}$. We obtain a $3\times 3$
matrix:
\begin{equation} \label{ch1:def33pol}
Q_{ij} = \left< E_i(t) E_j^*(t)\right>, \quad\quad (i,j) \in \{x,y,z\}
\ .
\end{equation}
This matrix and its 4-dimensional generalization is one of the main
tools in our formalism. It differs from the usual $2\times 2$ coherency
matrix in that it is $3\times 3$, the extra dimension corresponding to
the direction of photon propagation $\bm{n}$.  Adding the extra
dimension (and a fourth, when we introduce the covariant form in the
next section) makes it much easier to handle the computation of the
polarization of photons after general rotations, Lorentz boosts, and
scattering.

To our knowledge, only \cite{2000PhRvD..62d3004C} and
\cite{2000PhRvE..61.2024C} have systematically explored a similar
approach previously.  The matrix $Q_{ij}$ is denoted the
\emph{polarization matrix} or \emph{polarization tensor} (whether the
3-dimensional or 4-dimensional version is being talked about ought to
be clear from the context).  The polarization information is contained
in the normalized version of $Q_{ij}$, termed the \emph{normalized
  polarization tensor}:
\begin{equation}
\phi_{ij} = \frac{Q_{ij}}{\mbox{Tr}[\bm{Q}]} \ .
\end{equation}
For a given photon direction $\bm{n}$, the polarization vector is
transverse, implying
\begin{equation}
n^i Q_{ij} = 0 \ .
\end{equation}
It is useful to define a matrix with dimensions of specific intensity,
also called the polarization tensor or matrix:
\begin{equation}
I_{ij} = I \phi_{ij} \ ,
\end{equation}
where the specific intensity $I$ and the components of $\phi_{ij}$ are
associated with some mean frequency $\omega$ as discussed in the last
section.  The transition from the quasi-monochromatic case to the
general polychromatic case may be taken as discussed in the previous
section, and the components become functions of photon frequency. In
general, the polarization matrix is a function of photon frequency (or
momentum) and direction as well as space and time:
\begin{equation}
I_{ij} = I_{ij}(\nu,\bm{n},\bm{x},t) \ .
\end{equation}
Other conventions are also useful -- in the computation of the
Sunyaev-Zeldovich effects (SZE), it will be convenient to work with
polarization matrices whose trace is either the occupation number
$n(\nu,\bm{n},\bm{x},t)$ or the phase space distribution function
$f(\nu,\bm{n},\bm{x},t)$ (associated with a particular photon momentum
state and spatial position).  Since the Stokes parameters are usually
taken to have dimensions of specific intensity, we usually work with
$I_{ij}$, but it is occasionally useful to use the other forms.

Now in the usual description of polarized light, the Stokes parameters
are defined with respect to a particular choice of ``polarization
basis''.  This is a pair of mutually orthogonal unit vectors
$\bm{e}^{(1)}, \bm{e}^{(2)}$, both orthogonal to the beam direction.
The Stokes parameters $Q$ and $U$ depend on the orientation of these
vectors.  By contrast the polarization matrix is a tensor and its
components in any basis contain all the information about the
polarization ellipse.  Its advantage is that there is no need to
rotate axes to define Stokes parameters.  The Stokes parameters are
given in terms of the polarization matrix and the polarization basis
vectors as:
\begin{eqnarray} \label{ch1:stokesbasis}
\frac{1}{2}\left(\begin{array}{cc} I+Q & U+iV \\ U-iV &
  I-Q \end{array}\right) &=& e^{(a)}_i e^{(b)}_j I_{ij} \ ,
\quad\quad(a,b)\in (1,2) \ ,
\end{eqnarray}
(the sum over the Cartesian indices $ij$ is implied) which is just the
previously defined coherency matrix $\bm{J}$ of equation
(\ref{ch1:def22coh}).

It is of interest to see how the Stokes parameters transform if we
choose a rotated set of basis vectors.  In the case of a beam
propagating in the $z$-direction for example, we have, choosing
polarization basis vectors $\bm{e}^{(1)}=\bm{x},
\;\bm{e}^{(2)}=\bm{y}$,
\begin{equation} \label{ch1:Qofzbeam}
I_{ij} = \frac{1}{2} \left( \begin{array}{ccc} I+Q & U+iV & 0 \\ U-iV
  & I-Q & 0 \\ 0 & 0 & 0
\end{array} \right) \ . 
\end{equation}
If the basis vectors are rotated clockwise (according to an observer
looking in the direction of propagation) through an angle $\chi$, the
new set of basis vectors is
\begin{eqnarray}
\bm{e}^{(1)}_R &=& \cos\chi \,\bm{e}^{(1)} + \sin\chi \,\bm{e}^{(2)} \
, \nonumber \\ \bm{e}^{(2)}_R &=& \cos\chi \,\bm{e}^{(2)} - \sin\chi
\,\bm{e}^{(1)} \ .
\end{eqnarray}
Forming the matrices $e^{(a)}_{R,i}e^{(b)}_{R,j}$, with
$(a,b)\in\{1,2\}$, the primed Stokes parameters according to
Eqn.~(\ref{ch1:stokesbasis}) are:
\begin{eqnarray}
I' &=& I \nonumber \\ Q' &=& Q \cos 2\chi + U \sin 2\chi \ , \nonumber
\\ U' &=& U \cos 2\chi - Q \sin 2\chi \ , \nonumber \\ V' &=& V \ .
\end{eqnarray}
These transformations are also obtained directly from $Q_{ij}$ by
forming the rotation matrix:
\begin{equation}
\bm{R}(\chi) = \left( \begin{array}{ccc} \cos\chi & -\sin\chi & 0 \\
  \sin\chi & \cos\chi & 0 \\ 0 & 0 & 1 \end{array} \right) \ .
\end{equation}
Then
\begin{equation}
\frac{1}{2} \left( \begin{array}{ccc} I'+Q' & U'+iV' & 0 \\ U'-iV' &
  I'-Q' & 0 \\ 0 & 0 & 0
\end{array} \right) = \bm{R}(\chi) \bm{Q} \bm{R}^T(\chi) \ .
\end{equation}
These factors of $\cos 2\chi, \,\sin 2\chi$ in the transformation law
are well known and associated with the fact that the linear
polarization is described by a ``headless vector'' which is invariant
under a rotation through $\pi$ radians.

Now, given a set of matrix elements $I_{ij}$, supposed to represent a
beam propagating in the direction $\bm{n}$, how do we go about
deciding if this matrix can represent a physical beam?  Clearly the
matrix must be Hermitian and satisfy $I_{ij} n^j = 0$.  This yields a
matrix whose elements contain four independent real quantities. In
addition, the elements must satisfy some analogue of the relation
between the Stokes parameters Eqns.~(\ref{ch1:stokes_con_mono}) or
(\ref{ch1:stokes_con_part}). The required condition is apparent from
Eqn.~(\ref{ch1:detJ}) --- the eigenvalues of the matrix $\bm{I}$ must
be non-negative.

Another obvious question to ask is, how does one construct the matrix
of an unpolarized beam propagating in a general direction $\bm{n}$?
The only quantities we have available to construct the matrix are the
intensity $I$, the components of the direction vector $\bm{n}$, and
the Kronecker delta $\delta_{ij}$.  The matrix must therefore be of
the form:
\begin{equation} \label{ch1:unpoltensorform}
I_{ij}(\bm{n}) = A \delta_{ij} + B n_i n_j \ .
\end{equation}
Now the matrix of an unpolarized beam propagating in the
$z$--direction is obviously
\begin{equation}
I_{ij} = \frac{I}{2} \left(\begin{array}{ccc} 1 & 0 & 0 \\ 0 & 1 & 0
  \\ 0 & 0 & 0 \end{array}\right) .
\end{equation}
Comparing this with the form of Eqn.~(\ref{ch1:unpoltensorform}) for
the special case $n_i = \delta_{iz}$, we see that $A=-B=I/2$.  Thus
the matrix of an unpolarized beam in a general direction $\bm{n}$ is
\begin{equation} \label{ch1:3times3unpol}
I_{ij}(\bm{n}) = \frac{I}{2} (\delta_{ij}-n_i n_j) = \frac{I}{2}
P_{ij}(\bm{n}) \ ,
\end{equation}
where we have defined the projection matrix $\bm{P}$ which will figure
prominently later.

The polarization magnitude (squared) of the beam described by a
general matrix $I_{ij}$ is given by
\begin{equation} \label{ch1:defpmag}
\Pi^2 = \frac{2 \mbox{Tr}[\bm{I}^2]}{\mbox{Tr}[\bm{I}]^2}-1 \ .
\end{equation}
This is readily checked with the matrix (\ref{ch1:Qofzbeam}) of a beam
propagating in the $\bm{z}$ direction.  To see that this relation is
true for any beam, we need only note that the matrix of a beam
propagating in a general direction is related to (\ref{ch1:Qofzbeam})
by a similarity transformation with an orthogonal rotation matrix,
which does not change the traces in Eqn.~(\ref{ch1:defpmag}). Note
also that reality of the right hand side of Eqn.~(\ref{ch1:defpmag})
follows automatically from the Hermiticity of $\bm{I}$ (since $\bm{I}$
and $\bm{I}^2$ are Hermitian, and the trace of a Hermitian matrix is
real).

It is useful to write the matrix $\bm{P}(\bm{n})$ as an expansion in
spherical harmonics. First we expand the matrix in powers of the
Cartesian components of $\bm{n}$, and then identify the result with
expressions for the spherical harmonics in terms of the same
components.  Any spherical harmonic can be expanded in terms of the
complex quantities $(z_1,z_2,z_3)=(\sin\theta e^{i\phi},\sin\theta
e^{-i\phi},\cos\theta)$.  In terms of these functions we may write
$\bm{n}=((z_1+z_2)/2,i(z_2-z_1)/2,z_3)$.  The $l=2$ spherical
harmonics may be written as products of pairs of $z$'s as follows
(note $z_2 = z_1^*, \;z_3^* = z_3, \;z_1 z_2 = 1-z_3^2$)
\begin{eqnarray}
Y_{2,0} &=& \sqrt{\frac{5}{4\pi}} \left(\frac{3}{2} z_3^2 -
\frac{1}{2}\right), \nonumber \\ Y_{2,1} &=& -\sqrt{\frac{15}{8\pi}}
\;z_1 z_3, \;\;\;\;\;\;\; Y_{2,-1} = \sqrt{\frac{15}{8\pi}}\; z_2 z_3,
\nonumber \\ Y_{2,2} &=& \frac{1}{4}\sqrt{\frac{15}{2\pi}}\; z_1^2,
\;\;\;\;\;\;\; Y_{2,-2} = \frac{1}{4}\sqrt{\frac{15}{2\pi}}\; z_2^2 \ .
\end{eqnarray}
The components of $\bm{P}(\bm{n})$ are:
\begin{equation}
\bm{P}(\bm{n}) = \left(\begin{array}{ccc} 1-n_x^{2} & -n_x n_y & -n_x
  n_z \\ -n_y n_x & 1-n_y^{2} & -n_y n_z \\ -n_z n_x & -n_z n_y &
  1-n_z^{2}
\end{array} \right) \ .
\end{equation}
Thus in terms of the spherical harmonics, the projection matrix becomes
\begin{eqnarray} \label{projspher}
  \bm{P}(\bm{n}) = \frac{2}{3} \bm{I} +
  \mbox{Re}\left[Y_{2,0}(\bm{n})\right] \bm{A}_0 + \sum_{m=1,2} \left(
  \mbox{Re}\left[Y_{2,m}(\bm{n})\right] \bm{A}_m +
  \mbox{Im}\left[Y_{2,m}(\bm{n})\right] \bm{B}_m \right) \ ,
\end{eqnarray}
where $\bm{I}$ is the identity matrix, and
\begin{eqnarray}
\bm{A}_0 &=& \frac{1}{3}\sqrt{\frac{4\pi}{5}}\left(\begin{array}{ccc}
  1 & 0 & 0 \\ 0 & 1 & 0 \\ 0 & 0 & -2
\end{array} \right) \ , \nonumber \\
\bm{A}_1 &=& 2\sqrt{\frac{2\pi}{15}} \left(\begin{array}{ccc} 0 & 0 &
  1 \\ 0 & 0 & 0 \\ 1 & 0 & 0
\end{array} \right) \ , 
\quad\;\;\; \bm{B}_1 = 2\sqrt{\frac{2\pi}{15}}
\left(\begin{array}{ccc} 0 & 0 & 0 \\ 0 & 0 & 1 \\ 0 & 1 & 0
\end{array} \right) \ , \nonumber \\
\bm{A}_2 &=& 2\sqrt{\frac{2\pi}{15}} \left(\begin{array}{ccc} -1 & 0 &
  0 \\ 0 & 1 & 0 \\ 0 & 0 & 0
\end{array} \right) \ , 
\quad \bm{B}_2 = 2\sqrt{\frac{2\pi}{15}} \left(\begin{array}{ccc} 0 &
  1 & 0 \\ 1 & 0 & 0 \\ 0 & 0 & 0
\end{array} \right)
\ .
\end{eqnarray}
Thus
\begin{eqnarray} \label{projspher}
  \bm{P}(\bm{n}) = \sum_{lm} \bm{A}_{lm} Y_{lm}(\bm{n}) \ ,
\end{eqnarray}
where
\begin{eqnarray}
\bm{A}_{00} &=& \frac{16\pi}{3} \bm{I} \ , \quad \bm{A}_{20} =
\bm{A}_0 \ , \nonumber \\ \bm{A}_{21} &=&
\frac{1}{2}\left(\bm{A}_1-i\bm{B}_1\right) \ , \quad \bm{A}_{2,-1} = -
\bm{A}^*_{21} \ , \nonumber \\ \bm{A}_{22} &=&
\frac{1}{2}\left(\bm{A}_2-i\bm{B}_2\right) \ , \quad \bm{A}_{2,-2} =
\bm{A}^*_{22} \ .
\end{eqnarray}
This way of writing the projection matrix comes in handy when
performing the angular integrals in the transfer equation,
and for numerical computation.

In the computation of the Sunyaev-Zeldovich effect in the single
scattering limit, derived in detail in Paper II, we have a
situation where the scattered beam consists of an unpolarized
component plus a small polarized perturbation proportional to the
optical depth to scattering, $\tau$.  It is useful to compute at this
point an expression for the polarization matrix of the total beam to
first order in the intensity of the perturbation.  From
Eqn.~(\ref{ch1:3times3unpol}), the beam has polarization matrix
\begin{eqnarray}
I_{ij}(\bm{n}) = I^0_{ij} + \tau\Delta I_{ij}(\bm{n}) \ ,
\quad\mbox{where}\quad I^0_{ij}=\frac{I_0}{2} P_{ij}(\bm{n})\ ,
\end{eqnarray}
or in matrix notation, $\bm{I}=\bm{I^0}+\tau\bm{\Delta I}$, and
$\bm{I^0}=(I_0/2)\bm{P}(\bm{n})$.  Substituting this into
Eqn.~(\ref{ch1:defpmag}) we find, in matrix notation
\begin{eqnarray} \label{ch1:pertpol1}
\Pi^2(\bm{I^0}+\tau\bm{\Delta I}) &=& \frac{2\tau\mbox{Tr}\left[
    2\bm{I^0} \bm{\Delta I} - I_0 \bm{\Delta I}\right] +
  \tau^2\left(2\mbox{Tr}[\bm{\Delta I}^2] - \mbox{Tr}[\bm{\Delta
      I}]^2\right)} {(I_0 + \tau \mbox{Tr}[\bm{\Delta I}])^2} \ .
\;\;\;\;
\end{eqnarray}
Now the unpolarized part of the beam is just a projection matrix
multiplied by a scalar, so it has the property:
\begin{equation}
\mbox{Tr}[\bm{I^0}\bm{\Delta I}] = \frac{I_0}{2}\;\mbox{Tr}[\bm{\Delta
    I}] \ .
\end{equation}
Therefore the first trace in the numerator in
Eqn.~(\ref{ch1:pertpol1}) vanishes. The second term in the denominator
can be ignored in the limit of a small perturbation intensity, and the
squared polarization magnitude reduces to
\begin{eqnarray} \label{ch1:pertpol2}
\Pi^2(\bm{I^0}+\tau\bm{\Delta I}) \approx \tau^2
\left(\frac{\mbox{Tr}[\bm{\Delta I}]}{\mbox{Tr}[\bm{I^0}]}\right)^2
\left[\frac{2 \mbox{Tr}[\bm{\Delta I}^2]}{\mbox{Tr}[\bm{\Delta
        I}]^2}-1\right] \ .
\end{eqnarray}
In other words, the polarization magnitude of the total beam is just
that of the polarized perturbation multiplied by the ratio of the
intensity of the polarized part relative to the unpolarized part:
\begin{eqnarray} \label{ch1:unpolpert}
\Pi(\bm{I^0}+\tau\bm{\Delta I}) \approx \tau
\left(\frac{\mbox{Tr}[\bm{\Delta I}]}{\mbox{Tr}[\bm{I^0}]}\right)
\Pi(\bm{\Delta I}) \ .
\end{eqnarray}

Finally in this section, we note that the polarization matrices of
incoherent beams associated with the same direction and frequency may
simply be summed, by an obvious extension of the proof for coherency
matrices mentioned in \S\ref{ch1:sec1}.

\section{Extension to a covariant polarization tensor \label{ch1:sec3}}

The discussion so far has been in terms of electric fields measured in
a particular Lorentz frame.  In treating problems involving scattering
from a moving medium, it is necessary to Lorentz transform the fields
between frames. This can be done explicitly by writing down the time
dependent electric and magnetic fields of the waves, and using the
transformation law of the fields. However it turns out to be much
simpler to use an extension of the matrix approach we have described
in which the beam is described by a second rank tensor on spacetime.
In this approach the Lorentz transformations become simple tensor (or
matrix) relations. Indeed a full development of the radiative transfer
of polarized light on a curved spacetime is possible with this
covariant formalism.  In this section we work in a curved spacetime
initially but eventually restrict to flat spacetime, which is adequate
for our application to the SZE.  We use the Minkowski metric with the
convention $g_{\mu\nu}=\mbox{diag}\{-1,1,1,1\}$.  The coordinates of a
point in spacetime will be denoted either abstractly as $x$, or as an
upper index quantity $x^{\mu}=(t,x,y,z)$.  Latin indices will denote
components in the orthonormal basis
$\{\vec{e}_x,\vec{e}_y,\vec{e}_z\}$.


A truly covariant description of the electromagnetic field requires
introduction of the field strength tensor $F_{\alpha\beta}$, and
indeed a covariant description of the polarization of light can be
accomplished entirely in terms of the field strength tensor
\citep{Dialetis1969}.  But we wish to maintain an explicit connection
with the Stokes parameters which are defined as time averaged
quadratic combinations of electric field amplitudes, as measured by an
observer at rest in some Lorentz frame.  Thus we must express the
electric field amplitudes measured in the rest frame of a given
observer in a Lorentz covariant manner. The rest frame of the observer
along the light beam can be defined by specifying a differentiable
time-like vector field $\vec{v}(x)$ giving the observer 4-velocity all
along the light cone (with $\vec v\cdot\vec v=-1$).

To generalize the coherency matrix of the previous sections, we need
to find a covariant way to describe the time averaged product of
electric fields. This must be done by constructing the electromagnetic
field strength tensor for a plane wave in the WKB (or shortwave)
approximation of geometrical optics (see
e.g. \cite{BornWolf,Schneider1992,MTW1973}).  In this approximation we
treat the antisymmetric electromagnetic field strength tensor
$F_{\mu\nu}$ as a test field (meaning that we may ignore the influence
of $F_{\mu\nu}$ on the gravitational field) and assume that there are
no charges or currents in the region we are considering.  The field
tensor thus obeys the source free Maxwell equations:
\begin{equation}
  \label{ch1:maxwell}
  \nabla_\mu F^{\mu\nu}=0\ ,\quad \nabla_\alpha F_{\mu\nu}+\nabla_\mu
  F_{\nu\alpha}+ \nabla_\nu F_{\alpha\mu}=0\ .
\end{equation}
The geometrical optics approximation consists in assuming that the
field strength tensor can be written as the product of a slowly
varying complex amplitude and a relatively rapidly varying phase
factor:
\begin{equation}
  \label{ch1:wkb}
  F_{\mu\nu}=\Re\hbox{e}\left\{\widetilde F_{\mu\nu}(x)\exp
  [i\varphi(x)/\epsilon]\right\}
\end{equation}
where $\epsilon\sim\lambda/L$ is a perturbation parameter with
$\lambda$ being the wavelength and $L$ the length-scale over which the
amplitude $\widetilde F_{\mu\nu}$ changes (roughly the local radius of
curvature of spacetime).  In the geometrical optics limit we expand
the Maxwell equations in an asymptotic series in $\epsilon$, take the
limit $\epsilon\to 0$, and read off the lowest order terms.  Then
$\epsilon$ is absorbed into $\varphi(x)$, by replacing
$\varphi(x)/\epsilon$ with $\widetilde\varphi(x)$ and then dropping
the tilde.  The lowest order terms describe the evolution of
electromagnetic waves which, on scales which are large compared to
$\lambda$ but small compared to $L$, are plane and monochromatic to an
excellent approximation.

Substituting equation (\ref{ch1:wkb}) into the Maxwell equations
(\ref{ch1:maxwell}) and working to lowest order in $\epsilon$, we
obtain
\begin{eqnarray}
  \label{ch1:geopt1}
  k_\mu \widetilde F^{\mu\nu}=0\ , \nonumber \\ k_\alpha \widetilde
  F_{\mu\nu}+k_\mu \widetilde F_{\nu\alpha}+ k_\nu \widetilde
  F_{\alpha\mu}=0\ .
\end{eqnarray}
where the wavevector $k_\mu$ is a one-form field normal to surfaces of
constant phase, defined by:
\begin{equation}
  \label{ch1:kmu}
  k_\mu(x)\equiv\nabla_\mu\varphi\ .
\end{equation}
It follows from this, and the fact that covariant derivatives commute
when applied to a scalar field, that
\begin{equation} \label{ch1:kcomm}
  \nabla_\mu k_\nu=\nabla_\nu k_\mu\ .
\end{equation}
Contracting the second equation in (\ref{ch1:geopt1}) with
$k^{\alpha}$, and assuming that $\widetilde F_{\mu\nu}$ vanishes only
on hyper-surfaces, we find
\begin{equation} \label{ch1:knull}
\quad k^\mu k_\mu=0\ .
\end{equation}
Thus the wavevector $\vec{k}$ is null.  If desired we may associate a
photon 4-momentum $\vec{p}=\hbar\vec{k}$ with the wavevector, and go
over to a particle description.  The frequency of the wave as measured
by a local observer with worldline $x^{\mu}(\tau)$ and 4-velocity
$u^{\mu}=dx^{\mu}/d\tau$ is given by
$\omega=-\vec{k}\cdot\vec{u}=d\varphi/d\tau$ (taking $\epsilon=1$).
Eqns.~(\ref{ch1:kcomm}) and (\ref{ch1:knull}) imply that the
wavevector is parallel transported:
\begin{equation}
  \label{ch1:geopt2}
  \nabla_k k_\mu\equiv k^\alpha\nabla_\alpha k_\mu=0\ .
\end{equation}

The curves $x^\mu(\lambda)$ with $dx^\mu/d\lambda =k^\mu$ are called
light rays ($\lambda$ is an affine parameter along the ray).  Note
$\nabla_k=k^{\mu}\nabla_{\mu}=d/d \lambda$ is the directional
derivative along the ray. As a consequence of Eqn.~(\ref{ch1:geopt2}),
the system of rays is equivalent to a Hamiltonian flow for particles
with Hamiltonian
\begin{equation}
  \label{ch1:hamilt}
  H(x,k)=\frac{1}{2}g^{\mu\nu}(x)k_\mu k_\nu.
\end{equation}
Hamilton's equation $dk_\mu/d\lambda=-\partial H/ \partial x^\mu$ is
equivalent to Eqn.~(\ref{ch1:geopt2}), which is the geodesic equation
for photons, while $dx^\mu/d\lambda=\partial H/\partial k_\mu$ gives
the advance of the wavefront along the ray.  The Hamilton-Jacobi
equation $H(x,\nabla\varphi)=0$ is also known as the eikonal equation
for the phase factor $\varphi(x)$.

Now we would like to express the components of $F_{\mu\nu}
=-F_{\nu\mu}$ in terms of the electric field.  Writing a propagation
equation for the electric field requires that we have a differentiable
time-like vector field $v^\mu$ ($\vec v\cdot\vec v=-1$) giving the
4-velocity (hence rest frame) of observers all along the light cone.
In other words, the electric field is defined with respect to a family
of observers with 4-velocity $\vec{v}(x)$.  In the local Lorentz frame
at point $x$ of the observer with 4-velocity $\vec{v}(x)$, the
electric field components are $E_{\hat i}=F_{\hat i\hat 0}$ and the
magnetic field components are $B_{\hat i}=\frac{1}{2}\epsilon_{\hat i
  \hat l\hat m}F^{\hat l\hat m}$ where Latin indices range over the
spatial components and the carets indicate an orthonormal basis, with
$\vec{e}_{\hat{0}}=\vec{v}$.  The transversality from equation
(\ref{ch1:geopt1}) implies $B_{\hat i}=\epsilon_{\hat i\hat l \hat
  m}\hat k^{\hat l}E^{\hat m}$ where $\hat k$ is a spatial unit vector
along the wavevector.  In a general basis, we promote the electric
field to a 4-vector
\begin{equation}
E_\mu\equiv v^\nu F_{\mu\nu} \ .
\end{equation}
By antisymmetry of $F_{\mu\nu}$, $E_{\mu}$ is orthogonal to the
4-velocity of the observer $v^\mu$:
\begin{equation}
v^{\mu} E_{\mu}=0 \ .
\end{equation}
In the geometrical optics limit we may define the complex amplitude of
the 4-vector electric field using the complex amplitude of the field
strength tensor:
\begin{equation} \label{ch1:etilde}
\widetilde E_\mu\equiv v^\nu \widetilde F_{\mu\nu} \ , \quad v^{\mu}
\widetilde E_{\mu}=0 \ .
\end{equation}
Thus
\begin{equation}
E_{\mu} = \Re e \{\widetilde E_\mu\exp[i\varphi(x)/\epsilon] \} \ .
\end{equation}
Eqns.~(\ref{ch1:geopt1}) imply $k^\mu E_\mu=k^{\mu}\widetilde
E_\mu=0$, which correspond to the transversality of the electric field
to the the wavevector.  The electric field 4-vector may be factored as
\begin{equation}
E_{\mu} = E \epsilon_{\mu} \ ,
\end{equation}
where $\vec{\epsilon}$ is a vector which satisfies
$g^{\mu\nu}\epsilon^\ast_\mu\epsilon_\nu=1$, called the \emph{electric
  polarization vector}.  In the rest frame of $\vec{v}$, this reduces
to a 4-vector with spatial parts equal to the usual polarization
3-vector.

Contracting the second of Eqns.~(\ref{ch1:geopt1}) with $v^{\alpha}$
and substituting Eqn.~(\ref{ch1:etilde}) yields an expression for the
field strength amplitude in terms of the electric field 4-vector
amplitude:
\begin{equation}
  \label{ch1:fgeopt}
  \widetilde F_{\mu\nu}=k^{-1}(k_\mu \widetilde E_\nu-k_\nu \widetilde
  E_\mu)\quad \hbox{where}\quad k\equiv-k_\mu v^\mu\ .
\end{equation}
Next we would like to know how the amplitudes $\widetilde F_{\mu\nu}$
and $\widetilde E_{\mu}$ change along a ray.  We proceed by computing
the divergence of the second of the Maxwell equations
(\ref{ch1:maxwell}).
\begin{equation}
g^{\alpha\beta}\nabla_{\beta}\left[\nabla_{\alpha}F_{\mu\nu} +
  \nabla_{\mu}F_{\nu\alpha} + \nabla_{\nu}F_{\alpha\mu}\right] = 0 \ .
\end{equation}
Note that swapping the order of the covariant derivatives in the
second two terms kills each term by Maxwell's equations.  Thus using
the following identity for the commutator of covariant derivatives in
terms of the Riemann tensor,
\begin{equation}
(\nabla_\mu \nabla_\nu-\nabla_\nu \nabla_\mu) S_{\alpha\beta} =
-S_{\sigma\beta}R^{\sigma}_{\
  \alpha\mu\nu}-S_{\alpha\sigma}R^{\sigma}_{\ \beta\mu\nu} \ ,
\end{equation}
we find a wave equation for $F_{\mu\nu}$ with curvature terms:
\begin{equation}
  \label{ch1:wavef}
  g^{\alpha\beta}\nabla_\alpha\nabla_\beta F_{\mu\nu}
  -R_{\mu\alpha}F^\alpha_{\ \,\nu}+R_{\nu\alpha}F^\alpha_ {\
    \,\mu}+R_{\mu\nu\alpha\beta}F^{\alpha\beta}=0\ .
\end{equation}
Substituting equation (\ref{ch1:wkb}) and working to the two lowest
orders in $\epsilon$, one finds the following equation for the
evolution of the field strength amplitude (the Riemann tensor terms do
not appear to this order):
\begin{equation}
  \label{ch1:emprop}
  \nabla_k\widetilde F_{\mu\nu}=-\frac{1}{2}\theta\widetilde
  F_{\mu\nu}\ , \quad\theta\equiv\nabla_\alpha k^{\alpha}\ .
\end{equation}
The amplitude of the electromagnetic field changes along rays due to
curvature of the wavefronts.  For example, diverging rays ($\theta>0$)
lead to a decrease in the electromagnetic field strength as the wave
propagates.
 
Substituting equation (\ref{ch1:fgeopt}) into equation
(\ref{ch1:emprop}) now gives an equation for the electric field
evolution along a ray,
\begin{equation}
  \label{ch1:eprop1}
  \nabla_k \widetilde E_\mu=\left(\nabla_k\ln
  k-\frac{1}{2}\theta\right) \widetilde
  E_\mu+\frac{k_\mu}{k}\left(\nabla_k v^\alpha\right) \widetilde
  E_\alpha\ .
\end{equation}
Factoring the electric field into its magnitude and direction
(polarization) vector, $\widetilde E_\mu=\widetilde E\epsilon_\mu$
where $g^{\mu\nu}\epsilon_\mu\epsilon_\nu=1$ and
$v^\mu\epsilon_\mu=0$, we obtain
\begin{subequations} \label{ch1:eprop2}
\begin{eqnarray}
  \nabla_k \widetilde E&=&\left(\nabla_k\ln
  k-\frac{1}{2}\theta\right)\widetilde E\ ,
    \label{ch1:eprop}\\
  \quad \nabla_k\epsilon_\mu&=&\frac{k_\mu}{k}\left(\nabla_k
  v^\alpha\right)\epsilon_\alpha\ .\label{ch1:polprop}
\end{eqnarray}
\end{subequations}
The first of these equations yields for example the $1/r$ fall off of
the electric field magnitude expected for a radiation field.  The
right-hand side of both equations vanishes for a plane wave in flat
space, but not for a curved wavefront (e.g. a spherical wave), or for
a wave propagating in a general curved space.

It is perhaps surprising that the electric polarization vector is not
parallel transported in a curved spacetime.  This fact leads to a
rotation of the polarization vector when a beam passes through a
strong gravitational field.  This effect has been noted before by
several authors
\citep{Skrotskii,1960PhRv..118.1396P,1999PhRvD..60b4013N,2002PhRvD..65f4025K},
and is important in considering for example the propagation of
polarized radiation in the vicinity of a black hole.  However it is
true that if one defines the polarization vector to be parallel to the
vector potential rather than the electric field of the electromagnetic
wave, then it is parallel transported in the geometrical optics limit
(see e.g. \cite{MTW1973}, \cite{Schneider1992}).  This turns out to be
consistent with the electric field evolution due to the enforcement of
the gauge choice of the vector potential all along the photon
worldline.  Thus in considering the propagation of polarized photons
on a curved spacetime it is more convenient to use a polarization
tensor constructed from the vector potential to evolve the
polarization state along the ray, and then make the transformation to
electric fields.
If the photon path does not pass through regions with an exceptionally
strong gravitational field however, the rotation of the polarization
vector resulting from this gravitational effect is small (but note
that, strictly speaking, the rotation arises from the acceleration of
the local observers, $\nabla_k v^{\alpha}=dv^{\alpha}/d\lambda$, which
can be large even in flat spacetime if a peculiar vector field of
observer 4-velocities is chosen).  In considering the propagation of
photons through a cluster of galaxies for example, the effect is
entirely negligible, and so henceforth we will restrict the discussion
to flat spacetime, and work with the more physical polarization
tensors we defined in terms of the electric fields.  In flat spacetime
we may drop the right hand sides in Eqns.~(\ref{ch1:eprop2}).

Having described the propagation of electromagnetic waves in the
geometrical optics approximation, and defined the electric field in a
covariant manner, we are equipped to construct the covariant version
of the coherency matrix.  We consider a plane electromagnetic wave
propagating in flat spacetime, in the geometrical optics limit, with
wavevector $\vec{k}$ and associated photon momentum $\vec{p}$.
Henceforth we will write the complex amplitude of the 4-vector
electric field of the wave as $\widetilde E^{\mu}$, dropping the tilde
for brevity.  The 4-vector $\vec{E}$ has the property that its spatial
components $E^i$ in the rest frame of the observer, in which
$v^{\mu}=(1,0,0,0)$, are equal to the measured electric field, and
also $E^{0}=0$ in this frame.  Thus by analogy with the $3\times 3$
polarization matrix $Q_{ij}$ defined in Eqn.~(\ref{ch1:def33pol}), we
are lead to define a complex valued rank $(0,2)$ tensor called the
\emph{polarization tensor}:
\begin{equation}
  \label{ch1:qmunu}
  Q_{\mu\nu}(x,\vec{p},\vec{v})\equiv\left\langle E_\mu
  E^*_\nu\right\rangle \ .
\end{equation}
The spatial components of this tensor in the rest frame of the
observer are entirely equivalent to the elements of the $3\times 3$
coherency matrix considered in the previous section.  It is related to
the stress-energy tensor $T^{\mu\nu}=F^\mu_{\
  \,\alpha}F^{\nu\alpha}-\frac{1}{4}g^{\mu\nu}
F^{\alpha\beta}F_{\alpha\beta}$ (in Heaviside-Lorentz units).  In
particular, the time-average energy density in the geometrical optics
limit is $v_\mu v_\nu\left\langle T^{\mu\nu}\right\rangle =Q\equiv
Q^\mu_{\ \,\mu}$ where angle brackets denote averaging over a few
periods.  Note that $Q^{\mu \nu} v_{\mu} = Q^{\mu \nu} p_{\mu} = 0$
(where $p^{\mu}$ is the four momentum of the photon).

To define Stokes parameters, we need to specify a set of polarization
basis vectors. The natural choice is the orthonormal tetrad basis
vectors $\{\vec \epsilon_a\}$:
\begin{equation}
  \label{ch1:t03}
  \vec \epsilon_0=\vec v\ ,\quad \vec \epsilon_3=p^{-1}\vec p-\vec v\
  , \quad \vec \epsilon_1 \ ,\, \;\;\vec\epsilon_2\,
\end{equation}
where $p\equiv-\vec v\cdot\vec p$. These vectors have the property
$\vec \epsilon_a\cdot\vec \epsilon_b=\eta_{ab}$.  Latin indices
$\{a,b,\ldots\}$ are tetrad indices; Greek indices
$\{\mu,\nu,\ldots\}$ are coordinate indices.  The spatial direction of
the photon momentum for an observer with 4-velocity $\vec \epsilon_0$
is $\vec \epsilon_3$. The remaining basis vectors, $\vec \epsilon_1$
and $\vec \epsilon_2$, give the physical polarization space.
We call this tetrad the polarization tetrad.  There are associated
basis $1$--forms $\{\tilde \epsilon^a\}$), which are dual to the basis
vectors: $\left\langle\tilde \epsilon^a,\vec
\epsilon_b\right\rangle=\delta^a_{\ \,b}$.  (Note that the
polarization tetrad depends on the photon momentum, i.e.  $\vec
\epsilon_a=\vec \epsilon_a(x,p)$.  Thus, in general a different basis
is needed for each photon momentum).  The coherency matrix of
Eqn.~(\ref{ch1:def22coh}) is then given by
\begin{equation}
J_{ab} = \epsilon^{\mu}_a \epsilon^{\nu}_b
\;Q_{\mu\nu}(x,\vec{p},\vec{v})
\ , \quad\quad (a,b)\in(1,2)\ .
\end{equation}

In the case of beam which has a definite polarization vector
$\vec\epsilon$ (lying in the polarization subspace spanned by $\{\vec
\epsilon_1,\vec \epsilon_2\}$) which does not vary with time, i.e. a
pure beam, the polarization tensor is given by
\begin{equation}
Q_{\mu\nu}(x,\vec{p},\vec{v}) = Q \;\epsilon_{\mu} \epsilon_{\nu} \ .
\end{equation}

So far we have used the tensor $Q^{\mu\nu}$ to describe a polarized EM
wave. However if we wish to consider energy transfer between photons
and some scattering medium, free electrons for example, we must
consider the trajectories of photons in phase space.
To describe an ensemble of polarized photons we must define a
distribution function on phase space.  The matrix $Q^{\mu\nu}$ is not
very useful because it describes a single classical system (the
classical counterpart of a pure state) with specified wavevector
$k_\mu(x)$.  Developing a kinetic theory requires an ensemble of
systems encompassing a continuous distribution of wave-vectors $k_\mu$
at each spacetime point.  We accomplish this heuristically by analogy
with the usual treatment of the unpolarized case.

 In general, the stress-energy tensor of a system of photons may be
 written (in flat space) as
 \begin{eqnarray} \label{ch1:defunpolf}
 T^{\mu\nu}(x) &=& \int \frac{d^3p}{p} \,p^{\mu} p^{\nu} f(x,\vec{p})
 \ ,
 \end{eqnarray}
 where $p=p^0$ and $f(x,\vec{p})$ is the (unpolarized) photon
 distribution function, which determines the total number of photons
 in the quantum state corresponding to phase space element $d^3x
 \,d^3p$ according to
 \begin{equation}
 dN = f(x,\vec{p}) \;d^3x \,d^3p \ .
 \end{equation}
The occupation number is given by $n(x,\vec{p})=h^3 f(x,\vec{p})$.  It
is not hard to prove that $n$ and $f$ are Lorentz scalars (see for
example \cite{1975STIA...7626675L} for a proof).

To incorporate polarization we define the \emph{distribution function
  polarization tensor} $f_{\mu\nu}(x,\vec{p},\vec{v})$ in a manner
similar to the scalar distribution function $f(x,\vec{p})$.  The
polarization tensor of a general superposition of waves at a given
spacetime point, according to the local observer with 4-velocity
$\vec{v}$, may be defined as $Q_{\mu\nu}(x,\vec{v})\equiv\left\langle
E_\mu E^*_\nu\right\rangle$.  Then by analogy with
(\ref{ch1:defunpolf}) we have
\begin{equation} \label{ch1:def_fmunu}
  Q^{\mu\nu}(x,\vec{v}) = \int \frac{d^3p}{p} \,p^2\,
  f^{\mu\nu}(x,\vec{p},\vec{v}) \ .
\end{equation}
This obviously does not uniquely define
$f_{\mu\nu}(x,\vec{p},\vec{v})$. A rigorous definition requires a more
sophisticated discussion (as in \cite{1989A&A...219...25B}). However
we do not run into any difficulties if we simply regard
$f_{\mu\nu}(x,\vec{p},\vec{v})$ as a tensor generalization of the
scalar distribution function which satisfies
Eqn.~(\ref{ch1:def_fmunu}).  $f^{\alpha\beta}$ has the property that
$f^{\alpha\beta} \epsilon_{\alpha} \epsilon_{\beta}^* \; d^3p \;d^3x$
is proportional to the number of photons in the phase space element
$d^3p \;d^3x$ passing per unit time through a polarizer oriented to
select polarization $\epsilon^{\alpha}$ (where this 4-vector must lie
in the polarization subspace spanned by the vectors $\vec\epsilon_1,
\,\vec\epsilon_2$ of Eqn.~(\ref{ch1:t03})).  Contraction with the
metric yields the scalar distribution function:
\begin{equation}
g_{\mu\nu} f^{\mu\nu}(\vec{p},\vec{v})=f(\vec{p}) \ .
\end{equation}
The distribution function tensor also has the properties
\begin{eqnarray}
v_{\alpha} f^{\alpha\beta}(\vec{p},\vec{v}) = p_{\alpha}
f^{\alpha\beta}(\vec{p},\vec{v}) = 0 \ , \quad f^{\beta\alpha} =
\left(f^{\alpha\beta}\right)^* \ .
\end{eqnarray}
The generalization of Eqn.~(\ref{ch1:defpmag}) for the polarization
magnitude is
\begin{eqnarray}
\Pi^2(\vec{p}) = \frac{2 f^{\mu\nu}(\vec{p},\vec{v})
  f_{\mu\nu}(\vec{p},\vec{v})}{\left[g_{\alpha\beta}f^{\alpha\beta}(\vec{p},\vec{v})\right]^2}-1
\ ,
\end{eqnarray}
which is manifestly a Lorentz scalar.

Similarly to the case for coherency matrices and $3\times 3$
polarization matrices, the polarization tensors
$f_1^{\mu\nu}(\vec{p},\vec{v})$ and $f_2^{\mu\nu}(\vec{p},\vec{v})$ of
two incoherent beams associated with the same photon momentum may be
summed to yield the total polarization tensor.  We shall always make
the assumption that two separate beams are incoherent and have
polarization tensors that may be superimposed in this manner.

It is useful to define a covariant polarization tensor with dimensions
of specific intensity (whose components are combinations of Stokes
parameters).  In the unpolarized case, the specific intensity
$I(x,\vec{p})$ is introduced by defining
\begin{equation}
  T^{00} = \int d^3 p \,p \,f(x,\vec{p}) = \int d\nu \,d\Omega
  \;I(x,\vec{p}) \ ,
\end{equation}
where $d\Omega$ is the solid angle element about the photon direction.
It follows from (\ref{ch1:defunpolf}) that
\begin{equation}
  I = h p^3 \,f = h^4 \nu^3\, f = h\nu^3 n \ .
\end{equation}
We define a specific intensity (or brightness) tensor by analogy with
the unpolarized case:
 \begin{equation} \label{ch1:defImunu}
 I^{\mu\nu}(x,\vec{p},\vec{v}) \equiv h p^3
 f^{\mu\nu}(x,\vec{p},\vec{v}) \ .
\end{equation}
The intensity polarization matrix is zero outside from the
two-dimensional polarization space spanned by $\vec \epsilon_1$, $\vec
\epsilon_2$, where it may be written in terms of the usual Stokes
parameters $I$, $Q$, $U$, and $V$:
\begin{equation}
  \label{ch1:22stokesmat}
  f_{ab}=\frac{1}{2hp^3}\left(
    \begin{array}{cc}
      I+Q& U+iV \\ U-iV & I-Q
    \end{array}
  \right)
\end{equation}
The Stokes parameters are functions of frequency (photon energy);
$I=I_\nu$ is the spectral intensity.  In an arbitrary basis the
intensity is $I=(h\nu^3/c^2)g^{\mu\nu}f_{\mu\nu}$.  The normalization
factor is chosen so that $f_{11}$ is the photon occupation number
(phase space density divided by $h^3$) for photons passed by a linear
polarizer oriented along $\vec \epsilon_1$ (and similarly for other
directions). In terms of the total spectral intensity, we may write
the polarization magnitude as $\Pi^2=2(I^{-2}) I^{\mu\nu}I_{\mu\nu}-1$.
 
Now we wish to obtain an equation for the evolution of $f^{\mu\nu}$ in
time. In the absence of scattering, photons follows geodesics (free
stream) and the distribution function evolves according to the
Liouville equation.  The Liouville equation for the unpolarized
distribution function is simply $Df/d\lambda=0$, where $\lambda$ is an
affine parameter along the ray:
\begin{eqnarray}\label{ch1:evolf}
  \frac{dx^\mu}{d\lambda}\frac{\partial f}{\partial x^\mu}+
  \frac{dp_\mu}{d\lambda}\frac{\partial f}{\partial p_\mu} &=&0\ .
\end{eqnarray}
This may also be written in the more familiar form (valid in curved
spacetime)
\begin{eqnarray}
  \nabla_p f-\Gamma^i_{\,\mu\nu}p^\mu p^\nu\frac{\partial f} {\partial
    p^i}&=&0 \quad \mbox{where $\nabla_p=p^\alpha\nabla_\alpha$}
\label{lindf}\ ,
\end{eqnarray}
provided one regards $f$ as a function of the 3-momentum in some
frame, $f=f(\bm{p})$ (not $f=f(\vec{p})$), by enforcing the mass shell
constraint $p^{\mu}p_{\mu}=0$.

Now we consider the generalization to the polarized case.  From
Eqns.~(\ref{ch1:eprop2}) it follows that the evolution equation for
$Q_{\mu\nu}$ in the geometrical optics approximation in flat spacetime
is
\begin{equation}\label{ch1:qmunuprop}
  \nabla_p Q_{\mu\nu}= 0 \ .
\end{equation}
This is suggestive that the Liouville equation for $f_{\mu\nu}$ in
flat spacetime is simply
\begin{equation}
  \label{ch1:fmunuprop}
   p^{\alpha}\partial_{\alpha} f_{\mu\nu}=0 \ .
\end{equation}
This obviously reduces to the correct evolution of the
unpolarized distribution function on taking the trace.  This is in
fact the correct Liouville equation in the polarized case.
The proof is easy and goes along the following lines.
[give proof]

Generalizing this to a curved spacetime is more difficult. The
evolution equations for the 4-vector electric field produce unusual
terms.
But it can be shown that using a polarization matrix based on the
vector potential, the Liouville equation is simply given 
by Eqn. (\ref{ch1:evolf}) with tensor indices added. We will not
derive this result here, but refer to the discussions in 
\cite{1978AN....299...13D,
  1989A&A...219...25B,1989CQGra...6.1171B,1990CQGra...7.2367B,
1980RSPSA.370..389B,1981RSPSA.374...65B}.

So in flat spacetime, the Boltzmann (or kinetic) equation for the
distribution function polarization tensor is
\begin{equation}
  \label{ch1:fmunuprop}
      p^{\alpha}\partial_{\alpha} f_{\mu\nu}=C_{\mu\nu} \ ,
\end{equation}
where the effect of scattering is contained in the \emph{scattering
  term} $C_{\mu\nu}$.  The form of the scattering term in the
case of Compton scattering is derived in the limit of negligible
  electron recoil in \S\ref{ch3:sec2} and in the general case in \S\ref{ch3:sec3}.

\section{Lorentz transformation properties of the polarization tensor
\label{ch1:sec4}}

On performing a Lorentz transformation between inertial frames, it is
well known that the propagation direction of an EM wave (or
equivalently, photon) is aberrated and its frequency (or momentum)
Doppler boosted.  The transformation of the polarization state of the
beam is less well known.  Here we derive the transformation law of the
polarization tensor between frames.  This leads to the transformation
law for the Stokes parameters, which turns out to be very simple (in
fact, they are invariant) provided a certain choice of polarization
basis is made.

First, we find the transformation of the 4-vector electric field
$E^{\mu}(\vec{v})$ under a change of the local observer vector field
from $\vec{v}(x)$ to $\vec{v}'(x)$.  The spatial components of
$E^i(\vec{v})$ are the electric field (3-vector) components measured
by the observer with four-velocity $\vec{v}$ (in her rest frame).  Let
us find the relationship between the electric fields
$E^{\mu}(\vec{v})$ and $E^{\mu}(\vec{v}')$.  To determine this, recall
from (\ref{ch1:fgeopt}) that the definition of $E^{\mu}$ implies the
following relation between $E_{\mu}$ and the field strength tensor for
a plane wave:
 \begin{equation} \label{ch1:FintermsE}
 F_{\mu\nu} = p^{-1}(p_{\mu}E_{\nu}-p_{\nu}E_{\mu}) \ , \quad p \equiv
 -\vec{p} \cdot \vec{v} \ .
 \end{equation}
Therefore, since $E^{\mu}(\vec{v})=F^{\mu\nu}v_{\nu}$,
 \begin{eqnarray}\label{ch1:etrans}
 E^{\mu}(\vec{v}')
 &=& \frac{p'}{p} \left(g^{\mu}_{\;\;\;\nu} +
\frac{p^{\mu}v'_{\nu}}{p'}\right) E^{\nu}(\vec{v}) \ ,
 \end{eqnarray}
where $p'\equiv -\vec{v}'\cdot\vec{p}$.  These relations suggest
introduction of a tensor $P_{\mu\nu}(\vec p,\vec v\,)$ which projects
onto the physical polarization plane $\vec \epsilon_1$-$\vec
\epsilon_2$ by eliminating components in the surface spanned by
$\vec{\epsilon}_0$ and $\vec{\epsilon}_3$ (or $\vec{v}$ and $\vec{p}$):
\begin{eqnarray}\label{ch1:projtens}
  P_{\mu\nu}(\vec p,\vec v\,) &=& g_{\mu\nu} +
  \epsilon_{0\mu}\epsilon_{0\nu} - \epsilon_{3\mu}\epsilon_{3\nu}
  \nonumber \\ &=& g_{\mu\nu}+{1\over p}(p_\mu v_\nu+ p_\nu
  v_\mu)-{p_\mu p_\nu\over p^2} \quad\quad\mbox{where}\;\;\;
  p\equiv-v^\mu p_\mu \ .
\end{eqnarray}
This satisfies the idempotency relation $P^\mu_{\ \ \alpha}P^\alpha_{\
  \ \nu}= P^\mu_{\ \ \nu}$, so that $P^\mu_{\ \ \nu}$ is a projection
tensor, henceforth denoted the \emph{screen projection tensor} which
will prove to be important.  The transformation law for measured
electric fields, equation (\ref{ch1:etrans}), may be written in terms
of the screen projection tensor as follows:
\begin{equation}
  \label{ch1:ltranse}
  E_\mu(\vec v')={v'_\alpha p^\alpha\over v_\beta p^\beta}\, P^\nu_{\
    \ \mu}(\vec p,\vec v')E_\nu(\vec v)\ ,
\end{equation}
since the second and the last term in $P^{\nu}_{\;\;\;\mu}$ vanish
when contracted with $E^{\nu}$.  In the geometrical optics limit,
taking components in the appropriate Lorentz frame,
Eqn.~(\ref{ch1:ltranse}) reproduces the usual relativistic
transformation of electromagnetic fields.
The dependence on the four-momentum appears because the boosted
electric field depends on the magnetic field, which in the geometrical
optics limit is $\hat p\times\vec E$.

The transformation law of $Q^{\mu\nu}(x,\vec{v})$ follows from
Eqn.~(\ref{ch1:qmunu}):
\begin{eqnarray} \label{ch1:ftransform}
 Q^{\mu^{\prime}\nu^{\prime}}(\vec{p},\vec{v}') =
 \left(\frac{\vec{v}'\cdot\vec{p}}{\vec{v}\cdot\vec{p}}\right)^2
 P^{\mu^{\prime}}_{\;\;\;\mu}(\vec{p},\vec{v}')
 P^{\nu^{\prime}}_{\;\;\;\nu}(\vec{p},\vec{v}')
 \;Q^{\mu\nu}(\vec{p},\vec{v}) \ .
\end{eqnarray}
Since the integration measure in Eqn.~(\ref{ch1:def_fmunu}) is Lorentz
invariant, the transformation of $Q^{\mu\nu}$ implies that
$f^{\mu\nu}$ transforms in the following way under change of the local
observer 4-velocity:
\begin{eqnarray} 
 f^{\mu^{\prime}\nu^{\prime}}(\vec{p},\vec{v}') =
 P^{\mu^{\prime}}_{\;\;\;\mu}(\vec{p},\vec{v}')
 P^{\nu^{\prime}}_{\;\;\;\nu}(\vec{p},\vec{v}')
 f^{\mu\nu}(\vec{p},\vec{v}) \ .
\end{eqnarray}
Note that the following transformation property of the specific
intensity tensor follows immediately from the transformation
properties of the distribution function and Eqn.~(\ref{ch1:defImunu}):
\begin{equation} \label{ch1:Imunutrans}
  I^{\mu^{\prime}\nu^{\prime}}(\vec{p},\vec{v}') =
  \left(\frac{\vec{v}'\cdot\vec{p}}{\vec{v}\cdot\vec{p}}\right)^3
  P^{\mu^{\prime}}_{\;\;\;\mu}(\vec{p},\vec{v}')
  P^{\nu^{\prime}}_{\;\;\;\nu}(\vec{p},\vec{v}')
  I^{\mu\nu}(\vec{p},\vec{v}) \ .
\end{equation}
In the unpolarized limit, the trace of this reduces to the familiar
statement that $I/\nu^3$ is Lorentz invariant.  In the general
polarized case, one sees that all of $I/{\nu^3}$, $Q/{\nu^3}$,
$U/{\nu^3}$, $V/{\nu^3}$ are invariant under a boost along the photon
direction.

The transformation properties of the Stokes parameters under a boost
in a general direction are clearly dependent on the polarization basis
chosen in each frame.  To work out the general case, we consider the
transformation from frame $K'$ (the rest frame) with 4-velocity
$\vec{v}_e$ into frame $K$ (the lab frame) with 4-velocity
$\vec{v}_l$.  In lab frame coordinates, let
$v^{\mu}_e=\gamma(1,\bm{v})$.  In the rest frame, the brightness
tensor $I^{\mu'\nu'}(\vec{p},\vec{v}_e)$ contains all polarization and
intensity data of a photon with 4-momentum $\vec{p}$.  We denote the
photon momentum in rest frame coordinates, as
$p^{\mu'}=p'(1,\bm{n}')$, and in lab frame coordinates as
$p^{\mu}=p(1,\bm{n})$.

The Stokes parameters measured in $K'$ are defined by specifying a set
of orthonormal polarization basis vectors $\{\vec{s}_1,\vec{s}_2\}$,
where $\vec{s}_1\cdot\vec{p} = \vec{s}_2\cdot\vec{p}=0$, and
$\vec{s}_1\cdot\vec{v}_e=\vec{s}_2\cdot\vec{v}_e=0$.  Since the
vectors $\vec{s}_a$ are purely spatial in the rest frame, we may write
$s_a^{\mu'}=(0,\bm{\epsilon}'_a)$, $a\in\{1,2\}$ with
$\bm{\epsilon}'_a\cdot\bm{n}'=0,
\;\bm{\epsilon}'_a\cdot\bm{\epsilon}'_b=\delta_{ab}$.  The Stokes
parameters measured in $K'$ are determined by the quantities:
\begin{equation}\label{ch1:stokesK'}
J'_{ab} = I^{\mu'\nu'}(\vec{p},\vec{v}_e) s_{a\mu'} s_{b\nu'} \ ,
\quad\quad (a,b)\in\{1,2\} \ .
\end{equation}

To determine the Stokes parameters measured in $K$, we must specify
lab frame basis vectors $\{\vec{t}_1,\vec{t}_2\}$ which satisfy
$\vec{t}_1\cdot\vec{p} = \vec{t}_2\cdot\vec{p}=0$, and
$\vec{t}_1\cdot\vec{v}_l=\vec{t}_2\cdot\vec{v}_l=0$.  We write
$t_a^{\mu'}=(0,\bm{\epsilon}_a)$, $a\in\{1,2\}$ with
$\bm{\epsilon}_a\cdot\bm{n}=0$.  The analogous quantities to those in
Eqn.~(\ref{ch1:stokesK'}) in $K$ are
\begin{equation} \label{ch1:stokesK}
J_{ab} = I^{\mu\nu}(\vec{p},\vec{v}_l) t_{a\mu} t_{b\nu} \ ,
\quad\quad (a,b)\in\{1,2\} \ .
\end{equation}
The vectors $\{\vec{t}_a\}$ are not uniquely determined, but there is
a natural choice of basis which keeps the transformation of the Stokes
parameters simple.  Applying the transformation law
(\ref{ch1:Imunutrans}) to Eqn.~(\ref{ch1:stokesK}), and replacing
\begin{equation}
I^{\alpha\beta}(\vec{p},\vec{v}_e) \rightarrow P^{\alpha}_{\ \
  \gamma}(\vec{p},\vec{v}_e)P^{\beta}_{\ \ \delta}(\vec{p},\vec{v}_e)
I^{\gamma\delta}(\vec{p},\vec{v}_e) \ ,
\end{equation}
we obtain
\begin{equation} \label{ch1:QK'toK}
J_{ab} = \left(\frac{p}{p'}\right)^3 P^{\mu\nu}_{\ \ \
  \gamma\delta}(\vec{p},\vec{v}_e,\vec{v}_l)
I^{\gamma\delta}(\vec{p},\vec{v}_e) t_{a\mu} t_{b\nu} \ .
\end{equation}
where
\begin{eqnarray} 
&&P^{\mu\nu}_{\ \ \ \alpha\beta}(\vec{p},\vec{v}_e,\vec{v}_l) \equiv
P^{\mu}_{\ \ \alpha}(\vec{p},\vec{v}_l) P^{\nu}_{\ \
  \beta}(\vec{p},\vec{v}_l) P^{\alpha}_{\ \ \gamma}(\vec{p},\vec{v}_e)
P^{\beta}_{\ \ \delta}(\vec{p},\vec{v}_e) \ ,
\end{eqnarray}
and
\begin{eqnarray}
&&p' \equiv -\vec{v}_e\cdot\vec{p} \ , \quad\quad p \equiv
-\vec{v}_l\cdot\vec{p} \ .
\end{eqnarray}

Comparing Eqn.~(\ref{ch1:QK'toK}) to Eqn.~(\ref{ch1:stokesK'}), it is
apparent that if we demand that the vectors $\vec{t}_a$ satisfy:
\begin{equation} \label{ch1:ttrans1}
P^{\alpha}_{\ \ \gamma}(\vec{p},\vec{v}_e) P^{\mu}_{\ \
  \alpha}(\vec{p},\vec{v}_l) \;t_{a\mu} = s_{a\gamma} \ ,
\end{equation}
then the transformation law of the quantities $J_{ab}$ reduces to
\begin{equation}
J_{ab} = \left(\frac{p}{p'}\right)^3 J'_{ab} \ ,
\end{equation}
and thus the Stokes parameter $Q$, for example, transforms simply as
\begin{equation} \label{ch1:stokestrans}
Q = \left(\frac{p}{p'}\right)^3 Q' \ ,
\end{equation}
and similarly for the other Stokes parameters.  Since $t_{a\mu}$ is
assumed to be purely spatial in $K$ ($\vec{t}_a\cdot\vec{v}_l=0$), it
follows that $P^{\mu}_{\ \ \alpha}(\vec{p},\vec{v}_l)
\;t_{a\mu}=t_{a\alpha}$, and the transformation (\ref{ch1:ttrans1})
simplifies to
\begin{equation}
P^{\alpha}_{\ \ \gamma}(\vec{p},\vec{v}_e) t_{a\alpha} = s_{a\gamma} \
.
\end{equation}
In 4-vector notation, using $\vec{p}\cdot\vec{t}_a=0$ we find
\begin{equation}
\vec{s}_a = \vec{t}_a + \frac{1}{p'} (\vec{v}_e\cdot\vec{t}_a)\vec{p}
\end{equation}
This manifestly satisfies $\vec{s}_a\cdot\vec{v}_e=0$.  Since
$\vec{t}_a$ must be purely spatial in $K$ we have
\begin{equation}
\vec{t}_a\cdot\vec{v}_e = \left(\frac{p'}{p}\right)
\vec{s}_a\cdot\vec{v}_l \ ,
\end{equation}
which yields
\begin{equation} \label{ch1:tbasistrans}
\vec{t}_a = \vec{s}_a - \frac{1}{p} (\vec{v}_l\cdot\vec{t}_a)\vec{p} \
.
\end{equation}

The transformation law of the polarization basis 3-vectors
$\bm{\epsilon}_a, \;\bm{\epsilon}'_a$ now follows.  Denoting
$\vec{s}_a$ in lab frame coordinates as $s^{\mu}_a=(s^0_a,\bm{s}_a)$,
and Lorentz transforming $s^{\mu'}_a$ into $K$ we obtain
\begin{eqnarray}
s^0_a &=& \gamma \bm{v}\cdot\bm{\epsilon}'_a \ , \nonumber \\ \bm{s}_a
&=& \bm{\epsilon}'_a + (\gamma-1)\frac{\bm{v}
  (\bm{v}\cdot\bm{\epsilon}'_a)}{v^2} \ .
\end{eqnarray}
Thus the spatial part of Eqn.~(\ref{ch1:tbasistrans}) yields
\begin{equation} \label{ch1:basistrans}
\bm{\epsilon}_a = \bm{\epsilon}'_a + (\gamma-1)\frac{\bm{v}
  (\bm{v}\cdot\bm{\epsilon}'_a)}{v^2} -
\gamma\bm{n}(\bm{v}\cdot\bm{\epsilon}'_a) \ .
\end{equation}
This transformation law was previously obtained by
\cite{2002PhRvD..65j3001C}.  One may check, using the transformation
law of $\bm{n}$ (see Eqn.~(\ref{ch3:nvectrans})), that the
polarization basis 3-vectors $\bm{\epsilon}_a$ are indeed orthonormal
and orthogonal to $\bm{n}$.  The fact that such a complicated
transformation of basis is needed to ensure that the Stokes parameters
transform in a simple fashion demonstrates that the tensor approach is
more convenient when dealing with relativistic transformations of
polarized beams.

The \emph{screen projection tensor} $P^{\mu\nu}(\vec{p},\vec{v})$
defined in Eqn.~(\ref{ch1:projtens}) is an important tool in this
polarization tensor formalism.  It projects onto the ``screen''
subspace orthogonal to the photon momentum $\vec{p}$ and local
observer 4-velocity $\vec{v}$, in the sense that it leaves
$f^{\mu\nu}(\vec{p},\vec{v})$ invariant:
\begin{eqnarray}
P^{\mu}_{\ \ \alpha}(\vec{p},\vec{v}) f^{\alpha\beta}(\vec{p},\vec{v})
= f^{\mu\beta}(\vec{p},\vec{v}) \ .
\end{eqnarray}
In a local Lorentz frame $P^{\mu\nu}$ is simply the $2\times2$
identity matrix in the subspace orthogonal to $v^\mu$ and $p^\mu$.  It
is appropriate now to discuss some of its properties, which will be
useful to refer to in later sections.
It may also be written in the form (used in \cite{2000PhRvD..62d3004C}
and \cite{1981MNRAS.194..439T} for example)
\begin{eqnarray}
P_{\mu\nu} = g_{\mu\nu} + v_{\mu}v_{\nu} - n_\mu n_\nu \ ,
\end{eqnarray}
where $n^{\mu}$ is a spacelike unit vector giving the propagation
direction of the photon with respect to the observer:
\begin{equation}
\vec{p} = p \left(\vec{v}+\vec{n}\right) \ , \quad
\vec{n}\cdot\vec{v}=0 \ , \quad \vec{n}\cdot\vec{n}=1 \ , \quad
\vec{n}\cdot\vec{p}=p \ .
\end{equation}

In the rest frame of the observer with 4-velocity $\vec{v}$, the $00$
and $0i$ components of $P_{\mu\nu}(\vec{p},\vec{v})$ vanish, and the
spatial components are given by
\begin{eqnarray} \label{ch1:restPspatial}
P_{ij}(\vec{p},\vec{v}) = \delta_{ij} - n_i n_j \ .
\end{eqnarray}
where $\bm{n}$ is the photon direction 3-vector in the rest frame.  By
an obvious generalization of the argument leading to
Eqn.~(\ref{ch1:3times3unpol}), the distribution function tensor of an
unpolarized beam is given by
\begin{equation}\label{ch3:unpol}
f_{\mu\nu}(\vec{p},\vec{v}) = \frac{1}{2} f(\vec{p})
P_{\mu\nu}(\vec{p},\vec{v}) \ .
\end{equation}
There are the simple properties:
\begin{eqnarray}
v^{\mu}P_{\mu\nu}(\vec{p},\vec{v}) &=& 0 \ , \quad
p^{\mu}P_{\mu\nu}(\vec{p},\vec{v}) = 0 \ , \nonumber \\
g_{\mu\nu}P^{\mu\nu}(\vec{p},\vec{v}) &=& 2 \ , \quad
P^{\mu\nu}(\vec{p},\vec{v})P_{\mu\nu}(\vec{p},\vec{v}) = 2 \ ,
\nonumber \\ P^{\mu}_{\ \ \alpha}(\vec{p},\vec{v})P^{\alpha}_{\ \
  \nu}(\vec{p},\vec{v}) &=& P^{\mu}_{\ \ \nu}(\vec{p},\vec{v}) \ .
\end{eqnarray}
Contracting two projection tensors with the same photon momentum but
different observer velocities yields, with
$p_1\equiv-\vec{p}\cdot\vec{v}_1, \;p_2\equiv-\vec{p}\cdot\vec{v_2}$:
\begin{equation}
P^{\beta}_{\ \
  \lambda}(\vec{p},\vec{v}_1)P^{\gamma\lambda}(\vec{p},\vec{v}_2) =
g^{\beta\gamma} + \frac{p^{\gamma}v_2^{\beta}}{p_2} +
\frac{p^{\beta}v_1^{\gamma}}{p_1} + \frac{\vec{v}_1\cdot\vec{v}_2}{p_1
  p_2} p^{\beta}p^{\gamma} \ .
\end{equation}
Another contraction yields:
\begin{equation}\label{ch3:unpoltrans}
P^{\alpha}_{\ \ \gamma}(\vec{p},\vec{v}_1) P^{\beta}_{\ \
  \lambda}(\vec{p},\vec{v}_1)P^{\gamma\lambda}(\vec{p},\vec{v}_2) =
P^{\alpha\beta}(\vec{p},\vec{v}_1) \ ,
\end{equation}
which proves that a beam that is unpolarized according to some
observer is also unpolarized according to any other observer.

We close this section with a demonstration of the Lorentz
transformation of the polarization state of a beam
using polarization matrix manipulations.  This will serve as an
introduction to the more complicated matrix manipulations used in the
derivation of the SZE later.

We consider a photon with a general polarization state propagating in
the $z$--direction with momentum $p$. We will compute the polarization
matrix of the beam measured by an observer moving in the
$x$--direction with velocity $v$.  We work in the inertial frame $K$
(unprimed) with basis 4-vectors $\vec{e}_i$, and observer 4-velocity
$\vec{v}_0 = \vec{e}_t$ with unprimed components $v_0^{\mu} =
(1,0,0,0)$.  Let us consider a partially polarized photon beam
propagating in the $\vec{e}_z$ direction, with 4-momentum $\vec{p}$
with unprimed components $p^{\mu}=p(1,0,0,1)$, and distribution
function polarization tensor $f^{\mu\nu}(\vec{v}_0)$ as measured by
observer $\vec{v}_0$.  We suppress the photon 4-momentum argument of
the polarization tensor since we deal here with a monochromatic beam.

We may perform tensor manipulations by defining $4\times 4$ matrices
with entries equal to tensor components, with no distinction between
raised and lowered indices, provided there is a separate matrix for
each combination of raised and lowered indices.  Thus we define
$f^{\mu\nu}(\vec{v}_0) = \left[ \bm{f}(\vec{v}_0) \right]_{\mu\nu}$
where
\begin{eqnarray}
\bm{f}(\vec{v}_0) = \left( \begin{array}{cccc} 0 & 0 & 0 & 0 \\ 0 & a
  & b & 0 \\ 0 & b^* & d & 0 \\ 0 & 0 & 0 & 0
\end{array}
\right) \ ,
\end{eqnarray}
where the row elements from left to right and the column elements from
up to down refer to the $(t,x,y,z)$ components.  Choosing polarization
basis vectors $\bm{\epsilon}_1=\bm{x}, \;\bm{\epsilon}_2=\bm{y}$, the
coefficients $(a,b,c,d)$ are related to the usual specific intensity
Stokes parameters (here $h = c = 1$):
\begin{eqnarray} \label{ch1:origstokes}
a &=& \frac{I+Q}{2p^3}, \quad\quad\;\; b = \frac{U+iV}{2p^3} \ ,
\nonumber \\ b^* &=& \frac{U-iV}{2p^3}, \quad\quad d =
\frac{I-Q}{2p^3} \ .
\end{eqnarray}

Now we consider the polarization tensor measured by an observer moving
perpendicular to the photon momentum in the unprimed frame, with
4-velocity $\vec{v}$.  The rest frame of this observer is $K'$.  We
take 4-velocity $\vec{v}$ to have unprimed velocity components $v^\mu
= \gamma (1,v,0,0)$, where $\gamma=1/\sqrt{1-v^2}$.  Then the
polarization tensor measured by the observer with 4-velocity $\vec{v}$
has components in the unprimed frame as follows:
\begin{eqnarray}
f^{\mu\nu}(\vec{v}) = P^{\mu}_{\;\;\;\alpha}(\vec{v},\vec{p})
\;P^{\nu}_{\;\;\;\beta}(\vec{v},\vec{p}) f^{\alpha\beta}(\vec{v}_0) \ ,
\end{eqnarray}
where the projection tensor $P^{\mu}_{\;\;\;\nu}$ is given by:
\begin{eqnarray}
P^{\mu}_{\;\;\;\nu}(\vec{p},\vec{v}) = \eta^{\mu}_{\;\;\;\nu} +
\frac{1}{p'} \left( p^{\mu}v_{\nu} + v^{\mu}p_{\nu} \right) -
\frac{p^{\mu} p_{\nu}}{p'^2} \ ,
\end{eqnarray}
and $p' \equiv - \vec{p} \cdot \vec{v} = \gamma p$. In matrix form
$P^{\mu}_{\;\;\;\nu} = \left[ \bm{P}_1 \right]_{\mu\nu}$, where
\begin{eqnarray}
\bm{P_1} = \left( \begin{array}{cccc} -v^2 & v & 0 & v^2 \\ -v & 1 & 0
  & v \\ 0 & 0 & 1 & 0 \\ -v^2 & v & 0 & v^2
\end{array} \right) \ .
\end{eqnarray}
The lowered index quantity $P_{\mu\nu}$ is represented by a matrix
$\bm{P}_2$ with different entries:
\begin{eqnarray}
\bm{P}_2 = \left( \begin{array}{cccc} v^2 & -v & 0 & -v^2 \\ -v & 1 &
  0 & v \\ 0 & 0 & 1 & 0 \\ -v^2 & v & 0 & v^2
\end{array} \right) \ .
\end{eqnarray}
The idempotency relation satisfied by the projection tensor,
$P^{\mu}_{\;\;\;\alpha} P_{\mu\beta} = P_{\alpha\beta}$, implies the
matrix relation
\begin{eqnarray}
\bm{P}_1^T \bm{P}_2 &=& \bm{P}_2 \ ,
\end{eqnarray} 
which is satisfied by the matrices above.  Using the projection
matrices we find $f^{\mu\nu}(\vec{v}) =
\left[\bm{f}(\vec{v})\right]^{\mu\nu}$ where
\begin{eqnarray}
\bm{f}(\vec{v}) = \bm{P}_1 \bm{f}(\vec{v}_0) \bm{P}_1^T =
\left( \begin{array}{cccc} av^2 & av & bv & av^2 \\ av & a & b & av \\
  b^* v & b^* & d & b^* v \\ av^2 & av & bv & av^2
\end{array} \right) \ .
\end{eqnarray}

Now we would like to obtain the Stokes parameters measured by the
observer with 4-velocity $\vec{v}$.  This is given by the Lorentz
transformed matrix $\bm{f'}=\bm{\Lambda} \bm{I} \bm{\Lambda}^T$, where
in this case the matrix $\bm{\Lambda}$ is a boost matrix in the
$x$-direction:
\begin{eqnarray}
\bm{\Lambda} = \left( \begin{array}{cccc} \gamma & -\gamma v & 0 & 0
  \\ -\gamma v & \gamma & 0 & 0 \\ 0 & 0 & 1 & 0 \\ 0 & 0 & 0 & 1
\end{array} \right) \ .
\end{eqnarray}
Thus the polarization matrix of the beam measured by the observer at
rest in $K'$, in primed coordinates, is
\begin{eqnarray}
\bm{f'}(\vec{v}) = \left( \begin{array}{cccc} 0 & 0 & 0 & 0 \\ 0 &
  a/\gamma^2 & b/\gamma & av/\gamma \\ 0 & b^*/\gamma & d & b^* v \\ 0
  & av/\gamma & bv & av^2
\end{array} \right) \ .
\end{eqnarray}
Adding the diagonal elements of this matrix yields $a+d$, since the
total photon occupation number $f^{'\mu\nu}\eta_{\mu\nu}=I/p^3=a+d$ is
a Lorentz invariant quantity. The $ti$ and $it$ elements are zero
since the electric field 4-vectors in this frame are purely spatial by
definition of the tensor $f^{\mu\nu}(\vec{v})$. A matrix of this form
is termed a \emph{physical polarization matrix}, since its elements
correspond to quantities measured by a polarimeter in this frame.

We now examine the Stokes parameters measured by an observer with
4-velocity $\vec{v}$ according to the polarization tensor derived, in
order to check that our formalism agrees with the known transformation
properties of the Stokes parameters.  To obtain the Stokes parameters
in the boosted frame, we need to define a photon polarization basis.
The basis given by Eqn.~(\ref{ch1:basistrans}) should guarantee that
the Stokes (divided by the cube of the momentum) are invariant under
the boost.  In the unprimed frame, the basis vectors were
$\bm{\epsilon}_1=\bm{x}, \;\bm{\epsilon}_2=\bm{y}$.  In the lab frame
(here primed), the photon momentum is $p^{\mu'}=p(\gamma,-\gamma
v,0,1)$, the photon direction is $\bm{n'}=(-v,0,1/\gamma)$, and the
lab frame velocity is $\bm{v'}=-v\bm{x}$.  Substituting these into
(\ref{ch1:basistrans}) yields
\begin{eqnarray}
\bm{\epsilon}_1' &=& (0,1/\gamma,0,v) \ , \nonumber \\
\bm{\epsilon}_2' &=& (0,0,1,0) \ ,
\end{eqnarray}
which are clearly orthonormal and orthogonal to the primed photon
momentum, and reduce in the limit $v \rightarrow 0$ to the unprimed
basis.  A more general polarization basis is obtained by rotating
these vectors through an angle $\chi$ about the photon momentum, as
follows:
\begin{eqnarray}
\bm{\epsilon_1'} &\rightarrow& \;\cos\chi \;\bm{\epsilon_1'} -
\sin\chi \;\bm{\epsilon_2'} = (0,\cos\chi/\gamma,-\sin\chi,v\cos\chi)
\ , \nonumber \\ \bm{\epsilon_2'} &\rightarrow& \;\cos\chi
\;\bm{\epsilon_2'} + \sin\chi \;\bm{\epsilon_1'} =
(0,\sin\chi/\gamma,\cos\chi,v\sin\chi) \ .
\end{eqnarray}
In this rotated basis, the Stokes parameters in the boosted frame are
given by the quantities
\begin{eqnarray}
a' &=& \bm{\epsilon_1'}^T \cdot \bm{f'} \bm{\epsilon_1'} = a\cos^2\chi
- (b+c) \cos\chi \sin\chi + d\sin^2\chi \ , \nonumber \\ b' &=&
\bm{\epsilon_1'}^T \cdot \bm{f'} \bm{\epsilon_2'} = b\cos^2\chi +
(a-d) \cos\chi \sin\chi - b^*\sin^2\chi \ , \nonumber \\
d' &=& \bm{\epsilon_2'}^T \cdot \bm{f'} \bm{\epsilon_2'} = d\cos^2\chi
+ (b+c) \cos\chi \sin\chi + a \sin^2\chi \ .
\end{eqnarray}
Thus the measured Stokes parameters in the primed frame are given by:
\begin{eqnarray}
\frac{I'}{(p')^3} &=& a' + d' = a + d = \frac{I}{p^{3}} \ , \nonumber
\\ \frac{Q'}{(p')^3} &=& a' - d' = (a-d) \cos 2\chi - (b+b^*) \sin
2\chi = \frac{Q}{p^3} \cos 2\chi - \frac{U}{p^3} \sin 2\chi \ ,
\nonumber \\ \frac{U'}{(p')^3} &=& b' + b'^* = (b+b^*) \cos 2\chi +
(a-d) \sin 2\chi = \frac{U}{p^3} \cos 2\chi + \frac{Q}{p^3} \sin 2\chi
\ , \nonumber \\ \frac{V'}{(p')^3} &=& i(b'-b'^*) = i(b-b^*) =
\frac{V}{p^3} \ .
\end{eqnarray}
We find that with the choice $\chi=0$, the Stokes parameters transform
as claimed in Eqn.~(\ref{ch1:stokestrans}), and with a general $\chi$
the Stokes $Q,\,U,\,V$ transform in the expected fashion under
rotation of the polarization basis vectors.

\begin{figure*}[tb]
\begin{center}
(a) \includegraphics[width=6cm]{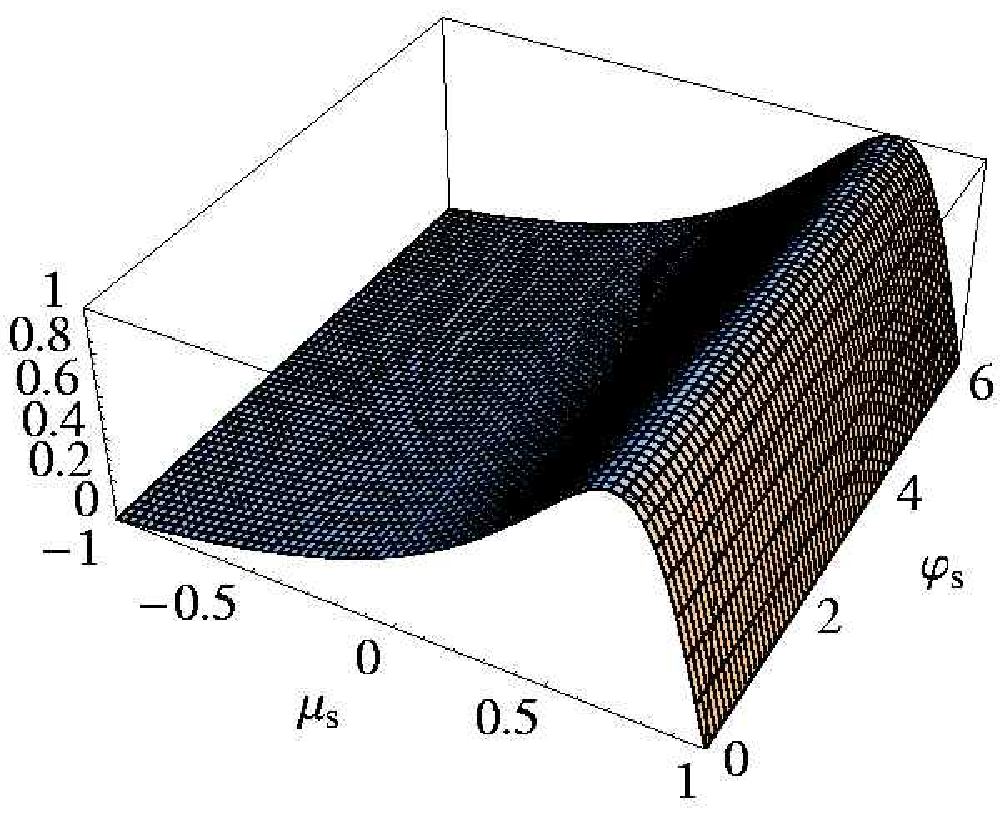} (b)
\includegraphics[width=6cm]{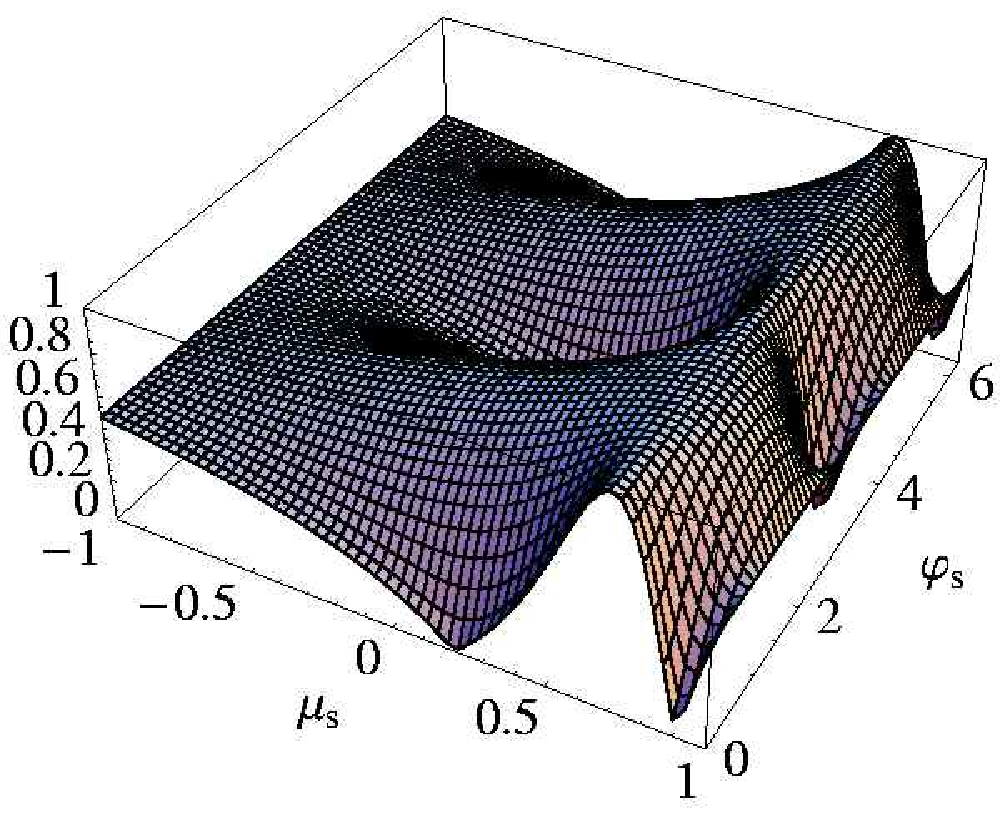}
\end{center}
\caption{\label{ch1:boost1} Polarization magnitude of the radiation
  field produced by Thomson scattering of a beam incident along the
  $z$-direction on an electron with velocity $0.7 c$ also along the
  $z$-direction, for the following two cases of the polarization state
  of the incident beam: (a) $Q=U=V=0$, (b) $Q/I=1/2, \;U=V=0$ (where
  the Stokes parameters are defined in the basis
  $\bm{\epsilon}_1=\bm{x}, \,\bm{\epsilon}_2=\bm{y}$).}
\label{phaseplot}
\end{figure*}

Fig. \ref{ch1:boost1}
illustrates the results of using this procedure to compute the
polarization magnitude $\Pi(\mu_s,\varphi_s)$ of the radiation field
produced by Thomson scattering of a polarized beam incident along the
$z$-direction (in lab) on a relativistic electron, as a function of
the polar angles ($\mu_s=\cos\theta_s, \,\varphi_s$) of the scattered
photon about the $z$-axis (see Eqn.~(\ref{ch3:genpmag})).  The Thomson
scattered radiation field in the electron rest frame is derived in the
next section.  The polarization matrix of the scattered beam in the
electron rest frame was computed, as given in
Eqn.~(\ref{ch3:fsscatt}), and then boosted to lab frame, where the
polarization magnitude was computed.  The polarization magnitude is a
scalar and thus changes under the boost due entirely to the aberration
of the photon direction.

\begin{figure*}[tb]
\begin{center}
(a) \includegraphics[width=6cm]{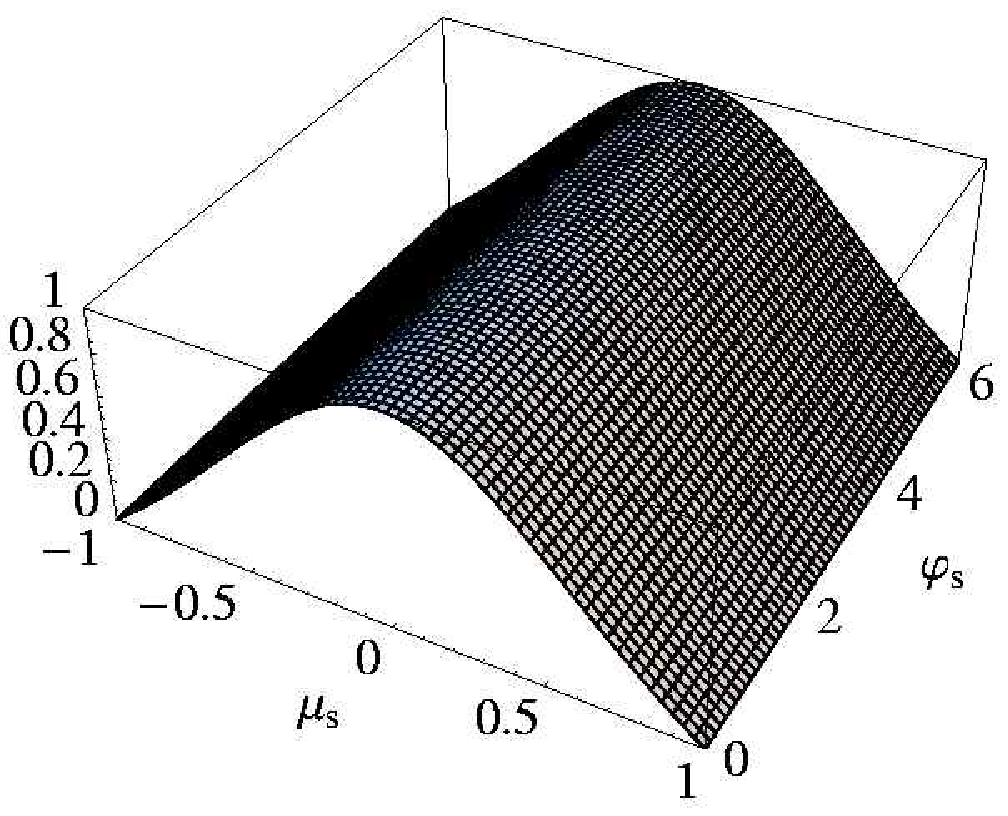} (b)
\includegraphics[width=6cm]{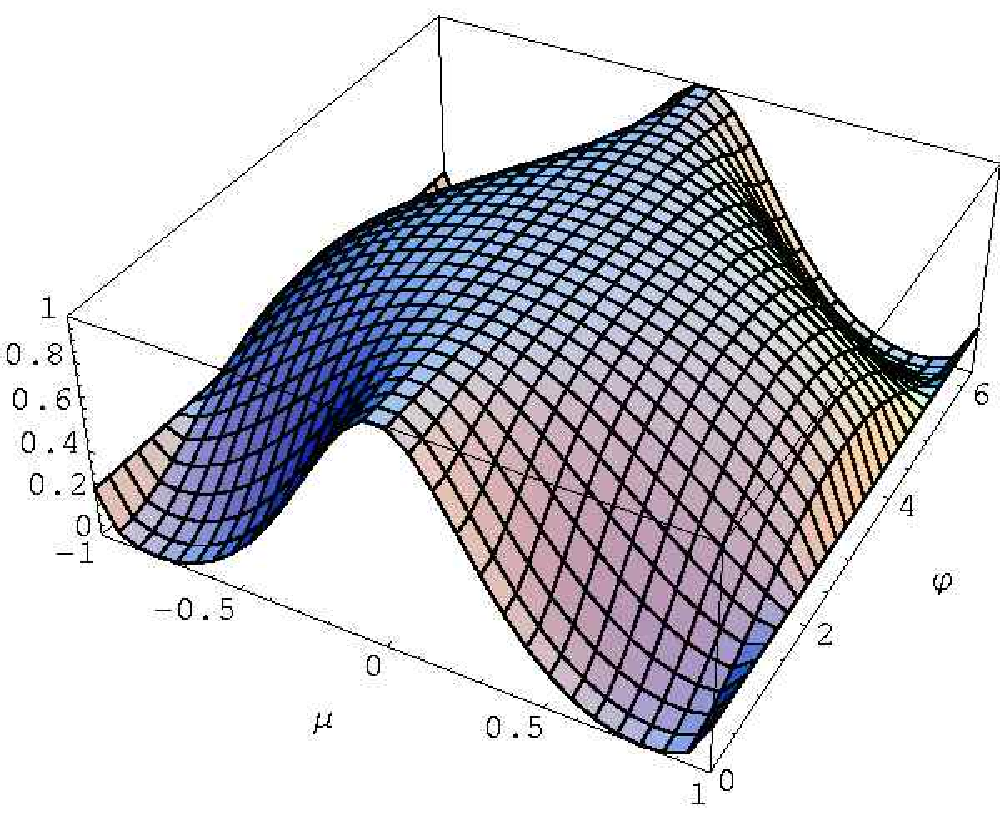}
\end{center}

\begin{center}
(c) \includegraphics[width=6cm]{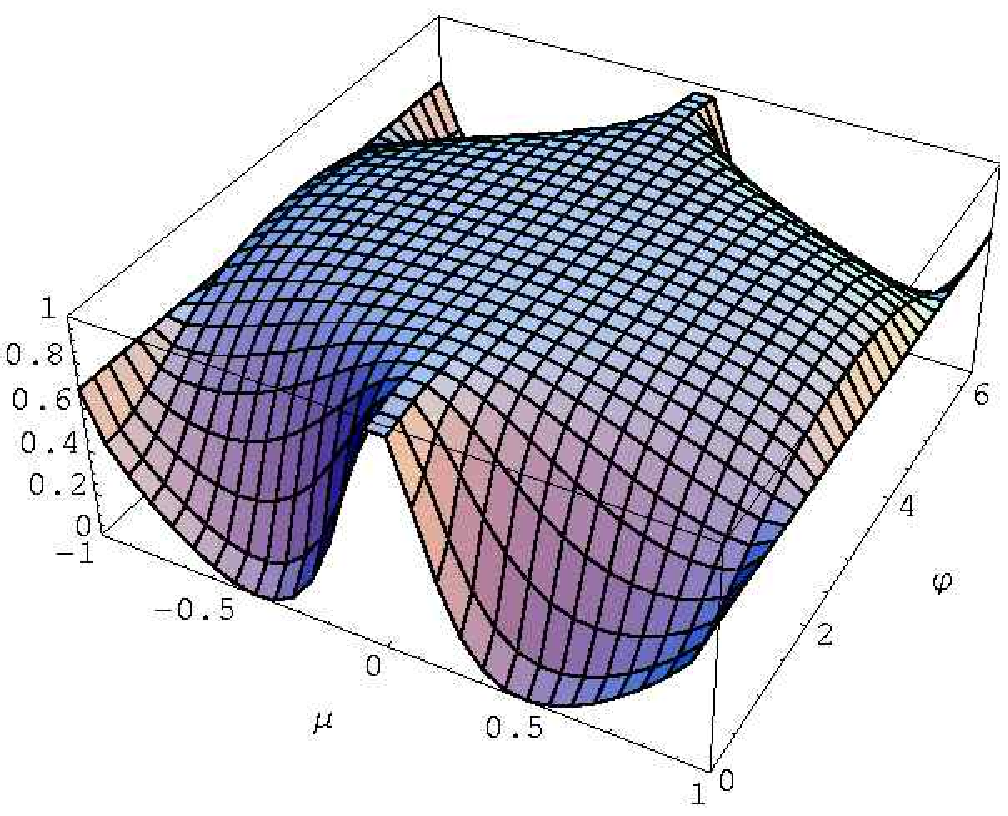} (d)
\includegraphics[width=6cm]{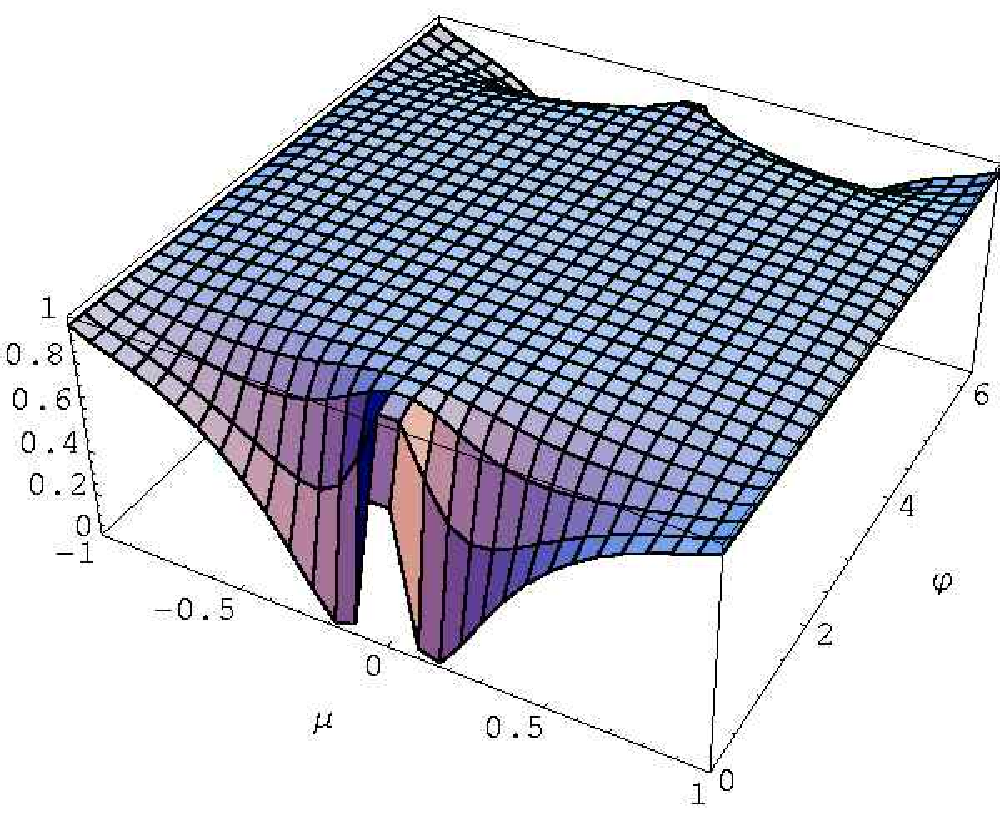}
\end{center}

\caption{\label{ch1:boost2} Polarization magnitude of the radiation
  field produced by Thomson scattering of an unpolarized beam incident
  along the $z$-direction scattering from an electron moving along the
  $x$-direction with velocity (a) zero, (b) $0.4c$, (c) $0.9c$, (d)
  $0.99c$.}
\end{figure*}

\section{Thomson scattering \label{ch3:sec1}}

\begin{figure}[tb]
\vspace{1.0cm}
\begin{center}
\includegraphics[width=8cm]{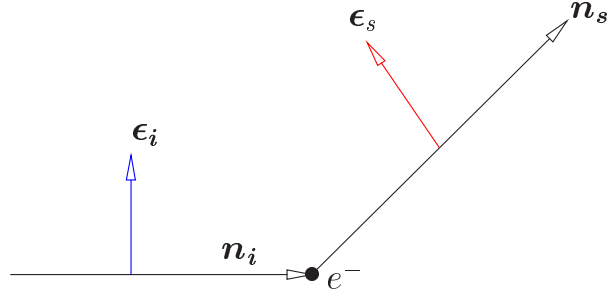}
\end{center}
\caption{Thomson scattering of a pure incident beam from an electron
  at rest into a specified final polarization
  state. \label{ch3:thompfig}}
\end{figure}
In this section we present a derivation of the equation for the time
evolution of the distribution function polarization tensor due to
Thomson scattering from a distribution of stationary electrons,
starting from the classical results for Thomson scattering, ignoring
the effects of electron recoil and induced scattering.
Note that throughout this and subsequent sections we work in flat
spacetime. 

Recall that for a completely linearly polarized beam,
$Q^{\mu\nu}\epsilon^\ast_\mu \epsilon_\nu$ is the time-average energy
density for electromagnetic radiation of polarization $\epsilon_\mu$,
where $\epsilon_\mu$ is spacelike and normalized, $\vec\epsilon^{\
  \ast} \cdot\vec\epsilon\equiv
g^{\mu\nu}\epsilon^\ast_\mu\epsilon_\nu =1$.  Consider a completely
polarized beam with polarization vector $\vec\epsilon_i$ and momentum
$\vec{p}_i$ incident upon an electron at rest
(Fig.~\ref{ch3:thompfig}).  The polarization matrix of the incident
beam is $Q_i\vec\epsilon_i\otimes\vec\epsilon_i^{\ast}$ where $cQ_i$
is the incident flux (we choose units such that $c=1$). Normalization
of the polarization vector implies $Q_i=Q_i^{\mu\nu}\epsilon^\ast_{i\,
  \mu}\epsilon_{i\,\nu}$. In the Thomson limit, in which the electron
recoil is negligible, the differential cross section for Thomson
scattering of a beam into final momentum $\vec p_s$ and polarization
xs$\vec\epsilon_s$ is \citep{Jackson}
\begin{eqnarray}\label{ch3:thomxsec}
  \frac{d\sigma}{d\Omega_s}=\frac{3\sigma_{\rm T}}{8\pi}
  \left\vert\vec\epsilon^ {\ \ast}_i\cdot\vec\epsilon_s\right\vert^2 \
  .
\end{eqnarray}
Thus the power per unit solid angle in the scattered beam is
\begin{equation}\label{ch3:pscatt}
  {dP_s\over d\Omega_s}={3\sigma_{\rm T}\over8\pi}
  \,Q_i\left\vert\vec\epsilon^ {\
    \ast}_i\cdot\vec\epsilon_s\right\vert^2
\end{equation}
where $d\Omega_s$ is the element of solid angle associated with the
direction of $\vec p_s$.  We may also write $Q_i\vert\vec \epsilon^{\
  \ast}_i\cdot\vec\epsilon_s\vert^2=Q_i^{\mu\nu}
\epsilon^\ast_{s\,\mu} \epsilon_{s\,\nu}$.

Next consider a gas of electrons all at rest with number density
$n_e$: we work in the rest frame of the electrons throughout this
section.  Assuming incoherent scattering, multiplying
Eqn.~(\ref{ch3:pscatt}) by $n_e d\Omega_s$ converts scattered power
per electron to the rate of change of energy density in final
polarization state $\vec\epsilon_s$:
\begin{equation}\label{tscatt1}
  {d Q_s^{\mu\nu}\over d t}\,\epsilon^\ast_{s\,\mu}
  \epsilon_{s\,\nu}={3\sigma_{\rm T}\over8\pi}n_e
  Q_i^{\mu\nu}\epsilon^\ast_{s\,\mu}\epsilon_{s\,\nu}\, d\Omega_s\ .
\end{equation}
Note that the time derivative $d/dt$ here should actually be
interpreted as a total derivative taken along the ray,
$d/dt=\partial/\partial t + \bm{n}\cdot\bm{\nabla}$, since eventually
the left hand side of the evolution equation will be replaced with the
left hand side of Eqn.~(\ref{ch3:genboltz}).  Using equation
(\ref{ch1:def_fmunu}), and setting $p_i=p_s$ since we are working in
the Thomson limit, we may convert this to the change in the phase
space density matrix, giving
\begin{equation}\label{tscatt2}
  {d f_s^{\mu\nu}\over dt}\,\epsilon^\ast_{s\,\mu} \epsilon_{s\,\nu}
  d\Omega_s={3\sigma_{\rm T}\over8\pi}n_e
  f_i^{\mu\nu}\epsilon^\ast_{s\,\mu}\epsilon_{s\,\nu}\,
  d\Omega_id\Omega_s\ .
\end{equation}

We would like an equation for the change in $f_{\mu\nu}$ due to
scattering, but Eqn.~(\ref{tscatt2}) gives the change only for a
particular (but arbitrary) polarization of the outgoing wave,
$\vec\epsilon_s$.  We cannot remove the polarization factors and
conclude $d f_s^{\mu\nu}\propto f_i^{\mu\nu}$ because the polarization
of the incoming wave does not lie in the same plane as the
polarization of the scattered wave.  For a given outgoing momentum
$\vec p_s$, the outgoing polarization is a linear combination of the
two basis vectors $\vec \epsilon_1$ and $\vec \epsilon_2$ (associated
with the photon of momentum $\vec{p}_s$) of \S \ref{ch1:sec3}.  Thus,
$f_i^{\mu\nu}\epsilon^\ast_{s\,\mu} \epsilon_{s\,\nu}$ projects out of
the incoming density matrix $f_i^{\mu\nu}$ only those components lying
in the $\vec \epsilon_1$-$\vec \epsilon_2$ plane.  This projection is
equivalent to first projecting $f_i^{\mu\nu}$ with $\vec
\epsilon_1\otimes\vec \epsilon_1+\vec \epsilon_2\otimes\vec
\epsilon_2$.  But this is exactly the projection tensor of
Eqn.~(\ref{ch1:projtens}), with $\vec p=\vec p_s$ being the outgoing
photon momentum and $\vec v$ being the electron 4-velocity.
Projecting the final polarization vector with $P_{\mu\nu}$ does not
change it: $P^\mu_{\ \ \alpha}\epsilon_{s\,\mu}=\epsilon_{s\,
  \alpha}$.  It follows that $f_i^{\mu\nu}\epsilon^\ast_{s\,\mu}
\epsilon_{s\,\nu}=f_i^{\alpha\beta}P^\mu_{\ \ \alpha}P^\nu_{\ \
  \beta}\epsilon^\ast_{s\,\mu}\epsilon_{s\,\nu}$.  Now it is safe to
remove the outgoing polarization vectors from Eqn.~(\ref{tscatt2}).

We conclude that, for any initial and final polarizations,
\begin{eqnarray}\label{ch3:fscatt1}
  {d f_s^{\mu\nu}(\vec{p}_s,\vec{v}_e)\over d t}= {3\sigma_{\rm
      T}\over8\pi}n_e \, P^\mu_{\ \ \alpha}(\vec p_s,\vec v_e\,)
  P^\nu_{\ \ \beta}(\vec p_s,\vec v_e\,)\, \int d\Omega_i
  \,f_i^{\alpha\beta}(\vec{p}_i,\vec{v}_e) \ .
\end{eqnarray}
Eqn.~(\ref{ch3:fscatt1}) is the key result for Thomson scattering in
the polarization tensor formalism.  It gives the photon scattering
rate per unit volume for given momenta and polarizations.  

If this argument seems a little dry, we note that the appearance of
these projection tensor follows straightforwardly 
from elementary classical electrodynamics. Consider scattering of a quasi-monochromatic beam with central
frequency $\omega$ incident in direction $\bm{n}^{(i)}$ by a single
electron at rest. Let the complex amplitude of the (analytic signal of
the) incident electric field be $\bm{E}^{(i)}(t)$, with
$\bm{E}^{(i)}\cdot\bm{n}^{(i)}=0$. The incident wavevector is
$\bm{k}=\omega \bm{n}^{(i)}/c$.
The complex dipole moment of the radiating electron induced by the
incoming wave is $\bm{p} = (e^2/m \omega^2)\bm{E}^{(i)}$. The analytic
signal of the electric field of the dipole radiation in the far field
produced by the oscillating electron (the scattered field) at position
$\bm{x}'=x'\bm{n}_s$ is, using the formula for electric dipole
radiation \citep{Jackson},
\begin{eqnarray}
\bm{E}_s &=& \frac{\mu_0}{4\pi} \frac{e^{i\omega x'/c}e^{i\omega
    t}}{x'} \omega^2 \left[\bm{p} - (\bm{n}_s\cdot \bm{p})\bm{n}_s\right] \ .
\end{eqnarray}
The matrix $Q_{ij}$ matrix of the scattered beam is thus given by
\begin{eqnarray} \label{ch1:thomsonscatt}
x'^2 d\Omega' \;Q_{ij}(\bm{n}_s) &=& x'^2 d\Omega' 
\;\langle E_{r,i} E_{r,j}^* \rangle \nonumber \\ 
&=& \frac{3}{4\pi} \sigma_T d\Omega' \;\left[\bm{P}(\bm{n_s})
  \langle\bm{Q}(\bm{n}^{(i)})\rangle \bm{P}(\bm{n_s})\right]_{ij} \ , 
\end{eqnarray}
where $\sigma_T = (e^2/mc^2)^2/12\pi\epsilon_0^2$ is the Thomson cross
section (in SI units), and we have defined
\begin{eqnarray}
\left[\bm{Q}(\bm{n}^{(i)})\right]_{ij} &=& \langle E_i^{(i)}(t)
E_j^{(i)}(t) \rangle  \nonumber \\
\left[\bm{P}(\bm{n}_s)\right]_{ij} &=& \delta_{ij} - [\bm{n}_s]_i [\bm{n}_s]_j \ .
\end{eqnarray}
The matrix $\bm{P}(\bm{n}_s)$ is the projection tensor we saw before,
which can be thought of as selecting the components of the incoming
field which are transverse to the wavevector of the scattered field.
The extension to a arbitrary ``polychromatic'' incident beam with
frequencies not restricted to a small waveband follows provided the
electric field components in seperate wavebands are completely
uncorrelated.  

If the integration time is sufficiently short that we can ignore
multiple scatterings, we may replace $n_e
\sigma_{\rm T}\,dt$ in Eqn.~(\ref{ch3:fscatt1}) with the optical depth
to Thomson scattering, $\tau$. Then we have
\begin{eqnarray}\label{ch3:fscatt2}
f_s^{\mu\nu}(\vec{p}_s,\vec{v}_e) = {3\tau\over8\pi} \, P^\mu_{\ \
  \alpha}(\vec p_s,\vec v_e\,) P^\nu_{\ \ \beta}(\vec p_s,\vec
v_e\,)\, \int d\Omega_i \, f_i^{\alpha\beta}(\vec{p}_i,\vec{v}_e) \ .
\end{eqnarray}
It follows that scattering of a photon with a given incident momentum
$\vec{p}_i$ leads to a scattered beam with normalized polarization
tensor
\begin{eqnarray}\label{ch3:phiscatt}
\phi_s^{\mu\nu}(\vec{p}_s,\vec{v}_e) = \frac{ P^\mu_{\ \ \alpha}(\vec
  p_s,\vec v_e\,) P^\nu_{\ \ \beta}(\vec p_s,\vec v_e\,)\,
  \phi_i^{\alpha\beta}(\vec{p}_i,\vec{v}_e)} {P_{\gamma\delta}(\vec
  p_s,\vec v_e\,)\phi_i^{\gamma\delta}(\vec{p}_i,\vec{v}_e)} \ .
\end{eqnarray}

Taking the trace of the scattering rate Eqn.~(\ref{ch3:fscatt1})
yields the evolution equation for the scalar distribution functions
$f_i(\vec{p}_i)=g_{\alpha\beta}f_i^{\alpha\beta}(\vec{p}_i)$,
$f_s(\vec{p}_i)=g_{\alpha\beta}f_s^{\alpha\beta}(\vec{p}_s)$, which
may be written in the following form (by definition of the
differential scattering cross section):
\begin{equation}\label{ch3:fscatt2}
  \frac{d}{dt}(f_s\,d^3x\,d^3p_s)=(n_e\,d^3x)\,\frac{d\sigma}
       {d\Omega_s}\,d\Omega_s\,(f_i\,d^3p_i) \ ,
\end{equation}
which yields the differential scattering cross section in the rest
frame (and the Thomson limit):
\begin{eqnarray}\label{ch3:thompxsec}
\frac{d\sigma}{d\Omega_s} = \frac{3\sigma_{\rm T}}{8\pi}
\;P_{\alpha\beta}(\vec{p}_s,\vec{v}_e)
\phi_i^{\alpha\beta}(\vec{p}_i,\vec{v}_e) \ .
\end{eqnarray}

In rest frame coordinates, we may deal with $3\times 3$ matrices rather
than tensors. Then we may write the normalized polarization matrix of
the scattered beam in terms of that of the incident beam as:
\begin{equation} \label{ch3:fsscatt}
\bm{\phi}_s(\bm{n}_s) =
\frac{\bm{P}(\bm{n}_s)\bm{\phi}_i(\bm{n}_i)\bm{P}(\bm{n}_s)
}{\mbox{Tr}[\bm{P}(\bm{n}_s)\bm{\phi}_i(\bm{n}_i)]} \ .
\end{equation}
The polarization magnitude of the beam reduces to a familiar form in
the case of an unpolarized incident beam. For example, consider the
case of an unpolarized beam incident in the $z$-direction in the rest
frame, and let the rest frame direction vector of the scattered beam
have components
\begin{equation} \label{ch3:ns}
\bm{n}_s=(\cos\varphi_s\sqrt{1-\mu_s^2},\sin\varphi_s\sqrt{1-\mu_s^2},\mu_s)
\ .
\end{equation}
The incident normalized polarization matrix is
$\bm{\phi}(\bm{n}_i)=\bm{P}(\bm{n}_i)/2$.
\begin{figure*}[tb]
\begin{center}
(a) \includegraphics[width=6cm]{Pmag1.eps} (b)
\includegraphics[width=6cm]{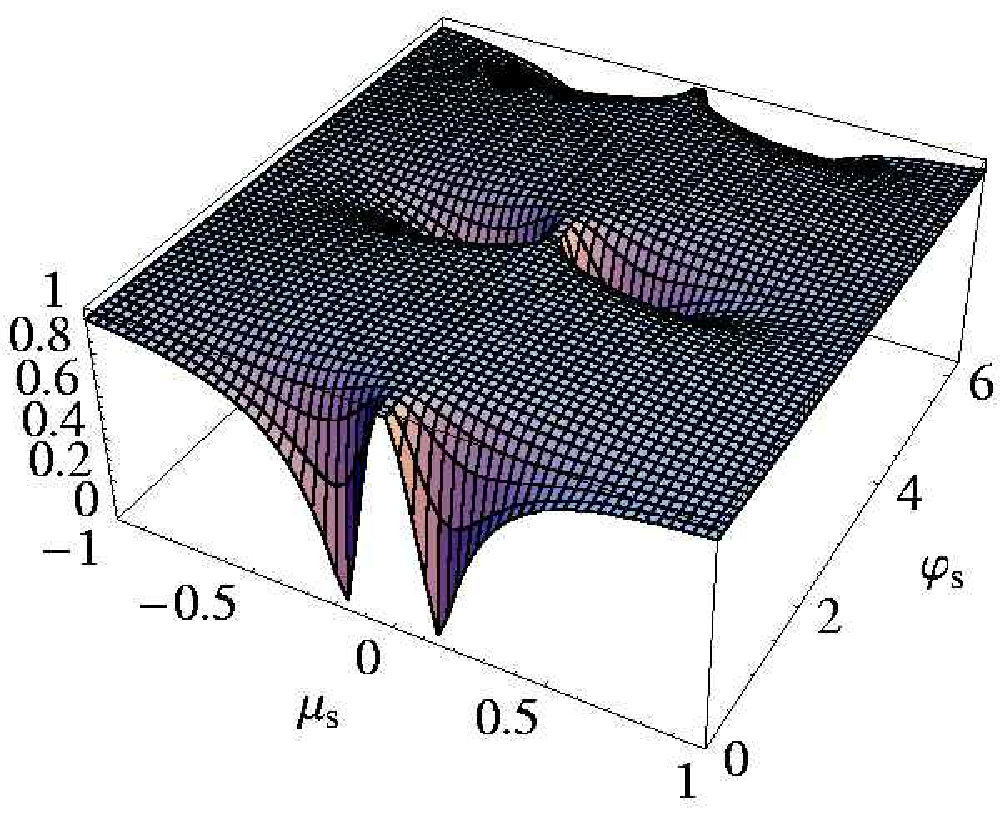}
\end{center}

\begin{center}
(c) \includegraphics[width=6cm]{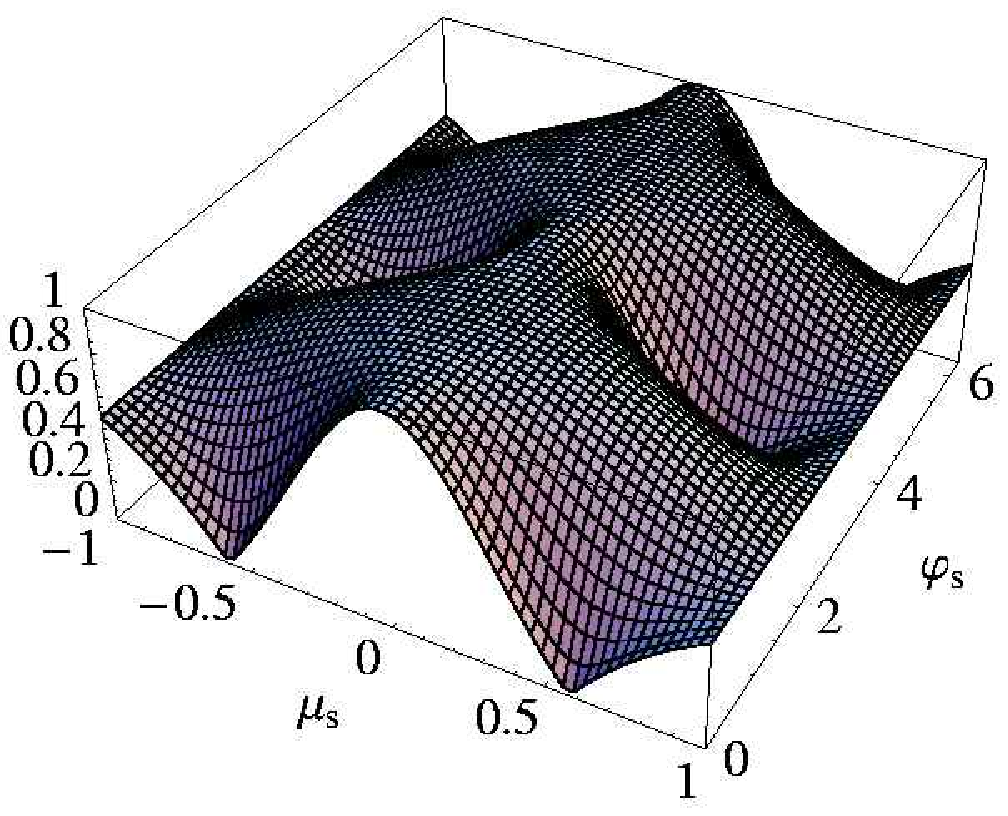} (d)
\includegraphics[width=6cm]{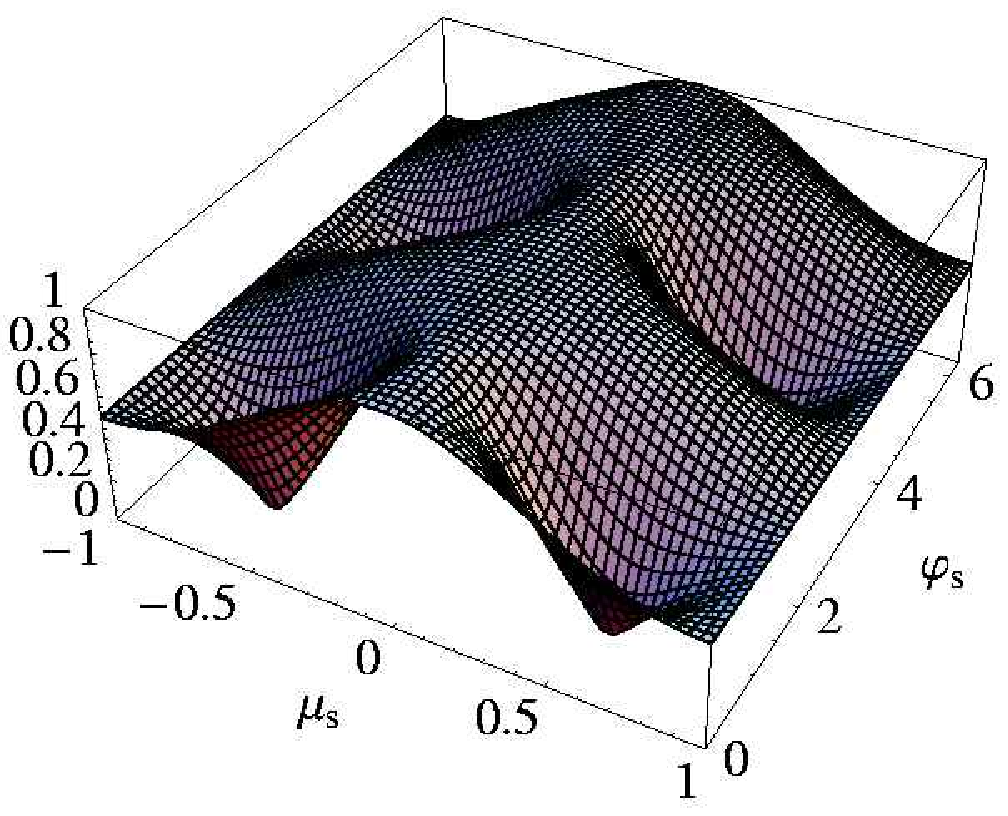}
\end{center}


\caption{Polarization magnitude versus rest frame scattering angles
  (in polar coordinates).  The plots for incident beams with Stokes
  parameters: (a) $Q=U=V=0$, (b) $Q/I=0.9, \;U=V=0$, (c) $Q/I=0.5,
  \;U=V=0$, (d) $U/I=0.5, \;Q=V=0$ .\label{ch3:pmagfig}}
\label{maxwellsample}
\end{figure*}
The polarization magnitude of the scattered beam is given by
\begin{equation}
\Pi^2(\mu_s,\varphi_s) =
2\mbox{Tr}[\bm{\phi}_s^2]/\mbox{Tr}[\bm{\phi}_s]^2-1=
\left(\frac{1-\mu_s^2}{1+\mu_s^2}\right)^2 \ ,
\end{equation}
which is independent of $\varphi_s$ since the incident unpolarized
beam picks out no preferred azimuth.  In the case of an incident beam
with a general polarization state, we may choose polarization basis
3-vectors
\footnote{The unit 3-vectors pointing along the Cartesian coordinate
  axes $x,y,z$ are denoted $\bm{e}_x, \bm{e}_y, \bm{e}_z$.}  for the
incident beam $\bm{\epsilon}_1=\bm{e}_x,\,\bm{\epsilon}_2=\bm{e}_y$
and write, in the $2\times 2$ polarization subspace,
\begin{equation}
\bm{\phi}_i = \frac{1}{2I}\left(\begin{array}{cc} I+Q & U+iV \\ U-iV &
  I-Q \end{array} \right) \ .
\end{equation}
Then we find the scattered polarization magnitude
\begin{equation} \label{ch3:genpmag}
\Pi^2(\mu_s,\varphi_s) = 1 + \frac{4\,\left( -I^2 + Q^2 + U^2 + V^2
  \right) \,{\mu_s }^2} {I^2{\left[ 1 + {\mu_s }^2 + \left( -1 +
      {\mu_s }^2 \right) \left(Q \,\cos (2\,\varphi_s ) + U
      \,\sin(2\,\varphi_s)\right)/I\right]}^2} \ .
\end{equation}
This may also be written as
\begin{equation} \label{ch3:genpmag2}
\Pi^2(\mu_s,\varphi_s) = 1 + \frac{\left( -I^2 + Q^2 + U^2 + V^2
  \right) \,{\mu_s }^2}
   {I^2{\left[1-n_{si}n_{sj}\phi_i^{ij}\right]}^2} \ .
\end{equation}
This function is plotted for incident beams with various polarization
states in Fig.~\ref{ch3:pmagfig}.

From Eqn.~(\ref{ch3:fsscatt}) we can determine the probability for a
photon to Thomson scatter into a particular solid angle element
$d\Omega_s$, which is conventionally termed the \emph{phase function}.
This is simply proportional to the differential cross section, which
in matrix notation is
\begin{eqnarray}
\frac{d\sigma}{d\Omega_s} = \frac{3\sigma_{\rm T}}{8\pi}
\;\mbox{Tr}[\bm{\phi}_i(\bm{n}_i)\bm{P}(\bm{n}_s)] \ .
\end{eqnarray}
Thus the phase function for Thomson scattering is a function of the
scattered direction vector $\bm{n}_s$ and the elements of the incident
polarization matrix $\bm{\phi}_i(\bm{n}_i)$ (the dependence on
$\bm{n}_i$ is implicit in $\bm{\phi}_i(\bm{n}_i)$).  Denoting the
phase function as $\Phi[\bm{n}_s,\bm{\phi}_i(\bm{n}_i)]$, we use the
normalization
\begin{equation}
\int \Phi[\bm{n}_s,\bm{\phi}_i(\bm{n}_i)] \;\frac{d\Omega_s}{4\pi} = 1
\ .
\end{equation}
Since $\int P_{ij}(\bm{n}_s) \;d\Omega_s/4\pi = 2\delta_{ij}/3$, we
have
\begin{equation} \label{ch3:phasefunc}
\Phi[\bm{n}_s,\bm{\phi}_i(\bm{n}_i)] = \frac{3}{2}
\;\mbox{Tr}[\bm{\phi}(\bm{n}_i)\bm{P}(\bm{n}_s)] \ .
\end{equation}

\begin{figure*}[tb]
\begin{center}
(a) \includegraphics[width=6cm]{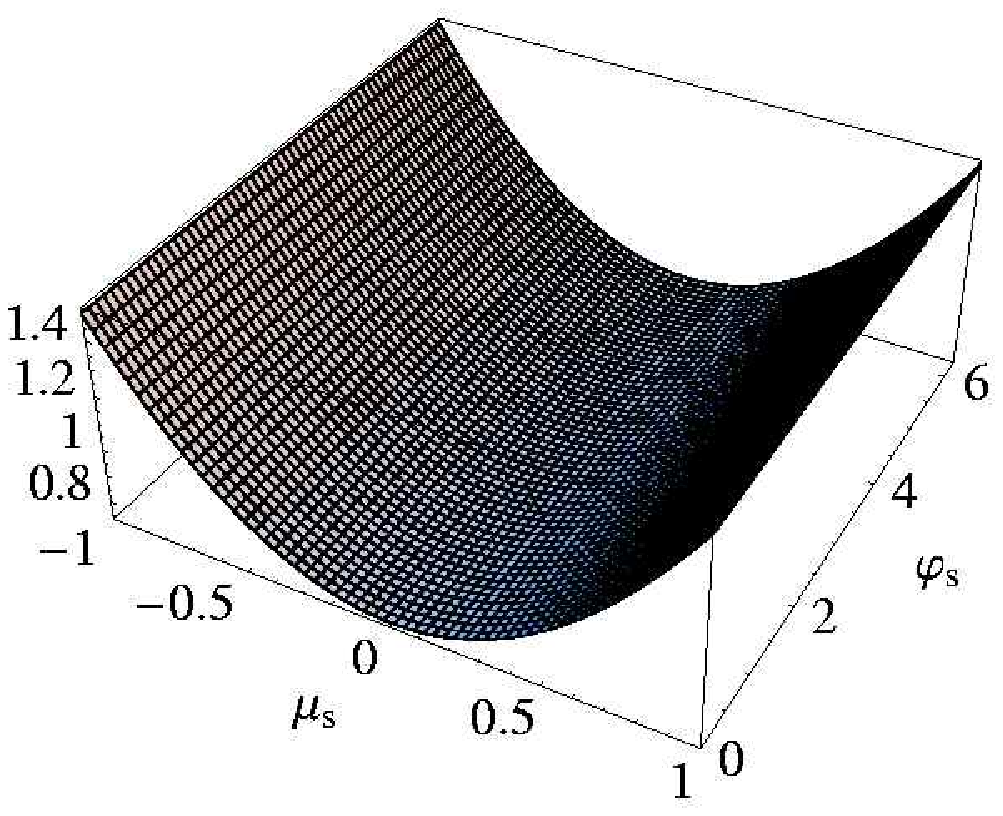} (b)
\includegraphics[width=6cm]{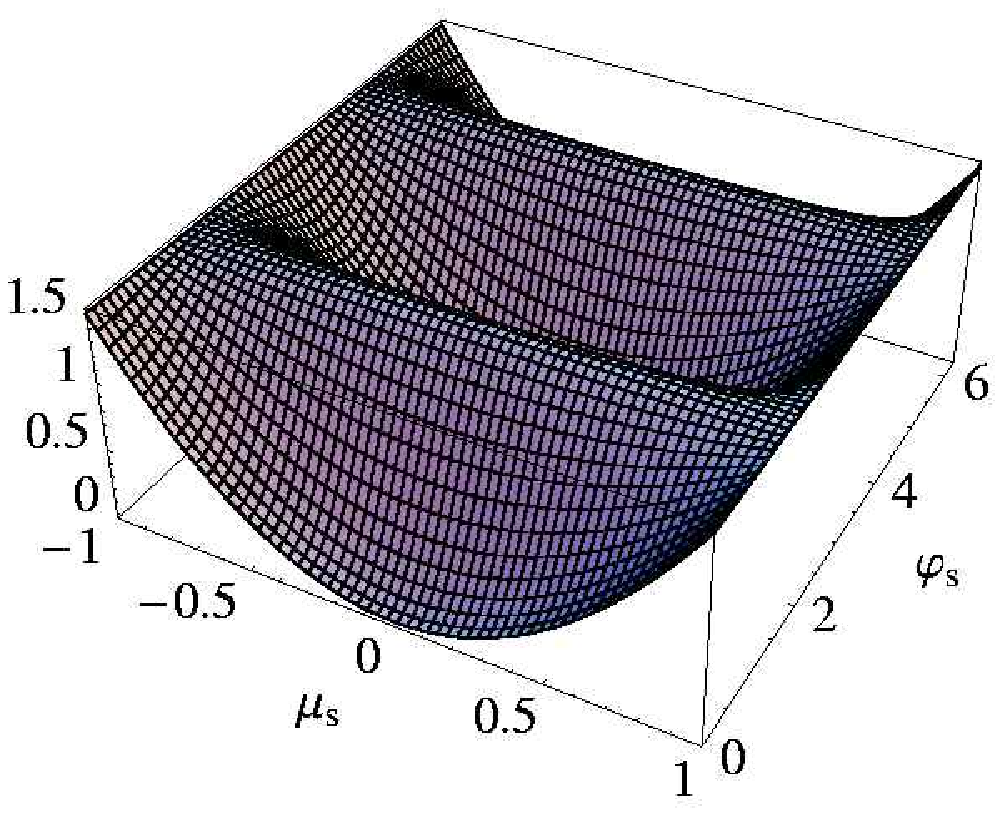}
\end{center}
\caption{Phase function $\Phi$ versus rest frame scattering angles (in
  polar coordinates, for a beam incident along the $z$-axis with
  Stokes parameters: (a) $Q=U=V=0$, (b) $Q/I=1, \;U=V=0$.}
\label{ch3:phaseplot}
\end{figure*}

For example, consider the case of an incident beam with
$\bm{n}_i=(0,0,1)$, and intensity polarization matrix with Stokes
parameters defined with respect to polarization basis vectors
(associated with the incident beam) $\bm{\epsilon}_1=\bm{e}_x,
\,\bm{\epsilon}_2=\bm{e}_y$:
\begin{eqnarray} \label{ch3:zbeam}
\bm{I}(\bm{n}) = \frac{1}{2} \left( \begin{array}{ccc} I+Q & U & 0 \\
  U & I-Q & 0 \\ 0 & 0 & 0
\end{array} \right) \ .
\end{eqnarray}
and let the scattered direction have the components (\ref{ch3:ns}).
Then we obtain the phase function \citep{1995ApJ...441..400C}:
\begin{eqnarray} \label{ch3:examplephase}
\Phi[\bm{n}_s,\bm{\phi}_i(\bm{n}_i)] &=& \frac{3}{4} \left[ 1+\mu_s^2
  - \;(1-\mu_s^2)\;\left(Q\cos 2\phi_s + U\sin 2\phi_s\right)/I
  \right] \ .
\end{eqnarray}
In Fig.~\ref{ch3:phaseplot} this function is compared for unpolarized
and completely polarized incident beams.  The polarization of the
incident beam destroys the azimuthal symmetry of the differential
cross section and phase function.

This completes our discussion of the generation of
polarization by classical Thomson scattering in the electron rest
frame. In the next section we use these results to construct
the photon Boltzmann equation for Thomson scattering.

\section{Kinetic equation in the Thomson limit\label{ch3:sec2}}

The evolution of the polarization matrix of the radiation field
due to Compton scattering is determined by the Boltzmann (or kinetic)
equation 
\begin{equation} \label{ch3:genboltz}
          p^{\alpha}\partial_{\alpha} f_{\mu\nu}=C_{\mu\nu} \ ,
\end{equation}
where $C_{\mu\nu}$ is the Compton scattering term.  In this section we
derive this scattering term for arbitrarily relativistic electrons and
polarized photons.  In fact, since the CMB photons have negligible
momentum in comparison to the electron rest mass, the SZE can be
calculated accurately with a simpler scattering term derived in the
Thomson limit, in which the electron recoil is ignored. However, we go
through the complete relativistic calculation in any case since there
are other applications in which the recoil effect cannot be ignored.

We do however ignore the effect of induced (or stimulated) scattering,
which is required for example to obtain the Kompaneets equation often
used to derive the thermal SZ distortion.  But the terms due to
induced scattering in the Kompaneets equation are negligible in the
case of cluster SZE, and in general in the unpolarized case it is
known that induced scattering is a negligible effect unless electron
energies are comparable to the electron rest mass. In any case a
rigorous derivation of the induced effects require a quantum treatment
\citep{2001A&A...379..664N}, which we have not developed here.
We also neglect electron polarization, assuming that
frequent Coulomb collisions destroy any spin alignments, and Pauli
blocking (which is irrelevant in the regimes of interest).

We now use the preceding results to derive the Boltzmann collision
term in the electron rest frame.  This is derived by the following
heuristic line of reasoning.  If we ignore polarization and assign a
scalar distribution function $f(\bm{p})$ to each photon, the
scattering rate is given by Eqn.~(\ref{ch3:fscatt2}) with the
cross-section for the transition from $\bm{ p}_i$ to $\bm{ p}_s$
replaced by its unpolarized form, which in the Thomson limit is
\begin{eqnarray}\label{ch3:thomxsecunpol}
  \frac{d\sigma}{d\Omega_s}=\frac{3\sigma_{\rm T}}{16\pi}
  \left[1+(\bm{n}_i\cdot\bm{n}_s)^2\right] \ .
\end{eqnarray}
We could then write the rate of change of phase space density by
subtracting from Eqn.~(\ref{ch3:fscatt2}) the rate of scattering out
of $\bm{ p}_s$.  That result is known as the \emph{master equation} or
Boltzmann equation for $f$ \citep{1987gady.book.....B,deGroot}:
\begin{eqnarray}\label{ch3:master00}
  \frac{d}{dt}f(\bm{ p})=n_e\int d\Omega_s\int
  d^3p_i\,\frac{d\sigma}{d\Omega_s}(\bm{ p}_s; \bm{ p}_i)\,f(\bm{
    p}_i) \left[\delta^3 (\bm{ p}_s-\bm{ p})- \delta^3(\bm{ p}_i-\bm{
      p})\right]\ .  \quad&&
\end{eqnarray}
The meaning of the master equation is that the rate of change of the
photon number in a given phase space element is given by summing over
all scatterings into and out of this element.  In this expression,
$\bm{p}_s$ is not a free variable, it is a function of the incident
photon momentum and scattering angles,
$\bm{p}_s=\bm{p}_s(\bm{p}_i,\bm{n}_s)$, determined by the scattering
kinematics.  In the Thomson limit,
$\vert\bm{p}_s\vert=\vert\bm{p}_i\vert$, so the scattered photon
momentum is simply given by
$\bm{p}_s(\bm{p}_i,\bm{n}_s)=\vert\bm{p}_i\vert\bm{n}_s$.  This allows
completion of the integral over the first delta function.

The first and second terms inside the square brackets correspond to
scatterings into and out of the beam (with momentum $\bm{p}$)
respectively, and are termed the \emph{gain} and \emph{loss} terms.
\begin{figure*}[bt]
  \centering
  \mbox{\subfigure[Gain]{\includegraphics[width=6.5cm]{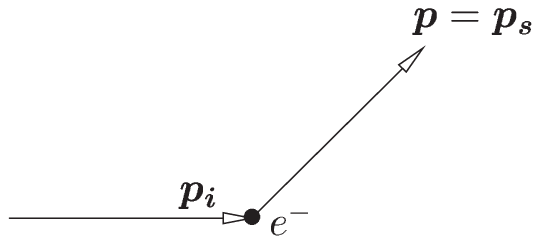}}
    \subfigure[Loss]{\includegraphics[width=5.5cm]{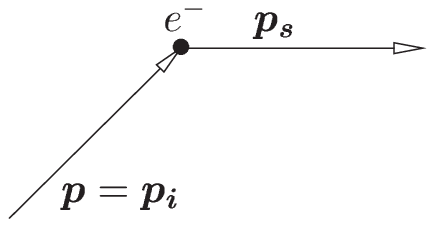}}}
\caption{ Gain and loss processes in kinetic/master equation for
  Thomson scattering.
\label{ch3:gainloss}}
\end{figure*}
The delta functions select the appropriate states, as indicated in
Fig.~\ref{ch3:gainloss}.  Eqn.~(\ref{ch3:master00}) is simply a
statement of photon number conservation combined with the rate of
scattering into the final momentum state (\ref{ch3:fscatt2}).

Now we wish to generalize this to the polarized case.  The
polarization tensor allows us to extend equation (\ref{ch3:master00})
to a general polarization state.  and write down the kinetic equation
for polarization corresponding to Eqn.~(\ref{ch3:fscatt1}).  Because
the transition rate is linear in $\bm{f}_i$ and $\bm{f}_s$ it is
possible to write the scattering rate for a linear superposition of
initial states to a linear superposition of final states.
Assuming linear superposition for incoherent light, we can write the
most general incident state as $f^{\alpha\beta}(\vec p_i,\vec v_e)$
and ask for the transition rate of each element of this matrix.
The transition rate is a linear transformation from
$f^{\alpha\beta}(\vec p_i,\vec v_e)$ to $f^{\mu\nu}(\vec p_s,\vec v_e
)$ and must therefore take the following form,
\begin{eqnarray}\label{ch3:fpolscatt_thomp}
  \frac{d}{dt}[f^{\mu\nu}(\vec p_s,\vec v_e)\,d^3p_s] = n_e
  \,\Phi^{\mu\nu}_{\alpha\beta}(\vec p_s,\vec v_e;\vec p_i,\vec v_e)\,
  f^{\alpha\beta}(\vec p_i,\vec v_e) \,d^3p_i\,d\Omega_s\ ,
\end{eqnarray}
with some matrix $\Phi^{\mu\nu}_{\alpha\beta}$ that we call the {\it
  polarization scattering tensor}.  It is convenient to write this as
$\Phi^{\mu\nu}_{\alpha\beta}(s;i)$, where the arguments $(i)$ and
$(s)$ are abbreviations for the pairs of 4-vectors $(\vec p_i,\vec
v_e)$ and $(\vec p_s,\vec v_e)$,

The polarization scattering tensor is effectively a $4\times4$ matrix
giving the transition rate between all possible initial and final
polarization states.
It follows from Eqn.~(\ref{ch3:fscatt1}) that in the Thomson limit the
polarization scattering tensor is given by
\begin{eqnarray} \label{ch3:Phithomp}
  \Phi^{\mu\nu}_{\alpha\beta}(s;i)= \frac{3\sigma_{\rm T}}{8\pi}
  P^{\mu}_{\ \ \alpha} (\vec{p}_s,\vec{v}_e) P^{\nu}_{\ \
    \beta}(\vec{p}_s,\vec{v}_e) \ .
\end{eqnarray}

Now we make the following ansatz for polarized analogue of the master
equation corresponding to Eqn.~(\ref{ch3:master00}):
\begin{eqnarray}\label{ch3:masterT}
  \frac{d}{dt}f^{\mu\nu}(\vec p,\vec v_e) = n_e
  \int\!d\Omega_s\int\!d^3p_i\, \Phi^{\alpha\beta}_{\gamma\delta}
  (\vec p_s,\vec v_e;\vec p_i,\vec v_e)\,f^{\gamma\delta}(\vec
  p_i,\vec v_e)\quad\quad&&\nonumber\\ \quad\quad\quad\quad
  \times\left[\delta^\mu_{\ \,\alpha}\delta^\nu_{\ \,\beta}\,
    \delta^3(\bm{ p}_s-\bm{ p})-g_{\alpha\beta}\, \phi^{\mu\nu}(\vec
    p,\vec v_e)\,\delta^3(\bm{ p}_i -\bm{ p})\right]\ .
\end{eqnarray}
This is the rest frame form of the scattering term $C^{\mu\nu}$ in
Eqn.~(\ref{ch3:genboltz}).  With this ansatz for the master equation,
it may be checked that for any two initial and scattered \emph{pure}
states $f^{\alpha\beta}(\vec p_i,\vec
v_e)=f_i\epsilon_i^\alpha(\epsilon_i^\beta)^\ast$ and $f^{\mu\nu}(\vec
p_s,\vec v_e)= f_s\epsilon_s^\mu(\epsilon_s^\nu)^\ast$,
Eqn.~(\ref{ch3:masterT}) reduces to Eqn.~(\ref{ch3:master00}) with the
Thomson cross section Eqn.~(\ref{ch3:thompxsec}).  Then since any
polarized beam can be written as some superposition of pure states, it
follows that Eqn.~(\ref{ch3:masterT}) is true for all polarization
states. This verifies that the ansatz (\ref{ch3:masterT}) is correct
in the Thomson limit.

The first term in the square brackets is exactly the gain term of
Eqn.~(\ref{ch3:fscatt1}).  The second term represents losses to any
final polarization state; the sum over polarizations is given by
$g_{\alpha\beta}$.  Each incident photon beam with polarization tensor
$\phi^{\mu\nu}(\vec p,\vec v_e)$ is lost by scattering implying that
the loss term must be proportional to $\phi^{\mu\nu}(\vec p,\vec
v_e)$.  Just as the loss term in Eqn.~(\ref{ch3:master00}) is
proportional to the same quantity occurring on the left-hand side of
the equation, the same is true here.  In fact this loss term is simply
the phase function multiplied by the incident beam, and is thus
proportional to the probability of a photon scattering from momentum
$\vec{p}_i$ to $\vec{p}_s$.  To see this, note that the loss term
contains the scalar obtained by contracting the projection tensors
which are orthogonal to the incident and scattered photons:
\begin{equation} \label{ch3:lossscalar}
  P^{\alpha\beta}(\vec{p}_i,\vec{v}_e)
  P_{\alpha\beta}(\vec{p}_s,\vec{v}_e) = 2 + \frac{2}{p_ip_s}
  \;\vec{p}_i\cdot\vec{p}_s +
  \frac{(\vec{p}_i\cdot\vec{p}_s)^2}{(p_ip_s)^2} \ .
\end{equation}
where $p_i\equiv -\vec{p}_i\cdot \vec{v}_e$, $p_s\equiv
-\vec{p}_s\cdot \vec{v}_e$.  In the rest frame of $\vec{v}_e$,
$p_i^{\mu}=p_i(1,\mbox{\boldmath $n$}_i)$,
$p_s^{\mu}=p_s(1,\mbox{\boldmath $n$}_s)$, giving
$\vec{p}_i\cdot\vec{p}_s = -p_ip_s(1-\mbox{\boldmath
  $n$}_i\cdot\mbox{\boldmath $n$}_s)$, and the loss term scalar has
the form
\begin{equation}
  P^{\alpha\beta}(\vec{p}_i,\vec{v}_e)
  P_{\alpha\beta}(\vec{p}_s,\vec{v}_e) = 1 + (\mbox{\boldmath
    $n$}_i\cdot\mbox{\boldmath $n$}_s)^2 \ ,
\end{equation}
which is the familiar angular dependence of the differential cross
section for Thomson scattering of unpolarized radiation.
The total loss term is thus simply proportional to the incident beam
multiplied by the total cross section (in this case, since we have
restricted to the Thomson limit, the Thomson cross section).

Note that the form of Eqn.~(\ref{ch3:masterT}) guarantees photon
number conservation (Compton scattering cannot change the overall
photon number):
\begin{equation}
\int d^3p \;g_{\mu\nu}\frac{df^{\mu\nu}(\vec{p},\vec{v}_e)}{dt}=0 \ .
\end{equation}

Integration over the delta functions yields the much simplified form
of the rest frame kinetic equation
\footnote{It is important to remember that on the left hand side, the
  derivative $d/dt$ stands for the operator $\partial/\partial t +
  \bm{n}\cdot\bm{\nabla}$, where $p^{\mu}=p(1,\bm{n})$ in a local
  Lorentz frame.}  :
\begin{eqnarray}\label{ch3:masterTsimp}
\frac{d}{dt}f^{\mu\nu}(\vec p,\vec v_e) = n_e \sigma_{\rm T} \Biggl[
  \frac{3}{2} \int\!\frac{d\Omega_i}{4\pi} \, P^{\mu}_{\ \ \alpha}
  (\vec{p},\vec v_e) P^{\nu}_{\ \ \beta}(\vec{p},\vec v_e)
  f^{\alpha\beta}(\vec p_i,\vec v_e) \Biggr.  - f^{\mu\nu}(\vec p,\vec
  v_e) \Biggr] \ .
\end{eqnarray}
This equation, in conjunction with the transformation to lab frame and
integration over electron velocities, discussed in the next sections,
is sufficient to compute all the Sunyaev-Zeldovich effects.
 
At this point we consider the form of Eqn.~(\ref{ch3:masterTsimp}) in
the case where the polarization tensors on the right hand side are
taken to be unpolarized. This will be used in Paper II in the
computation of the polarization generated by a single scattering of an
unpolarized radiation field.  In this case the integrand of the gain
term in Eqn.~(\ref{ch3:masterTsimp}) is proportional to the following
combination of projection tensors:
\begin{eqnarray} \label{ch3:gainprojcomb}
G^{\mu\nu}(\vec{p}_i,\vec{p}_s,\vec{v}_e) \equiv P^{\mu}_{\ \
  \alpha}(\vec{p}_s,\vec{v}_e) P^{\nu}_{\ \
  \beta}(\vec{p}_s,\vec{v}_e) P^{\alpha\beta}(\vec{p}_i,\vec{v}_e) \ .
\end{eqnarray}
In the electron rest frame, the spatial components of the integrand of the
gain term are given by
\begin{eqnarray} \label{ch3:simplegain}
G^{ij}(\vec{p}_i,\vec{p}_s,\vec{v}_e) &=& \left(\delta^i_{k} - n^i_s
n_{s,k}\right) \left(\delta^j_{l} - n^j_s n_{s,l}\right)
\left(\delta^{kl} - n^k_i n^l_i \right) \nonumber \\ &=&
(\delta^{ij}-n^i_i n^j_i) - n^i_s n_s^j[1+(\bm{n}_s\cdot\bm{n}_i)^2] +
(n^i_s n^j_s + n^j_s n^i_s) (\bm{n}_s\cdot\bm{n}_i) \ .
\quad\quad\quad
\end{eqnarray}
This form is convenient for the calculation of the
polarization generated by scattering of the CMB quadrupole, 
in the companion paper (and, for example, in \citep{2004PhRvD..70f3504P}).
%

We close this section with a demonstration that the rest frame form of
the kinetic equation, Eqn.~(\ref{ch3:masterTsimp}), yields the well
known results of \cite{Chandrasekhar1960} for the polarized radiative
transfer equations in the case of Thomson scattering from cold
(i.e. stationary) electrons in a slab geometry.  Since Chandrasekhar
used a different formalism based on transformations of the Stokes
parameters to derive his expressions, this is an important check of
the formalism we have developed. These equations also yield the
form of the scattering term for the polarization in the primary CMB
calculation. 

Consider a plane parallel atmosphere of cold electrons with uniform
density filling the half-space $z<0$, illuminated by an unpolarized
beam of monochromatic radiation incident along the normal to the plane
(we do not specify the boundary conditions here, since we are only
interested in deriving the form of the transfer equations and not
their solution). By symmetry the radiation field in this case is
clearly axisymmetric.  We work entirely in the electron rest frame,
where $I^{00}=I^{0i}=0$, and following \cite{Chandrasekhar1960} define
$\tau$ to be the optical depth from the surface $z=0$ along the
downward normal.
The Cartesian coordinates in the plane are $x$ and $y$, and The polar
and azimuthal angles about the $z$-axis are denoted $\theta$ and
$\varphi$ (and $\mu=\cos\theta$).

Then the kinetic equation for the lab frame intensity polarization
matrix $\bm{I}(\mu,\varphi,\tau)$ is given by
Eqn.~(\ref{ch3:masterTsimp}) (replacing distribution function tensors
by intensity tensors, which is trivially allowed here since the beam
is monochromatic):
 \begin{eqnarray}
 \mu \;\partial_{\tau} \bm{I}(\mu,\varphi,\tau) &=& - \frac{3}{2}
 \bm{P}(\bm{n}) \int_{-1}^{1} d\mu_s\; \int_0^{2\pi}
 \frac{d\varphi_s}{4\pi} \bm{I}(\mu_s,\varphi_s,\tau) \bm{P^T}(\bm{n})
 + \;\bm{I}(\mu,\varphi,\tau) \ . \nonumber \\
 \end{eqnarray}
 where
 $\bm{n}(\theta,\varphi)=(\cos\varphi\sin\theta,\sin\varphi\sin\theta,\cos\theta)$
 is the direction 3-vector of the beam.  Note that in this expression,
 following \cite{Chandrasekhar1960}, the gain and loss terms have
 picked up a minus sign since the direction of increasing optical
 depth (along the normal to the boundary $z=0$ directed into the
 half-space $z<0$) is defined to be opposite to the polar axis.

In contrast to the Stokes vector approach of \cite{Chandrasekhar1960},
the polarization matrix even in this azimuthally symmetric problem has
azimuthal dependence.  But we are free to exploit the azimuthal
symmetry here by choosing a convenient azimuth to evaluate the
projection factors, and then the result obtained for this azimuth may
be transformed into the other directions trivially. Choosing $\varphi
= 0$, we have $(n^x,n^y,n^z) = (\sqrt{1-\mu^2},0,\mu)$.  In the
$\varphi = 0$ direction, $I^{ij}$ has the form (since the $yx$ and
$yz$ cross-terms must vanish in order that the place of polarization
is parallel or perpendicular to the $y-z$ plane as required by
axisymmetry).
 \begin{equation}
 \bm{I}(\mu,0,\tau) = \left(\begin{array}{ccc} I_l \mu^2 & 0 & I_l \mu
   \sqrt{1-\mu^2} \\ 0 & I_r & 0 \\ I_l \mu \sqrt{1-\mu^2} & 0 & I_l
   (1-\mu^2)
 \end{array}\right) \ .
 \end{equation}

 Here $I_l(\mu,\tau)$ and $I_r(\mu,\tau)$ are the azimuth independent
 Stokes parameters, parallel and perpendicular respectively to the
 meridian plane, as defined in \cite{Chandrasekhar1960}.  The matrices
 inside the $d\varphi_s$ integrals range over all values of
 $\varphi_s$ though, so an expression for
 $I^{kl}(\mu_s,\varphi_s,\tau)$ is required.  By azimuthal symmetry,
 this is simply given by rotating $I^{kl}(\mu_s,0,\tau)$ through an
 angle $\varphi_s$ about the $\hat{z}$ axis (since under rotation
 polarization matrices transform according to the vector rotation of
 the electric field strength vectors):
 \begin{eqnarray}
 & &\bm{I}(\mu_s,\varphi_s,\tau) = \bm{R}(-\varphi_s\hat{z})
\bm{I}(\mu_s,0,\tau) \bm{R^T}(\varphi_s\hat{z}) \ .
 \end{eqnarray}
where $\bm{R}(\varphi_s\hat{z})$ is the matrix which rotates through
angle $\varphi_s$ about the $z$-axis.  
 \begin{eqnarray}
\bm{I}(\mu_s,\varphi_s,\tau)
 & &= \left[\begin{array}{ccc}
 I_l\mu_s^2\cos^2\varphi_s + I_r\sin^2\varphi_s &
 -I_l\mu_s^2\sin\varphi_s\cos\varphi_s+I_r\cos\varphi_s\sin\varphi_s & I_l
 \mu_s\sqrt{1-\mu_s^2}\cos\varphi_s \\
 -I_l\mu_s^2\cos\varphi_s\sin\varphi_s+I_r\cos\varphi_s\sin\varphi_s &
 I_l\mu_s^2\sin^2\varphi_s+I_r\cos^2\varphi_s &
 -I_l\mu_s\sqrt{1-\mu_s^2}\sin\varphi_s \\
 I_l\mu_s\sqrt{1-\mu_s^2}\cos\varphi_s & -I_l\mu_s\sqrt{1-\mu_s^2}\sin\varphi_s & I_l(1-\mu_s^2)
 \end{array}\right] \nonumber \\
\end{eqnarray}
 Evaluating the $I^{xx}$ component first, we obtain
 \begin{eqnarray}
 \mu \;\partial_{\tau} I^{xx}(\mu,0,\tau)
 &=& -\frac{3}{2}\int_{-1}^{1} d\mu_s\; \int_0^{2\pi} \frac{d\varphi_s}{4\pi} 
 S^{xx}(\mu,\mu_s,\varphi_s,\tau)  
 + \;I^{xx}(\mu,0,\tau) \ ,
 \end{eqnarray}
 where 
 \begin{eqnarray}
 S^{xx}(\mu,\mu_s,\varphi_s,\tau) &=& (\delta^x_{\;\;\;k}-n^x n_k)
 (\delta^x_{\;\;\;l}-n^x n_l) \;\;I^{kl}(\mu_s,\varphi_s,\tau) 
 \nonumber \\
 &=& \mu^4 I^{xx}(\mu_s,\varphi_s,\tau) 
 + \mu^2 (1-\mu^2) I^{zz}(\mu_s,\varphi_s,\tau) 
 - 2\mu(1-\mu^2)^{3/2} I^{xz}(\mu_s,\varphi_s,\tau) \ . \nonumber \\
 \end{eqnarray}
 Breaking this into three terms for clarity
 \begin{eqnarray}
 \mu \;\partial_{\tau} I^{xx}(\mu,0,\tau)
 &=& \mu^3 \;\partial_{\tau} I_l(\mu,\tau) \nonumber \\
 &=& -\frac{3}{2}\int_{-1}^{1} d\mu_s\; \int_0^{2\pi} \frac{d\varphi_s}{4\pi} 
 \;\; S^{xx}(\mu,\mu_s,\varphi_s,\tau)
 + \;I^{xx}(\mu,0,\tau) \nonumber \\
 &=& I_1(\mu,\tau) + I_2(\mu,\tau) + I_3(\mu,\tau) + \mu^2
 \;I_l(\mu,\tau) \ ,
 \end{eqnarray}
 where
 \begin{eqnarray}
 I_1(\mu,\tau) 
 &=& -\frac{3}{2}\mu^4 \int_{-1}^{1} d\mu_s\; \int_0^{2\pi} \frac{d\varphi_s}{4\pi} 
 \;\; \left( I_l(\mu_s,\tau)\mu_s^2\cos^2\varphi_s + I_r(\mu_s,\tau) \sin^2\varphi_s \right)
 \nonumber \\ 
 &=& -\frac{3}{8}\mu^4 \int_{-1}^{1} d\mu_s\; 
 \; \left( I_l(\mu_s,\tau)\mu_s^2 + I_r(\mu_s,\tau) \right)     \ ,
 \nonumber \\
 \nonumber \\
 I_2(\mu,\tau) 
 &=&- \frac{3}{2} \mu^2 (1-\mu^2) \int_{-1}^{1} d\mu_s\; \int_0^{2\pi} \frac{d\varphi_s}{4\pi} 
 \;\; I_l(\mu_s,\tau)(1-\mu_s^2) \nonumber \\
 \nonumber\\
 &=&- \frac{3}{4} \mu^2 (1-\mu^2) \int_{-1}^{1} d\mu_s\;
 I_l(\mu_s,\tau)(1-\mu_s^2) \ , \nonumber \\
 I_3(\mu,\tau) 
 &=& 3\mu(1-\mu^2)^{3/2} \int_{-1}^{1} d\mu_s\; \int_0^{2\pi} \frac{d\varphi_s}{4\pi} 
 \;\;\;I_l(\mu_s,\tau) 
 \mu_s\sqrt{1-\mu_s^2}\cos\varphi_s \nonumber\\
 &=& 0 \ .
 \end{eqnarray}
 Adding the three terms above and dividing by $\mu^2$ gives
 \begin{eqnarray} \label{ch_coupled1}
 \!\!\!\!\!\!\!\!\!\!\!\!\!\!\!\!\! \mu \;\partial_{\tau} I_l(\mu,\tau)
 &=& I_l(\mu,\tau)
 -\frac{3}{8} \int_{-1}^{1} d\mu_s\; 
 \; \biggl\{ I_l(\mu_s,\tau)\left[2(1-\mu_s^2)(1-\mu^2) + \mu^2\mu_s^2\right]
 + I_r(\mu_s,\tau) \mu^2 \biggr\} \ .
 \end{eqnarray}
The equation for $I_r$ is obtained similarly from the evolution of the
$I^{yy}$ component, 
 \begin{eqnarray}
 \mu \;\partial_{\tau} I^{yy}(\mu,0,\tau)
 &=& -\frac{3}{2}\int_{-1}^{1} d\mu_s\; \int_0^{2\pi} \frac{d\varphi_s}{4\pi} 
 S^{yy}(\mu,\mu_s,\varphi_s,\tau)  
 + \;I^{yy}(\mu,0,\tau) \ ,
 \end{eqnarray}
 where 
 \begin{eqnarray}
 S^{yy}(\mu,\mu_s,\varphi_s,\tau) &=& (\delta^y_{\;\;\;k}-n^y n_k)
 (\delta^y_{\;\;\;l}-n^y n_l) \;\;I^{kl}(\mu_s,\varphi_s,\tau)
 \nonumber \\
 &=& I^{yy}(\mu_s,\varphi_s,\tau) \ .
 \end{eqnarray}
 Therefore
 \begin{eqnarray} \label{ch_coupled2}
 \mu \;\partial_{\tau} I^{yy}(\mu,0,\tau)
 &=& \mu \;\partial_{\tau} I_r(\mu,\tau) \nonumber \\
 &=& I_r(\mu,0,\tau) -\frac{3}{2}\int_{-1}^{1} d\mu_s\; \int_0^{2\pi} \frac{d\varphi_s}{4\pi} 
 \;\; \left( I_l(\mu_s,\tau)\mu_s^2\sin^2\varphi_s + I_r(\mu_s,\tau) \cos^2\varphi_s \right)
 \nonumber \\
 &=& I_r(\mu,0,\tau) -\frac{3}{8}\int_{-1}^{1} d\mu_s\; 
 \biggl\{ I_l(\mu_s,\tau)\mu_s^2 + I_r(\mu_s,\tau) \biggr\} \ .
 \end{eqnarray}
Eqns.~(\ref{ch_coupled1}) and (\ref{ch_coupled2}) are
the coupled radiative transfer equations for the Stokes
parameters for Thomson scattering in slab geometry, first obtained by
Chandrasekhar \cite{Chandrasekhar1960}.


These equations also yield straightforwardly the form of the
polarization scattering term in the Boltzmann equation for the
evolution of the primary CMB anisotropies, in the form given (but not
derived) in \citep{2003moco.book.....D}, as follows.  In a frame in
which the  $\bm{k}$-mode considered is aligned with the local
$\bm{z}$-axis, and the observation direction is
$\bm{n}=(\sqrt{1-\mu^2},0,\mu)$, the form of the polarization matrix
of the CMB is:  
 \begin{equation}
 \bm{I}(\mu) = \frac{1}{2}\left(\begin{array}{ccc} (I-\Pi) \mu^2 & 0 & (I-\Pi) \mu
   \sqrt{1-\mu^2} \\ 0 & I+\Pi & 0 \\ (I-\Pi) \mu \sqrt{1-\mu^2} & 0 &
   (I-\Pi) (1-\mu^2)
 \end{array}\right) \ ,
 \end{equation}
where we have defined the total intensity $I=I_l+I_r$ and
total polarization intensity $\Pi=I_l-I_r$.
The polarization magnitude $\Pi$ is obvously independent
of $\varphi$ in this azimuthally symmetrical situation.
Using the previous equations for $I_l$ and $I_r$, we find
\begin{eqnarray}
\dot{\Pi} &=& \Pi - \frac{3}{8} (1-\mu^2) \int_{-1}^1
d\mu' \left[I_l(3\mu'^2-1)-I_l+I_r\right] \nonumber \\
&=& \Pi - \frac{1}{2}\left(1-P_2(\mu)\right) \int_{-1}^1
\frac{d\mu'}{2} \left[I P_2(\mu') + \Pi
  \left(P_2(\mu')-P_0(\mu')\right)\right] \ .
\end{eqnarray}
Following \citep{2003moco.book.....D}, we define
\begin{eqnarray}
\Delta_l \equiv (-i)^l \int_1^{-1} \frac{d\mu'}{2} I P_l(\mu') \ ,
\nonumber \\
\Delta_{Pl} \equiv (-i)^l \int_1^{-1} \frac{d\mu'}{2} \Pi P_l(\mu') \ . 
\end{eqnarray}
With $\Pi = \Delta_P$, we obtain the standard form of the CMB
polarization scattering term, 
\begin{eqnarray}
\frac{D}{D\tau}{\Delta}_P = -\dot{\tau_{\rm T}}
\left[-\Delta_P+\frac{1}{2}\left(1-P_2(\mu)\right)
  \left(\Delta_2+\Delta_{P2}+\Delta_{P0}\right) \right] \ .
\end{eqnarray}
(Note $\tau$ is now conformal time, $\tau_{\rm T}$ is Thomson optical
depth, $D/D\tau=\partial/\partial\tau+ik\mu$. The second term in the
total derivative comes from the other terms in the Liouville equation).

\section{Klein-Nishina scattering \label{ch3:sec3}}

Up to now we have worked in the Thomson limit.  In this section we
extend to the general case of Compton scattering with the full
Klein-Nishina form of the scattering cross section, and taking into
account the electron recoil.

For now we work still in the rest frame of the electron before
scattering (the ``initial'' electron).  To derive the polarization
tensor kinetic equation our starting point is the Klein-Nishina
differential cross-section \citep{Greiner1994}, in the initial
electron rest frame, for photons with 3-momentum $\bm{ p}_i$ and
polarization $\bm{\epsilon}_i$ to scatter into 3-momentum $\bm{ p}_s$
and polarization $\bm{\epsilon}_s$, generalized to allow for arbitrary
elliptical polarization
\footnote{This has been verified \citep{guthprivate} by
  A. H. Guth, by performing the explicit QED computation with complex
  polarization vectors using a computer algebra system.} 
\citep{1982PhRvD..26.2172S}:
\begin{eqnarray}\label{ch3:knxsec}
  \frac{d\sigma}{d\Omega_s}=\frac{3\sigma_{\rm T}}{8\pi}\left(
  \frac{p_s}{p_i}\right)^2\Biggl[\vert\bm{\epsilon}_i\cdot\bm{
      \epsilon}_s^{\,\ast}\vert^2
    +\frac{1}{4}\left(\frac{p_s}{p_i}+\frac{p_i}{p_s}-2\right)
    \left(1+\vert\bm{\epsilon}_i\cdot\bm{\epsilon}_s^{\,\ast}\vert^2
    -\vert\bm{\epsilon}_i\cdot\bm{\epsilon}_s\vert^2\right)\Biggr]\ .
\end{eqnarray}
where $p_i\equiv E(\bm{ p}_i)=\vert\bm{ p}_i\vert$ and $p_s\equiv
E(\bm{ p}_s)=\vert\bm{ p}_i\vert$.  We allow the polarization vector
to be complex in order to treat elliptical polarization; the
polarization vectors are normalized by
$\bm{\epsilon}_i\cdot\bm{\epsilon}_i^{\,\ast}=\bm{\epsilon}_s\cdot
\bm{\epsilon}_s^{\,\ast}=1$.  The factor $(1+\vert\bm{
  \epsilon}_i\cdot\bm{\epsilon}_s^{\,\ast}\vert^2-\vert\bm{
  \epsilon}_i\cdot\bm{\epsilon}_s\vert^2)$ is usually not given as it
reduces to unity for linearly polarized light, but we allow for light
of arbitrary polarization.  Eqn.~(\ref{ch3:knxsec}) assumes the
transverse gauge condition $\bm{\epsilon}_i \cdot\bm{
  p}_i=\bm{\epsilon}_s\cdot\bm{ p}_s=0$ and that the time component of
both polarization 4-vectors vanishes in the initial electron rest
frame. The factor $(p_s/p_i)^2$ in the cross section is a phase space
factor.

Conservation of energy-momentum relates the initial and final momenta
and scattering angle:
\begin{equation}\label{ch3:pcons}
  \frac{p_s}{p_i}=\left[1+\frac{p_i}{m_e}(1-\bm{ n}_i\cdot\bm{ n}_s)
    \right]^{-1}=1-\frac{p_s}{m_e}(1-\bm{ n}_i\cdot\bm{ n}_s)\ ,
\end{equation}
where $\bm{ n}_i$ and $\bm{ n}_s$ are unit three-vectors along the
spatial parts of the photon momenta $\bm{ p}_i=p_i\bm{ n}_i$ and $\bm{
  p}_s=p_s\bm{ n}_s$.  Note that for fixed directions,
$dp_s/dp_i=(p_s/p_i)^2$.

Equations (\ref{ch3:knxsec}) and (\ref{ch3:pcons}) both assume that
all quantities are given in the rest frame of the incident electron.
We note, incidentally, that the Klein-Nishina formula should be
symmetric under the interchange of the initial and final states, but
the gauge condition which was imposed to derive this form of the cross
section required that the incident and scattered photon polarization
basis 4-vectors be orthogonal to the \emph{incident} electron, and it
thus appears rather asymmetric. However, it is true that there is
nothing special about this choice of gauge.  In Appendix \ref{appc},
it is demonstrated that the cross section is manifestly symmetric
under interchange of the initial and final states.

The 4-velocity of the incident electron will from now on be denoted
$\vec{v}_i$, rather than $\vec{v}_e$ (since now we have to distinguish
between the incident and scattered electron momenta).  Now following
the procedure in the previous section, we write the transition rate of
the scattered polarization matrix $f^{\mu\nu}(\vec p_s,\vec v_i)$ as a
linear transformation of the incident matrix $f^{\alpha\beta}(\vec
p_i,\vec v_i)$, as in Eqns.~(\ref{ch3:fpolscatt_thomp}) and
(\ref{ch3:Phithomp}), except we pull the phase space factor out for
convenience:
\begin{eqnarray}\label{ch3:fpolscatt}
  \frac{d}{dt}[f^{\mu\nu}(\vec p_s,\vec v_i)\,d^3p_s] = n_e
  \,\Phi^{\mu\nu}_{\alpha\beta}(\vec p_s,\vec v_i;\vec p_i,\vec v_i)\,
  f^{\alpha\beta}(\vec p_i,\vec v_i)\left(\frac{p_s}{p_i}\right)^2
  d^3p_i\,d\Omega_s\ ,
\end{eqnarray}
The scattering polarization tensor $\Phi^{\mu\nu}_{\alpha\beta}$ for
the Klein-Nishina cross-section is then given by the modified form:
\begin{eqnarray}\label{ch3:Phikn}
  \Phi^{\mu\nu}_{\alpha\beta}(s;i)&=& \frac{3\sigma_{\rm
      T}}{8\pi}\Biggl\{P^{\mu\nu}_{\alpha\beta}
  (s)+\frac{1}{4}\left(\frac{p_s}{p_i}+\frac{p_i}{p_s}-2\right)
  \nonumber\\ &&
  \times\left[P^{\mu\nu}(s)P_{\alpha\beta}(i)+P^{\mu\nu}_{\alpha\beta}
    (s)-P^{\mu\nu}_{\beta\alpha}(s)\right]\Biggr\}\ ,
\end{eqnarray}
where
\begin{equation}\label{ch3:p2def}
  P^{\mu\nu}_{\alpha\beta}(s)\equiv P^\mu_{\ \,\alpha}(s)P^\nu_{\
    \,\beta}(s) \ ,
\end{equation}
and the arguments $(i)$ and $(s)$ are abbreviations for the pairs of
4-vectors $(\vec p_i,\vec v_i)$ and $(\vec p_s,\vec v_i)$, and
$p_s=-\vec v_i\cdot\vec p_s$ and $p_i=-\vec v_i\cdot\vec p_i$.
As a check, if we consider an unpolarized incident beam with
$\phi^{\alpha\beta}(i)=\frac{1}{2} P^{\alpha\beta}(i)$ and sum over
final polarizations, we get the usual spin-summed result for the
Klein-Nishina cross section:
\begin{eqnarray}\label{ch3:knunpol}
  &&\frac{1}{2}P_{\mu\nu}(\vec p_s,\vec v_i)\Phi^{\mu\nu}_
{\alpha\beta}(\vec p_s,\vec v_i;\vec p_i,\vec v_i)
P^{\alpha\beta}(\vec p_i,\vec v_i) =\frac{3\sigma_{\rm T}}{16\pi}
\left(\frac{p_s}{p_i}+\frac{p_i}{p_s}-\sin^2\theta\right)\ , \quad
\end{eqnarray}
where $\theta$ is the scattering angle, $\cos\theta=\bm{ n}_i
\cdot\bm{ n}_s$.

Now, as in the derivation in the Thomson limit case, we make an ansatz
for the rest frame kinetic equation:
\begin{eqnarray}\label{ch3:master1}
  \frac{d}{dt}f^{\mu\nu}(\vec p,\vec v_i) = n_e
  \int\!d\Omega_s\int\!d^3p_i\,
  \left(\frac{p_s}{p_i}\right)^2\Phi^{\alpha\beta}_{\gamma\delta}
  (\vec p_s,\vec v_i;\vec p_i,\vec v_i)\,f^{\gamma\delta}(\vec
  p_i,\vec v_i)&&\nonumber\\ \times\left[\delta^\mu_{\
      \,\alpha}\delta^\nu_{\ \,\beta}\,
    \delta^3(\bm{p}_s-\bm{p})-g_{\alpha\beta}\, \phi^{\mu\nu}(\vec
    p,\vec v_i)\,\delta^3(\bm{ p}_i -\bm{ p})\right]\ .
\end{eqnarray}
That this expression is correct can be verified as in the Thomson
limit case by substituting the matrices of pure states and checking
that Eqn.~(\ref{ch3:master00}) is regained with the the Klein-Nishina
cross section (\ref{ch3:knxsec}).  It is important to understand that,
as in Eqn.~(\ref{ch3:master00}), the scattered electron 3-momentum
$\bm{p}_s$ in the expression above is \emph{not} a free variable ---
it is determined by the scattering kinematics as $\bm{p}_s=p_s
\bm{n}_s$, where $p_s$ is given as a function of $\bm{p}_i$ and
$\bm{n}_s$ by Eqn.~(\ref{ch3:pcons}).

Having obtained the equation for polarized radiation transfer in the
rest frame of the scattering electron, we now consider the general
case of scattering from a distribution of electrons with varying
velocity.  To obtain this, first it is necessary to transform the
kinetic equation to a common lab frame.  
Henceforth, components of 4-vectors in the rest frame of the initial
electron are denoted with primes, and those in the lab frame without
primes. The incident particles are denoted with a subscript $i$, and
the scattered particle with a subscript $s$.  Photon momenta are
denoted by $p$, and electron momenta by $q$. As usual, quantities with
arrows are 4-vectors, and boldface quantities are 3-vectors.  The
4-momenta satisfy the mass shell conditions $\vec{p}_i\cdot\vec{p}_i =
\vec{p}_s\cdot\vec{p}_s = 0$, $\vec{q}_i\cdot\vec{q}_i =
\vec{q}_s\cdot\vec{q}_s = -m_e^2$.  The electron energies are given by
$E(\mbox{\boldmath $q$}) = \sqrt{m_e^2 + |\mbox{\boldmath $q$}|^2}$.

In lab frame, the initial and scattering electron 4-momenta are
written in terms of the lab frame electron 3-velocities as follows:
\begin{eqnarray}
\vec{q}_i &=& m_e \vec{v}_i = \gamma_i m_e(1,\bm{v}_i) \ , \quad
\gamma_i\equiv{1\over\sqrt{1-v_i^2}} \nonumber \\ \vec{q}_s &=& m_e
\vec{v}_s = \gamma_s m_e(1,\bm{v}_s) \ , \quad
\gamma_s\equiv{1\over\sqrt{1-v_s^2}} \ .
\end{eqnarray}

The Lorentz transformation into the rest frame of the initial electron
is given by the matrix:
\begin{eqnarray}
\Lambda^0_0 &=& \gamma_i, \quad \Lambda^0_i = -\gamma_i
\left[\bm{v}_i\right]_i \, \;\; \Lambda^i_j = \delta^i_j +
(\gamma_i-1) \frac{\left[\bm{v}_i\right]_i
  \left[\bm{v}_i\right]_j}{v_i^2} \ .
\end{eqnarray}
This yields the transformation of the photon direction vector between
frames:
\begin{equation} \label{ch3:nvectrans}
\bm{n}'_i = \left[\gamma_i(1-\bm{n}_i\cdot\bm{v}_i)\right]^{-1}
\left[\bm{n}_i+\frac{\gamma^2_i}{\gamma_i^2+1}\bm{v}_i(\bm{n}_i\cdot\bm{v}_i)-\gamma_i\bm{v}_i
  \right] \ .
\end{equation}
The relationship between the rest and lab frame momenta of the
incident and scattered photons is
 \begin{eqnarray}
 p_s^{\prime} &=& - \vec{p}_s\cdot\vec{q}_i/m_e =
 \frac{1}{m_e}\left[p_s q_i - \mbox{\boldmath
     $p$}_s\cdot\mbox{\boldmath $q$}_i\right] \ , \nonumber \\
 p_i^{\prime} &=& - \vec{p}_i\cdot\vec{q}_i/m_e =
 \frac{1}{m_e}\left[p_i q_i - \mbox{\boldmath
     $p$}_i\cdot\mbox{\boldmath $q$}_i\right] \ .
  \end{eqnarray}
Or in terms of the $\gamma_i$ factor,
\begin{eqnarray}\label{ch3:a5}
\frac{p'_i}{p_i} &=& \gamma_i\left(1-\bm{n}_i\cdot\bm{v}_i\right) \ ,
\nonumber \\ \frac{p'_s}{p_s} &=&
\gamma_i\left(1-\bm{n}_s\cdot\bm{v}_i\right) \ .
\end{eqnarray}

The scattered and incident photon energies in the rest frame are
related by the familiar Compton scattering formula:
\begin{equation}\label{ch3:restcompt}
\frac{p'_s}{p'_i} =
\frac{1}{1+\left(p'_i/m_e\right)\left[1-\bm{n}'_i\cdot\bm{n}'_s\right]}
\ .
\end{equation}
The lab frame version of this is
\begin{equation} \label{ch3:labcompt}
 p_s = \frac{p_i(1-\mbox{\boldmath $n$}_i\cdot\mbox{\boldmath
     $v$}_i)}{1 - \mbox{\boldmath $n$}_s\cdot\mbox{\boldmath $v$}_i +
   \left(p_i/\gamma_i m_e\right)(1-\bm{n}_i\cdot\bm{n}_s)} \ .
\end{equation}

We are now in a position to Lorentz transform Eqn.~(\ref{ch3:master1})
to a lab frame in which the electrons have 3-velocity $\bm{v}_i$.
Four quantities need to be transformed: $n_e'$, $dt'$, the 3-vector
$\bm{p}'$ of the beam on the left hand side of the master equation,
and $f^{\mu\nu}(\vec p,\vec v_i)$ itself because of its dependence on
$\vec v_i$.
Let the four-vector $\vec p$ have spatial components $p\bm{n}$ in the
lab frame and $p' \bm{n}'$ in the electron rest frame.  The
transformation laws of $p$ and $\bm{n}$ have already been derived:
\begin{eqnarray} \label{ch3:ptransf}
  p'&=&\gamma_i p(1-\bm{n}\cdot\bm{v}_i)\ ,\nonumber\\ \bm{n}' &=&
  \left[\gamma_i(1-\bm{n}\cdot\bm{v}_i)\right]^{-1}
  \left[\bm{n}+\frac{\gamma^2_i}{\gamma_i^2+1}\bm{v}_i(\bm{n}\cdot\bm{v}_i)-\gamma_i\bm{v}_i
    \right] \ .
\end{eqnarray}
The Lorentz transformation of the electron density is
\begin{eqnarray}\label{ltrans1}
  n_e'=\gamma_i^{-1}n_e \ .
\end{eqnarray}
This is simply due to the Lorentz contraction of the volume element.

The Lorentz transformation of the time element is more subtle; $dt$
transforms like $p$, not like $n_e$:
\begin{eqnarray}\label{ltrans2}
dt'=\gamma_i(1-\bm{n}\cdot\bm{v}_i)\, dt\ .
\end{eqnarray}
From a mathematical point of view, this is because the transport
operator on the left-hand side of the Boltzmann equation is actually
the directional derivative $d/d\lambda\equiv(dx^\mu/d\lambda)\partial
/\partial x^\mu+(dp^\mu/d\lambda)\partial/\partial p^\mu=p^\mu
\partial/\partial x^\mu$ (in flat space).
In a local Lorentz frame, $d/d\lambda=pd/dt$ where
$d/dt\equiv\partial/\partial t+\bm{n} \cdot\bm{\nabla}$. Physically,
the transformation of $dt$ arises due to the enhancement of the rate
of scattering of photons from electrons which are approaching compared
to that from electrons which are receding, which is due to the
dependence of the flux of photons incident on the electrons on their
relative velocity.  Thus we refer to the factor
$\gamma_i(1-\bm{n}\cdot\bm{v}_i)$ which appears in
Eqn.~(\ref{ltrans2}) as the \emph{flux factor}. This factor is crucial
in the derivation of the SZ effects!

It follows from the transformation of the time element and the
invariance of the trace of the distribution function tensor, that the
left-hand sides of the Boltzmann equation in rest and lab frame are
related by
\footnote{Recall that $\vec{p}$ is a Lorentz covariant 4-vector in
  these expressions.}
\begin{equation} \label{ch3:ftracetrans1}
\frac{df'(\vec{p})}{dt'} = \frac{1}{\gamma_i(1-\bm{n}\cdot\bm{v}_i)}
\frac{df(\vec{p})}{dt} \ ,
\end{equation}
or alternatively
\begin{equation} \label{ch3:ftracetrans2}
\frac{df(\vec{p})}{dt} = \frac{1}{\gamma_i(1+{\bm{n}'}\cdot\bm{v}_i)}
\frac{df'(\vec{p})}{dt'} \ .
\end{equation}
(The photon occupation numbers in rest and lab frames also satisfy
these equations, of course).

We now have all the ingredients needed to transform equation
(\ref{ch3:master1}) to the rest frame of an observer with 4-velocity
$\vec v_l$ (the subscript standing for ``lab'').  The transformation
of the polarization tensors from lab to rest frame was derived in \S
\ref{ch1:sec4}:
\begin{eqnarray} 
 f^{\mu^{\prime}\nu^{\prime}}(\vec{p},\vec{v}_i) =
 P^{\mu^{\prime}}_{\;\;\;\mu}(\vec{p},\vec{v}_i)
 P^{\nu^{\prime}}_{\;\;\;\nu}(\vec{p},\vec{v}_i)
 f^{\mu\nu}(\vec{p},\vec{v}_l) \ .
\end{eqnarray}
It is convenient to insert the tensors which project into the electron
rest frame into the scattering tensor, by redefining the tensor
$P^{\mu\nu}_{\alpha\beta}(s)$ which appears in Eqn.~(\ref{ch3:Phikn})
as
\begin{equation}\label{ch3:p3def}
  P^{\mu\nu}_{\alpha\beta}(s;i)\equiv P^\mu_{\ \,\gamma}(s)P^\nu_{\
    \,\delta}(s)P^\gamma_{\ \,\alpha}(i) P^\delta_{\ \,\beta}(i) \ .
\end{equation}
where the arguments $(i)$ and $(s)$ are abbreviations for the pairs of
4-vectors $(\vec p_i,\vec v_i)$ and $(\vec p_s,\vec v_i)$, and
$p_s=-\vec v_i\cdot\vec p_s$ and $p_i=-\vec v_i\cdot\vec p_i$.

Finally, we can generalize the electron density to a distribution of
electrons, $n_e=\int d^3q_i\,g_e(\bm{ q}_i)$ where $\bm{ q}_i$ is the
electron 3-momentum, and $g_e$ is the scalar phase space distribution
function for the electrons. Putting everything together, we obtain
\begin{eqnarray}\label{ch3:master22}
  p\frac{d}{dt}f^{\kappa\lambda}(\vec p,\vec v_l) =P^\kappa_{\
    \,\mu}(\vec p,\vec v_l)P^\lambda_{\ \,\nu} (\vec p,\vec v_l)\int
  \frac{d^3q_i}{E(\bm{ q}_i)}\, g_e(\bm{ q}_i\,)\, m_e p'
  \quad\quad\quad\quad&&\nonumber\\ \times\int d\Omega'_s\int
  d^3p'_i\,
  \left(\frac{p'_s}{p'_i}\right)^2\Phi^{\alpha\beta}_{\gamma\delta}
  (\vec p_s,\vec v_i;\vec p_i,\vec v_i)\,f^{\gamma\delta}(\vec
  p_i,\vec v_l)\quad\quad&&\nonumber\\ \times\left[\delta^\mu_{\
      \,\alpha}\delta^\nu_{\ \,\beta}\, \delta^3(\bm{ p}'_s-\bm{
      p}')-g_{\alpha\beta}\, \phi^{\mu\nu}(\vec p_i,\vec
    v_l)\,\delta^3 (\bm{ p}'_i-\bm{ p}')\right]\ ,
\end{eqnarray}
where $E(\bm{ q}_i)=\gamma_i m_e$.  Primes denote components in the
rest frame of $\vec v_i=\vec q_i/m_e$, e.g.  $p'\equiv-\vec
v_i\cdot\vec p$.  The flux factor is contained in the $p'$ factor in
the first line on the right hand side (the Lorentz invariant measure
$d^3q_i/E(\bm{q}_i)$ is pulled out to facilitate the derivation of the
covariant form to follow).  Note that the projection operators which
project $f^{\gamma\delta}$ into the rest frame of $\vec v_i$ are
already present in $\Phi^{\alpha\beta}_{\gamma\delta}$.  Similarly, it
does not matter whether $\phi^{\mu\nu}$ is evaluated in the rest frame
of $\vec v_i$ or $\vec v_l$, because of the projection operators in
front of the integral.  It follows that we may, without loss of
generality, drop the 4-velocity argument from $f^{\mu\nu}$ and
$\phi^{\mu\nu}$ provided it is understood that the final results must
always be projected into the physical polarization space of the
observer.

Eqn.~(\ref{ch3:master22}) looks complicated and is not the most
convenient form for computation.  The delta functions can be
integrated over resulting in a simpler expression.  This requires the
Jacobian of $(\bm{ p}'_i,\bm{ n}'_s)$ and $(\bm{ p}'_s,\bm{ n}'_i)$,
which follows from Eqns.~(\ref{ch3:restcompt}) and
~(\ref{ch3:ptransf}), yielding:
\begin{equation}\label{ch3:measure}
  d\Omega'_s\,d^3p'_i\left(\frac{p'_s}{p'_i}\right)^2=
  d\Omega'_i\,d^3p'_s\left(\frac{p'_i}{p'_s}\right)^2\ .
\end{equation}
The Boltzmann equation now becomes
\begin{eqnarray}\label{ch3:master3}
  \frac{d}{dt}f^{\mu\nu}(\vec p\,)=\int d^3q_i \,g_e(\bm{
    q}_i)\,(1-\bm{ n}\cdot\bm{ v}_i)
  \quad\quad\quad\quad\quad\quad\quad\quad\quad\quad&&\nonumber\\
  \times\Biggl[\int d\Omega'_i\left(\frac{p_i'}{p'}\right)^2
    \Phi^{\mu\nu}_{\alpha\beta}(\vec p,\vec v_i;\vec p_i, \vec
    v_i)\,f^{\alpha\beta}(\vec p_i)\quad\quad\quad\quad
    \quad&&\nonumber\\ -\int
    d\Omega'_s\left(\frac{p'_s}{p'}\right)^2\phi^{\mu\nu} (\vec
    p\,)g_{\alpha\beta}\Phi^{\alpha\beta}_{\gamma\delta} (\vec
    p_s,\vec v_i;\vec p,\vec v_i)f^{\gamma\delta} (\vec p\,)\Biggr]\
  .&&\
\end{eqnarray}
This form is convenient for both analytic and Monte Carlo
calculations. The flux factor is explicit in the integration over the
electron momenta. (Recall that $f^{\mu\nu}$ must be projected into the
observer frame at the end of the calculation.)

The covariant kinetic equation for Compton scattering was derived for
unpolarized photons by \cite{1995ApJ...439..503D}.  Their expression
for the time evolution of the photon phase space distribution function
has the form (Eqn.~(2.3) of \cite{1995ApJ...439..503D}):
\begin{eqnarray}\label{ch3:master4}
p_1\frac{d}{dt}f(\vec p_1) = \int\frac{d^3q_1}{E(\bm{
    q}_1)}\int\frac{d^3q_2}{E(\bm{ q}_2)} \int\frac{d^3p_2}{E(\bm{
    p}_2)}\, \vert M \vert^2 \;\delta^4(\vec p_1+\vec q_1-\vec
p_2-\vec q_2) \quad\quad\quad\quad &&\nonumber \\ \Biggl[f(\vec
  p_2)(1+f(\vec p_1))g_e(\bm{ q}_2) -f(\vec p_1)(1+f(\vec
  p_2))g_e(\bm{ q}_1)\Biggr]. \;\;\;\quad\quad &&
\end{eqnarray}
where $|M|^2$ is the squared matrix element for unpolarized Compton
scattering. Note that this contains stimulated emission factors, which
we ignore in our treatment of the polarized case.  The integration
measures are Lorentz invariant: $\int d^3q_1/ E(\bm{ q}_1)=\int
d^4q_1\delta[\frac{1}{2}(\vec q_1\cdot\vec q_1 +m_e^2)]$.  $E(\bm{
  q})\equiv q^0$ is set by the mass shell condition.
We complete our discussion of the Boltzmann equation by checking that
the lab frame kinetic equation can be recast in such a manifestly
covariant form.  We will need the following identities:
\begin{eqnarray}\label{ch3:ident}
  \int\frac{d^3q_1}{E(\bm{ q}_1)}\int\frac{d^3q_2}{E(\bm{ q}_2)}
  \int\frac{d^3p_2}{E(\bm{ p}_2)}\,\delta^4(\vec p_1+ \vec q_1-\vec
  p_2-\vec q_2)\quad&&\nonumber\\ =\int\frac{d^3q_1}{E(\bm{
      q}_1)}\int\frac{d\Omega_2\,p_2^2} {(-\vec p_2\cdot\vec
    q_2)}=\int\frac{d^3q_2}{E(\bm{ q}_2)}
  \int\frac{d\Omega_2\,p_2^2}{(-\vec p_2\cdot\vec q_1)}\ .  \quad&&
\end{eqnarray}
These follow from writing the time part of the delta function as
$\delta[E(\bm{ p}_1)+E(\bm{ q}_1)-E(\bm{ p}_2)-E(\bm{ p}_1+\bm{
    q}_1-p_2\bm{ n}_2)]$ or $\delta[E(\bm{ p}_1) +E(p_2\bm{ n}_2+\bm{
    q}_2-\bm{ p}_1)-E(\bm{ p}_2)-E(\bm{ q}_2)]$.  To get the forms
that we finally need, we replace the denominators of the angular
integrals using the identities $\vec p_1\cdot\vec q_1=\vec
p_2\cdot\vec q_2$ and $\vec p_1\cdot\vec q_2=\vec p_2 \cdot\vec q_1$,
which follow from conservation of total 4-momentum.

With these identities, the Boltzmann equation finally takes a
manifestly covariant form,
\begin{eqnarray}\label{ch3:master4}
  p_1\frac{d}{dt}f^{\mu\nu}(\vec p_1\,) =\int\frac{d^3q_1}{E(\bm{
      q}_1)}\int\frac{d^3q_2}{E(\bm{ q}_2)} \int\frac{d^3p_2}{E(\bm{
      p}_2)}\,\delta^4(\vec p_1+ \vec q_1-\vec p_2-\vec q_2)
  \quad\quad\quad\quad &&\nonumber\\
  \times\Biggl[\Phi^{\mu\nu}_{\gamma\delta}(\vec p_1,\vec v_2; \vec
    p_2,\vec v_2)f^{\gamma\delta}(\vec p_2)g_e(\bm{ q}_2)
    -\phi^{\mu\nu}(\vec p_1)g_{\alpha\beta}\Phi^{\alpha\beta}_
    {\gamma\delta}(\vec p_2,\vec v_1;\vec p_1,\vec
    v_1)f^{\gamma\delta} (\vec p_1)g_e(\bm{ q}_1)\Biggr]. \;\;\; &&
\end{eqnarray}
The integration measures are Lorentz invariant: $\int d^3q_1/ E(\bm{
  q}_1)=\int d^4q_1\delta[\frac{1}{2}(\vec q_1\cdot\vec q_1 +m_e^2)]$.
It may be checked that working backwards from this equation,
integration over the 4-dimensional delta function yields the rest
frame form of the master equation Eqn.~(\ref{ch3:master22}).

\appendix

\section{Symmetry of Klein-Nishina Matrix Element \label{appc}}

It is apparent that the square of the invariant amplitude for the
Klein-Nishina formula should be symmetric under the interchange of
the initial and final states, but it is written in a way that is
very asymmetric. Here we show that it is possible to write the K-N
invariant matrix element in a way that is manifestly symmetric
between the initial and final states. 
\footnote{This appendix is based on a private communication from
  A. H. Guth \citep{guthprivate}.}

\def\openup{\afterassignment\xopenup\dimen0=}
\def\xopenup{\advance\lineskip\dimen0
\advance\baselineskip\dimen0 \advance\lineskiplimit\dimen0}
\def\eqalign#1{\null\,\vcenter{\openup 4pt \mathsurround=0pt
\ialign{\strut\hfil$\displaystyle{##}$&$\displaystyle{{}##}$\hfil
\crcr#1\crcr}}\,}
\def\line{\hbox to\hsize}

The matrix element for Compton scattering is usually written in an
asymmetric way. One can call the initial and final electron
4-momenta $\vec q_i$ and $\vec q_f$, and the initial and final photon
4-momenta $\vec p_i$ and $\vec p_f$.  Using conventions for which the
Lorentz dot-product is defined as time minus space, one defines
$p_1=\vec p_i \cdot \vec q_i$ and $p_2=\vec p_f \cdot \vec q_i$.  That is, in the
initial rest frame of the electron, $p_1$ and $p_2$ are the
initial and final photon energies, multiplied by $m_e$, the mass
of an electron.  One also defines polarization vectors for the
photons to have zero time-components in this frame, so $\vec\epsilon_i
\cdot \vec q_i=0$, $\vec\epsilon_f \cdot \vec q_i=0$.  The differential cross
section is then 
\footnote{Note that in this Appendix we use the standard Klein-Nishina
  cross section for linearly polarized photons, not the general
formula in Eqn.~(\ref{ch3:knxsec}) which is valid for circularly
polarized photons too \citep{1982PhRvD..26.2172S}.}
  $${d \sigma \over d \Omega} = {\alpha^2 \over 4 m_e^2}
     \,\left(p_2 \over p_1\right)^2 \, M^2 \ , \eqno({\rm A}.1)$$
where
  $$M^2= {p_2 \over p_1} + {p_1 \over p_2} - 2 + 4 (\vec\epsilon_i
     \cdot \vec\epsilon_f)^2 \ .\eqno({\rm A}.2)$$
The invariant matrix element $M^2$ should be symmetric under the
interchange of initial and final states, $i \leftrightarrow f$,
but it does not look that way.  However, it really is.
One might think that the
non-invariance of the $\vec\epsilon_i \cdot \vec\epsilon_f$ term is
perhaps canceled by the noninvariance of the rest, but it turns
out that it is much simpler than that.  Each part is separately
symmetric. To see this, note that conservation of 4-momentum implies that
  $$\vec p_i+\vec q_i=\vec p_f+\vec q_f \ .\eqno({\rm A}.3)$$
Squaring both sides, and using the fact that $\vec q_i^2=\vec q_f^2$ and
$\vec p_i^2=\vec p_f^2$, one has immediately that
  $$\vec q_i \cdot \vec p_i=\vec q_f \cdot \vec p_f \ ,\eqno({\rm A}.4)$$
so in fact $p_1$ is invariant under $i \leftrightarrow f$.
Similarly conservation of 4-momentum implies that
  $$\vec q_i-\vec p_f=\vec q_f-\vec p_i \ ,\eqno({\rm A}.5)$$
and squaring implies that
  $$\vec q_i \cdot \vec p_f=\vec q_f \cdot \vec p_i \ ,\eqno({\rm A}.6)$$
so $p_2$ is invariant under $i \leftrightarrow f$.

The only remaining problem is the $(\vec\epsilon_i \cdot
\vec\epsilon_f)^2$ term, which is not manifestly invariant, since the
$\vec\epsilon$'s were both defined to have vanishing 4th components in
the initial rest frame of the electron, so $\vec\epsilon_i \cdot
\vec q_i=0$, $\vec\epsilon_f \cdot \vec q_i=0$.  One can use an arbitrary gauge
for the polarization vectors, however, if one explicitly
constructs the gauge transformation satisfying $\vec\epsilon \cdot
\vec q_i=0$ before calculating the dot product.  That is, if
$\vec\epsilon_i$ does not satisfy $\vec\epsilon_i \cdot \vec q_i=0$, then one
constructs
  $$\vec\epsilon_i' = \vec\epsilon_i - {\vec\epsilon_i \cdot \vec q_i \over \vec p_i
     \cdot \vec q_i} \, \vec p_i \eqno({\rm A}.7)$$
and
  $$\vec\epsilon_f' = \vec\epsilon_f - {\vec\epsilon_f \cdot \vec q_i \over \vec p_f
     \cdot \vec q_i} \, \vec p_f \ ,\eqno({\rm A}.8)$$
so $\vec\epsilon_i' \cdot \vec q_i = \vec\epsilon_f' \cdot \vec q_i = 0$.
To continue, it is useful to define a more compact notation.  Let
  $$A_{\alpha \beta} \equiv \vec\epsilon_\alpha \cdot \vec q_\beta \eqno({\rm A}.9)$$
  $$B_{\alpha \beta} \equiv \vec\epsilon_\alpha \cdot \vec p_\beta \ .\eqno({\rm A}.10)$$
where $\alpha$ and $\beta$ can be either $i$ or $f$.  Each
polarization vector is orthogonal to its corresponding momentum,
so $B_{ii} = B_{ff} = 0$.  In this notation Eqns.~(A.7) and (A.8)
become
  $$\vec\epsilon_i' = \vec\epsilon_i - {A_{ii} \over p_1} \, \vec p_i \eqno({\rm A}.11)$$
and
  $$\vec\epsilon_f' = \vec\epsilon_f - {A_{fi} \over p_2} \, \vec p_f \ .\eqno({\rm A}.12)$$
So, for polarization vectors $\vec\epsilon_i$ and $\vec\epsilon_f$ written
in an arbitrary gauge, the equation for $M^2$ must be written by
replacing $\vec\epsilon_i \cdot \vec\epsilon_f$ with
  $$\vec\epsilon_i' \cdot \vec\epsilon_f' = \vec\epsilon_i \cdot \vec\epsilon_f -
     {A_{fi} B_{if} \over p_2} - {A_{ii} B_{fi} \over p_1} +
     {A_{ii} A_{fi} \over p_1 p_2} \vec p_i \cdot \vec p_f \ .\eqno({\rm A}.13)$$
To proceed, we want to use some identities that follow from
energy-momentum conservation.  Dotting both sides of Eqn.~(A.3) with
$\vec\epsilon_i$, one finds
  $$A_{ii}=B_{if}+A_{if} \ ,\eqno({\rm A}.14)$$
and dotting both sides with $\vec\epsilon_f$ (and reversing the sides
of the equation) gives
  $$A_{ff}=B_{fi}+A_{fi} \ .\eqno({\rm A}.15)$$
Since one has 6 dot products --- $A_{ii}$, $A_{if}$, $A_{fi}$,
$A_{ff}$, $B_{if}$, and $B_{fi}$ --- and two constraints
(Eqns.~(A.14) and (A.15)), one can eliminate two of the dot products
from all expressions.  The simplest result seems to arise from
eliminating the $B$'s.  One also needs to simplify $\vec p_i \cdot \vec p_f$,
which can be done by dotting Eqn.~(A.3) with $\vec p_f$:
  $$p_2 + \vec p_i \cdot \vec p_f = p_1 \ ,$$
so
  $$\vec p_i \cdot \vec p_f = p_1 - p_2 \ .\eqno({\rm A}.16)$$
Finally, substituting into Eqn.~(A.13),
  $$\eqalign{\vec\epsilon_i' \cdot \vec\epsilon_f' &= \vec\epsilon_i \cdot
     \vec\epsilon_f - {A_{fi} (A_{ii} - A_{if}) \over p_2} - {A_{ii}
     (A_{ff}-A_{fi}) \over p_1} + {A_{ii} A_{fi} \over p_1 p_2}
     (p_1 - p_2) \cr
  &= \vec\epsilon_i \cdot \vec\epsilon_f + {A_{fi} A_{if} \over p_2} -
     {A_{ii} A_{ff} \over p_1} \ .\cr}\eqno({\rm A}.17)$$
Written in this form, the result is manifestly symmetric under the
$i \leftrightarrow f$ interchange, and it is valid for
polarization vectors $\vec\epsilon_i$ and $\vec\epsilon_f$ written in an
arbitrary gauge.  One can check that the expression vanishes if
$\vec\epsilon_i$ is replaced by $\vec p_i$, or if $\vec\epsilon_f$ is replaced
by $\vec p_f$.

\bibliographystyle{apj} \bibliography{main,apj-jour,mn-jour}

\end{document}